\documentclass[prd,tightenlines,floatfix,showpacs,preprintnumbers,nofootinbib,eqsecnum,superscriptaddress,twocolumn]{revtex4-1}

\usepackage{color}      
\usepackage{graphicx}    
\usepackage{amsbsy}     
\usepackage{amsfonts}       
\usepackage{amsmath}        
\usepackage{amssymb}        
\usepackage{xcolor}
\usepackage{wasysym}
\usepackage{multirow}
\usepackage{mciteplus}
\usepackage{psfrag}
\usepackage{slashed}
\usepackage{cancel}
\usepackage{graphicx}
\usepackage{ulem}
\usepackage{subfig}

\newcommand{\DMNLO}{{\tt DM@NLO}}
\newcommand{\MO}{{\tt micrOMEGAs}}

\newcommand{\FeynArts}{{\tt FeynArts}}
\newcommand{\FORM}{{\tt Form}}

\newcommand{\FeynCalc}{{\tt FeynCalc}}

\newcommand{\MSbar}{{$\overline{\tt MS}$}}
\newcommand{\DRbar}{{$\overline{\tt DR}$}}
\newcommand{\dd}{{\rm d}}

\begin{document}

\preprint{LAPTH-008/19}
\preprint{MS-TP-19-07}
\preprint{TUM-HEP-1194-19}

\title{Squark-pair annihilation into quarks at next-to-leading order}

\author{S.~Schmiemann}
 \email{saskia.schmiemann@uni-muenster.de}
 \affiliation{
	Institut f\"ur Theoretische Physik, Westf\"alische Wilhelms-Universit\"at M\"unster, Wilhelm-Klemm-Stra{\ss}e 9, D-48149 M\"unster, Germany
  }
  
\author{J.~Harz}
 \email{julia.harz@tum.de}
 \affiliation{
	Physik Department T70, James-Franck-Stra{\ss}e, Technische Universit\"at M\"unchen, D-85748 Garching, Germany
  }
  
\author{B.~Herrmann}
 \email{herrmann@lapth.cnrs.fr}
 \affiliation{
	Univ.\ Grenoble Alpes, Univ.\ Savoie Mont Blanc, CNRS, LAPTh, F-74000 Annecy, France
  }

\author{M.~Klasen}
 \email{michael.klasen@uni-muenster.de}
 \affiliation{
	Institut f\"ur Theoretische Physik, Westf\"alische Wilhelms-Universit\"at M\"unster, Wilhelm-Klemm-Stra{\ss}e 9, D-48149 M\"unster, Germany
  }

\author{K.~Kova\v{r}\'ik}
 \email{karol.kovarik@uni-muenster.de}
 \affiliation{
	Institut f\"ur Theoretische Physik, Westf\"alische Wilhelms-Universit\"at M\"unster, Wilhelm-Klemm-Stra{\ss}e 9, D-48149 M\"unster, Germany
  }

\date{\today} 

\begin{abstract}
The Minimal Supersymmetric Standard Model (MSSM) is under intense scrutiny at the LHC and in dark matter searches. Interestingly, scenarios with light squarks of the third generation remain not only viable, but also well motivated by the observed Standard-Model-like Higgs boson mass and dark matter relic density. The latter often requires important contributions from squark pair annihilation. Following up on previous work, we present in this paper a precision analysis of squark pair annihilation into quarks at next-to-leading order of QCD including Sommerfeld enhancement effects. We discuss all technical details of our one-loop, real emission and resummation calculations, their implementation in the precision tool DM@NLO, as well as the numerical impact on the annihilation cross section and cosmological relic density in phenomenological MSSM scenarios respecting in particular current LHC constraints. We demonstrate that including these radiative corrections leads to substantial shifts in the preferred parameter regions by up to 20 GeV.
\end{abstract}

\maketitle



%
\section{Introduction}
\label{Sec:Intro}

Strong evidence for the existence of dark matter, along with the fact that neutrinos are massive, are compelling signs for the need of physics beyond the Standard Model. Even though dark matter still evades direct detection in Earth-based experiments such as LUX \cite{LUX:2016} and XENON1T \cite{XENON:2019}, there is overwhelming evidence from cosmological data such as the cosmic microwave background that dark matter exists in the Universe. Moreover, the relic density of cold dark matter (CDM) has been determined to the unprecedented precision of
\begin{align}
    \Omega_{\rm CDM}h^2 ~=~ 0.1200 \pm 0.0012
    \label{eq:omh2}
\end{align}
as measured by the Planck satellite and interpreted within the $\Lambda$CDM cosmological model \cite{Aghanim:2018eyx}. The indicated uncertainty corresponds to the 68\% confidence level, and $h$ stands for the present Hubble expansion rate $H_0$ in units of $100\,{\rm km}\,{\rm s}^{-1}\,{\rm Mpc}^{-1}$.

The Standard Model does not contain any suitable candidate for dark matter with the required properties.
The leading candidate therefore remains a weakly interacting massive particle (WIMP), which leads to the correct relic density via the freeze-out mechanism. However, alternative candidates and mechanisms do exist, e.g.\ in the form of the freeze-in mechanism \cite{Hall:2009bx,Klasen:2013ypa,Bernal:2017kxu,Belanger:2018sti}).
The Standard Model is therefore extended to include new particles which provide the required dark matter candidate. The new particles are usually protected from decaying by introducing an ad-hoc $Z_2$-symmetry, where all new particles are $Z_2$-odd and the Standard Model particles are $Z_2$-even. One such model, which was actually not introduced to address the existence of dark matter, is the Minimal Supersymmetric Standard Model (MSSM), where the conserved $Z_2$-symmetry is the $R$-parity. In most MSSM scenarios, the lightest supersymmetric particle (LSP) is the lightest neutralino $\tilde{\chi}^0_1$, which is stable and interacts only weakly. The lightest neutralino is an extremely well-studied candidate for cold dark matter.

The theory prediction for its relic abundance, $\Omega_{\tilde{\chi}^0_1}h^2$, is related to the number density $n_{\chi}$ of the neutralino, which can be computed solving the Boltzmann equation \cite{Gondolo:1990dk, Griest:1990kh, Edsjo:1997bg}
\begin{align}
    \frac{\dd n_{\chi}}{\dd t} ~=~ 
        -3 H n_{\chi} - \langle \sigma_{\rm ann}v \rangle \Big[ n^2_{\chi} - (n_{\chi}^{\rm eq})^2 \Big] \,,
\end{align}
where $H$ denotes the (time-dependent) Hubble parameter, $n_{\chi}^{\rm eq}$ the number density in thermal equilibrium, and $v $ the M\o ller velocity of the annihilating particles \cite{Cannoni:2013bza}. All specifics about the interaction of dark matter with other particles in the chosen particle physics model is contained in the annihilation cross section $\sigma_{\rm ann}$, which accounts for all possible annihilation and co-annihilation processes. Its thermal average can be expressed as
\begin{align}
    \langle \sigma_{\rm ann} v \rangle ~=~ \sum_{i,j} \langle \sigma_{ij} v_{ij} \rangle \frac{n_i^{\rm eq}}{n_{\chi}^{\rm eq}} \frac{n_j^{\rm eq}}{n_{\chi}^{\rm eq}} \,,
\end{align}
where the double sum runs over all $Z_2$-odd particles of the theory. The ratios of the equilibrium number densities are proportional to
\begin{align}
    \frac{n_i^{\rm eq}}{n_{\chi}^{\rm eq}} ~\sim~ \exp \left\{ -\frac{m_i - m_{\chi} }{T} \right\} \,,
\end{align}
$T$ being the temperature. From this last equation, it becomes obvious that annihilations involving particles other than the lightest neutralino are suppressed if these particles are heavy compared to the neutralino. On the other hand, co-annihilations with the next-to-lightest supersymmetric particle (NLSP) will be important or even dominant if the mass difference is rather small. Typical examples in the MSSM are co-annihilations of the neutralino with a scalar top quark or a scalar tau lepton. For smaller mass differences between the LSP and NLSP, even pair-annihilations of the next-to-lightest particle contribute in a sizeable manner, and can even become dominant in the total annihilation cross section. In case there are more than two almost mass degenerate particles, (co-)annihilations between all particles have to be taken into account.

In the present paper, we focus on the case, where the masses of one or two squarks, the lightest stop and/or the lightest sbottom, are close to the neutralino mass. A case of a light scalar top quark is very well motivated. First, a light scalar top is necessary to achieve the electroweak baryogenesis in the MSSM \cite{Delepine:1996vn, Liebler:2015ddv}. Second, scenarios with light scalar tops satisfy the experimental constraints from LHC searches and can also contribute to a successful prediction of the mass of the lightest Higgs boson in the MSSM \cite{Arbey:2012bp}.

The relic density of dark matter in scenarios with a light stop which is almost mass degenerate with the lightest neutralino is very sensitive to the mass difference of the two particles. Any small shift in the predicted relic density can cause a large shift of the parameter region where the relic density is compatible with the experimental limits given by Eq.\ (\ref{eq:omh2}). In this analysis we focus on next-to-leading supersymmetric QCD (SUSY-QCD) corrections to the corresponding annihilation and co-annihilation cross sections in scenarios with a light scalar quark. These corrections have the potential to significantly modify the annihilation cross section and thereby also the relic density.

The impact of such radiative corrections of order $\alpha_s$ on the relic density has been demonstrated for gaugino pair annihilation into quarks \cite{Herrmann:2007ku, Herrmann:2009wk, Herrmann:2009mp, Herrmann:2014kma}, gaugino-squark co-annihilation into final states containing a quark \cite{Freitas:2007sa, Harz:2012fz, Harz:2014tma}, and squark-antisquark annihilation into electroweak final states \cite{Harz:2014gaa}. Moreover, electroweak corrections to neutralino annihilation have been presented \cite{Boudjema:2005hb, Baro:2007em, Baro:2009na}, leading to similar conclusions concerning their impact on the relic density.

Including radiative corrections to the total annihilation cross section not only shifts the parameter regions corresponding to the correct relic density, but it also reduces the theoretical uncertainty of the relic density prediction. The theoretical uncertainty from scale and scheme variations on the annihilation cross section and the neutralino relic density has been evaluated for specific subclasses of processes in Ref.\ \cite{Harz:2016dql}. 

After the work presented in Refs.\ \cite{Herrmann:2007ku, Herrmann:2009wk, Herrmann:2009mp, Herrmann:2014kma, Harz:2012fz, Harz:2014tma, Harz:2014gaa}, with the present paper, we make a first step towards completing the missing processes sensitive to radiative corrections of order $\alpha_s$. More precisely, we present such corrections for squark-squark annihilation into quark-quark pairs. The discussion of squark-antisquark annihilation into quarks and gluons is left for forthcoming publications. 

\begin{table*}[t!]
    \begin{center}
    \begin{tabular}{|c|ccccccccccc|c|}
    \hline
           & $M_1$ & $M_2$ & $M_3$ & $M_{\tilde{q}_L}$ & $M_{\tilde{t}_R}$ & $M_{\tilde{b}_R}$ & $A_t$ & $A_b$ & $\mu$ & $m_{A^0}$ & $\tan\beta$ & $Q_{\rm SUSY}$ \\
    \hline
    \hline
        Scenario I & 1278.5 & 2093.5 & 1267.2 & 2535.1 & 1258.7 & 3303.8 & 2755.3 & 2320.9 & -3952.6 & 3624.8 & 15.5 & 1784.64\\
        Scenario II & 1629.2 & 3613.4 & 1720.8 & 1513.2 & 3964.9 & 3871.5 & -4434.9 & 2201.7 & 2615.4 & 3451.3 & 53.1 & 2447.96\\
    \hline
    \end{tabular}
    \\[2mm]
    \begin{tabular}{|c|cccccccc|c|}
    \hline
           & ~$m_{\tilde{\chi}^0_1}$~ & ~$m_{\tilde{\chi}^0_2}$~ & ~$m_{\tilde{\chi}^{\pm}_1}$~ & ~$m_{\tilde{t}_1}$~ & ~$m_{\tilde{b}_1}$~ & ~$m_{\tilde{g}}$~ & ~$m_{h^0}$~ & ~$m_{H^0}$~ & ~~$\Omega_{\tilde{\chi}^0_1}h^2$~~ \\
    \hline
    \hline
        Scenario I & 1279.7 & 2153.6 & 2153.5 & 1301.9 & 2554.2 & 1495.5 & 125.8 & 3625.6 & 0.1200 \\
        Scenario II & 1624.4 & 2606.6 & 2606.6 & 1652.0 & 1654.9 & 1944.9 & 127.8 & 3451.2 & 0.1200 \\
    \hline
    \end{tabular}~~~~~~~~~~~~~~~~~~~~~~~~~~~~
    \end{center}
    \caption{Reference scenarios within the phenomenological MSSM for our numerical study. Note that only the parameters which are relevant for our analysis are given here. All dimensionful quantities are given in GeV.}
    \label{tab:scenarios}
\end{table*}

In the following, in Sec.\ \ref{Sec:Pheno}, we start by discussing the phenomenological importance of the processes under consideration in this work. We also present two reference scenarios featuring important contributions of the processes of our interest. In Sec.\ \ref{Sec:NLO}, we then detail the analytical calculation of the radiative corrections. We discuss in particular points that are beyond the discussion presented in Refs.\ \cite{Herrmann:2007ku, Herrmann:2009wk, Herrmann:2009mp, Herrmann:2014kma, Harz:2012fz, Harz:2014tma, Harz:2014gaa} and analyze the impact that the radiative corrections have on the corresponding cross sections. The impact of the corrections on the relic density in the two reference scenarios is presented in Sec.\ \ref{Sec:Results}. Our conclusions are given in Sec.\ \ref{Sec:Conclusion}.

%
\section{Phenomenology of squark annihilation}
\label{Sec:Pheno}
The analysis presented in this paper concentrates on the contributions from squark-pair annihilation to the total annihilation cross section $\sigma_{\rm ann}$ of neutralino dark matter. We investigate scenarios in the phenomenological MSSM (pMSSM), where the processes
\begin{align}
    \tilde{t}_1 \tilde{t}_1 ~&\longrightarrow~ t t \,, \label{eq:process_tt} \\
    \tilde{b}_1 \tilde{b}_1 ~&\longrightarrow~ b b \,, \label{eq:process_bb} \\
    \tilde{t}_1 \tilde{b}_1 ~&\longrightarrow~ t b \,, \label{eq:process_tb} 
\end{align}
play an important role. Supersymmetry and the MSSM in particular have been extensively tested by searches at the Large Hadron Collider (LHC) and at experiments aiming at the detection of direct signals from elastic collisions of dark matter with heavy nuclei such as XENON1T. In order to take into account the most important experimental constraints from the searches for supersymmetry, we use the results of an analysis performed by the ATLAS collaboration in the light of recent searches at the LHC \cite{ATLAS2015pMSSM}.\footnote{A similar study of the pMSSM has been conducted by the CMS collaboration \cite{CMS2016pMSSM}.} The ATLAS analysis is performed in the pMSSM with 19 parameters (defined at the SUSY scale) and is based on a sample of $5\cdot 10^8$ parameter points. Applying constraints from ATLAS SUSY searches, electroweak precision observables such as $\Delta \rho$ and $(g-2)_\mu$, flavor observables such as $b\rightarrow s\gamma$, and requiring that the neutralino is the LSP and a dark matter candidate with the relic density less than $0.1208$ (for details on further contraints see Ref.\ \cite{ATLAS2015pMSSM}) leads to a subset of about 300.000 viable points. We have analyzed this subset in order to examine in which regions of parameter space the above processes contribute significantly.

\begin{table}
    \begin{tabular}{|c|cc|}
    \hline
         Contributing processes & ~Scenario I ~  & ~Scenario II ~ \\
    \hline
    \hline
         $\tilde{t}_1 \, \tilde{t}_1 ~\to~ t \, t$ & 30.5\% & 8.8\% \\
         $\tilde{b}_1 \, \tilde{b}_1 ~\to~ b \, b$ & -- & 7.4\% \\
         $\tilde{t}_1 \, \tilde{b}_1 ~\to~ t \, b$ & -- & 34.0\% \\
    \hline
         $\tilde{\chi}^0_1 \, \tilde{\chi}^0_1 ~\to~ q \, \bar{q}$ & -- & -- \\
    \hline
         $\tilde{\chi}^0_1 \, \tilde{t}_1 ~\to~ t \, g$ & 9.3\% & -- \\
         $\tilde{\chi}^0_1 \, \tilde{t}_1 ~\to~ q \, V \,, q \, \phi$ & 5.8\% & -- \\
         $\tilde{\chi}^0_1 \, \tilde{b}_1 ~\to~ q \, V \,, q \, \phi$ & -- & -- \\
    \hline
    \hline
         $\tilde{t}_1 \, \tilde{t}^*_1 ~\to~ g\,g$ & 38.7\% & 9.8\% \\
         $\tilde{t}_1 \, \tilde{t}^*_1 ~\to~ q \, \bar{q}$ & 5.1\% & 2.4\% \\
         $\tilde{t}_1 \, \tilde{b}^*_1 ~\to~ q \, \bar{q}'$ & -- & 4.0\% \\
         $\tilde{b}_1 \, \tilde{b}^*_1 ~\to~  q \, \bar{q} \,, g\,g$ & -- & 8.1\% \\
         $\tilde{\chi}^0_1 \, \tilde{g} ~\to~  X $ & 3\% & -- \\
         $\tilde{g} \, \tilde{g} ~\to~  X $ & -- & -- \\
    \hline
    \hline
         ~\DMNLO\ current analysis~ & 30.5\% & 50.2\% \\
         ~\DMNLO\ total \cite{Herrmann:2007ku, Herrmann:2009wk, Herrmann:2009mp, Herrmann:2014kma, Harz:2012fz, Harz:2014tma, Harz:2014gaa}~ & 45.6\% & 50.2\% \\
    \hline
    \end{tabular}
    \caption{Dominant annihilation channels contributing to $\sigma_{\rm ann}$ and thus to the neutralino relic density in the two reference scenarios given in Table \ref{tab:scenarios}. Here, $V=\gamma, Z^0, W^{\pm}$ and $\phi = h^0, H^0, A^0, H^{\pm}$. Further contributions below 1\% are omitted.}     
    \label{tab:channels}
\end{table}

In order for the contribution from the annihilation of third-generation squarks to the total dark matter cross section to be significant, one (or more) scalar quarks have to be almost mass degenerate with the lightest neutralino. This is not an unnatural requirement because a light scalar top quark is necessary to explain the measured mass of the Standard Model Higgs boson within the MSSM. Moreover, scenarios where a scalar top is almost mass degenerate with the lightest neutralino are quite frequent, as the mass degeneracy gives rise to different topologies in collider searches making their testing more challenging and their exclusion less likely. Another aspect of scenarios where scalar quarks are the next-to-lightest supersymmetric particle (NLSP) is that the lightest neutralino is mostly bino-like. Higgsino-like and wino-like lightest neutralinos mostly lead to scenarios with other gauginos being the NLSP. A consequence of the lightest neutralino being bino-like is that the annihilation of dark matter is typically not efficient enough for the relic density to reach the value determined by the {\it Planck} collaboration given in Eq.\ \eqref{eq:omh2}. Therefore in scenarios with bino-like neutralino dark matter, some enhancement mechanism is needed for them to be consistent with the relic density measurements. As we will discuss below, in the scenarios analyzed here, the enhancement comes from the presence of LSP-NLSP co-annhilations as well as NLSP annihilations. 
\begin{figure*}[t!]
	\begin{center}
	    \includegraphics[width=0.485\textwidth]{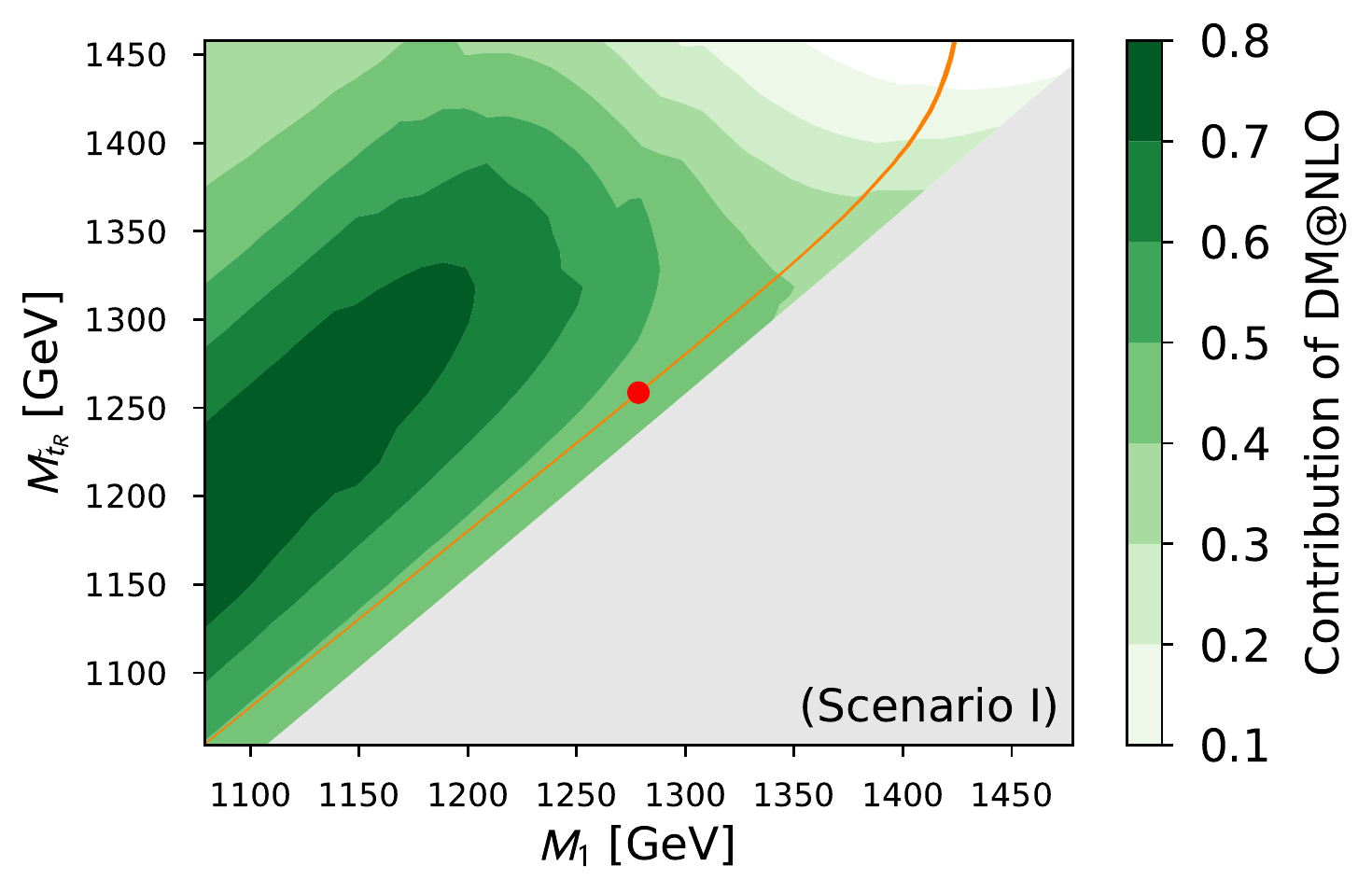}~~~~~
	    \includegraphics[width=0.495\textwidth]{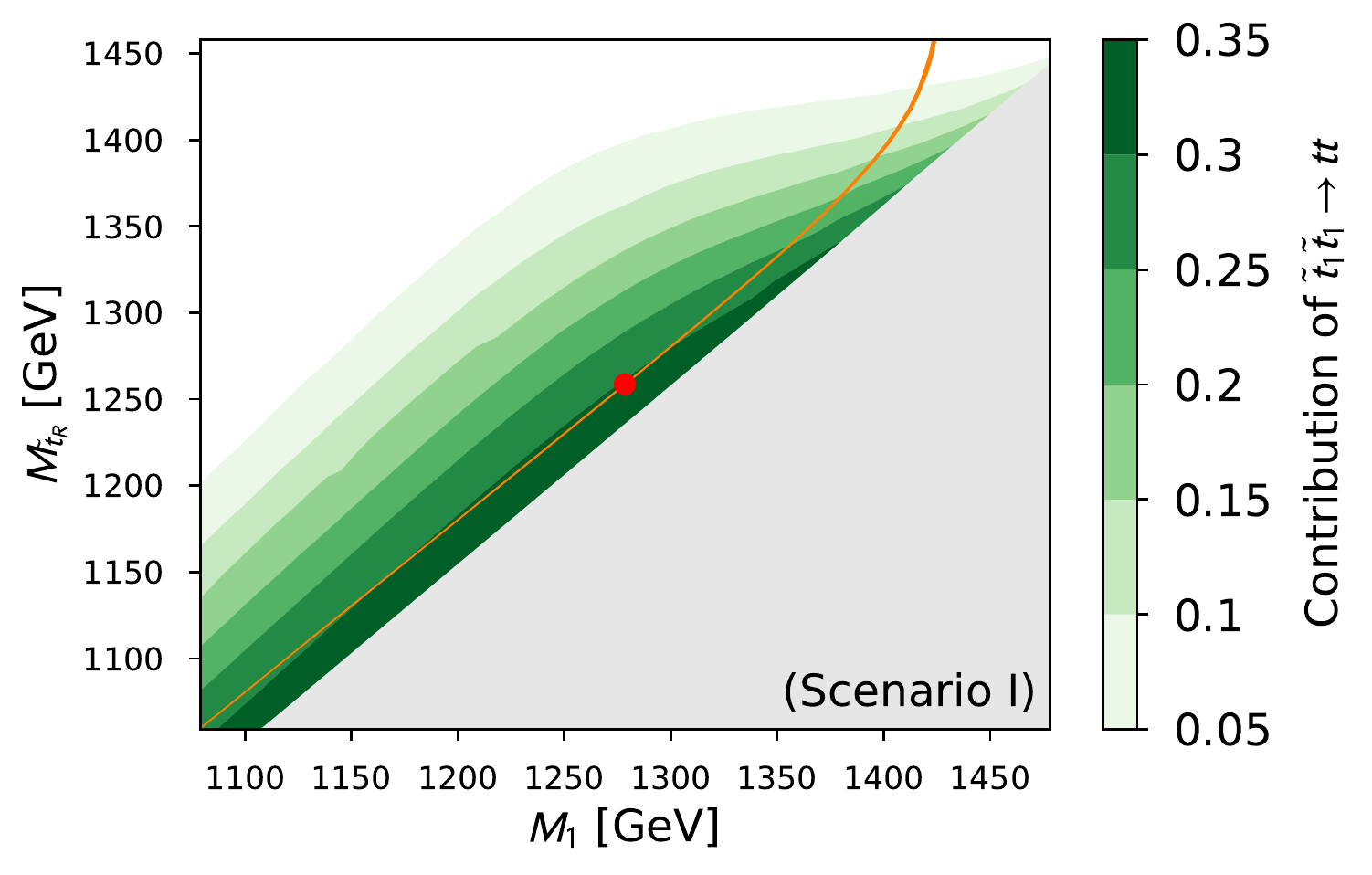}\\
	    \includegraphics[width=0.495\textwidth]{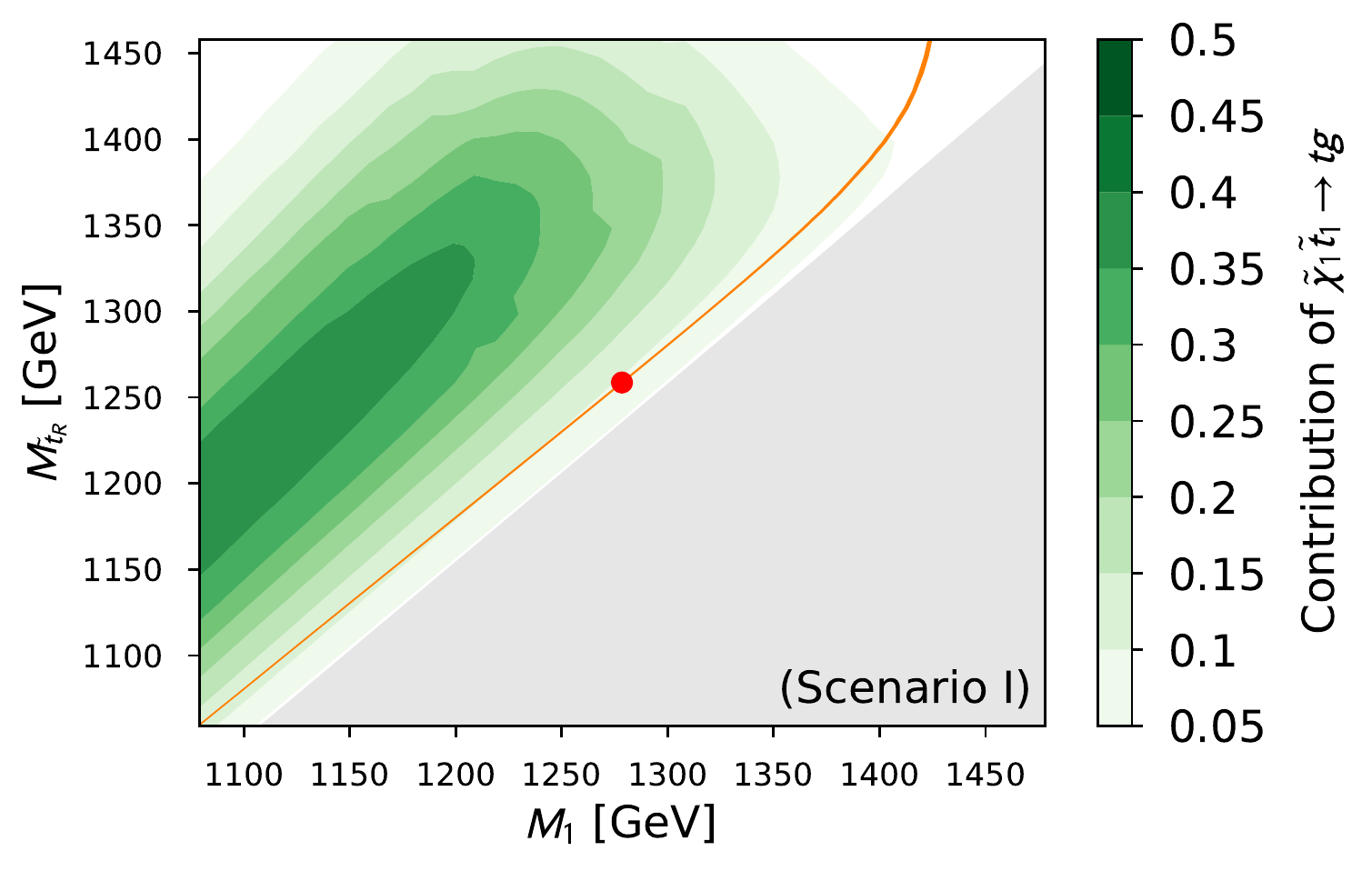}~~~
	    \includegraphics[width=0.495\textwidth]{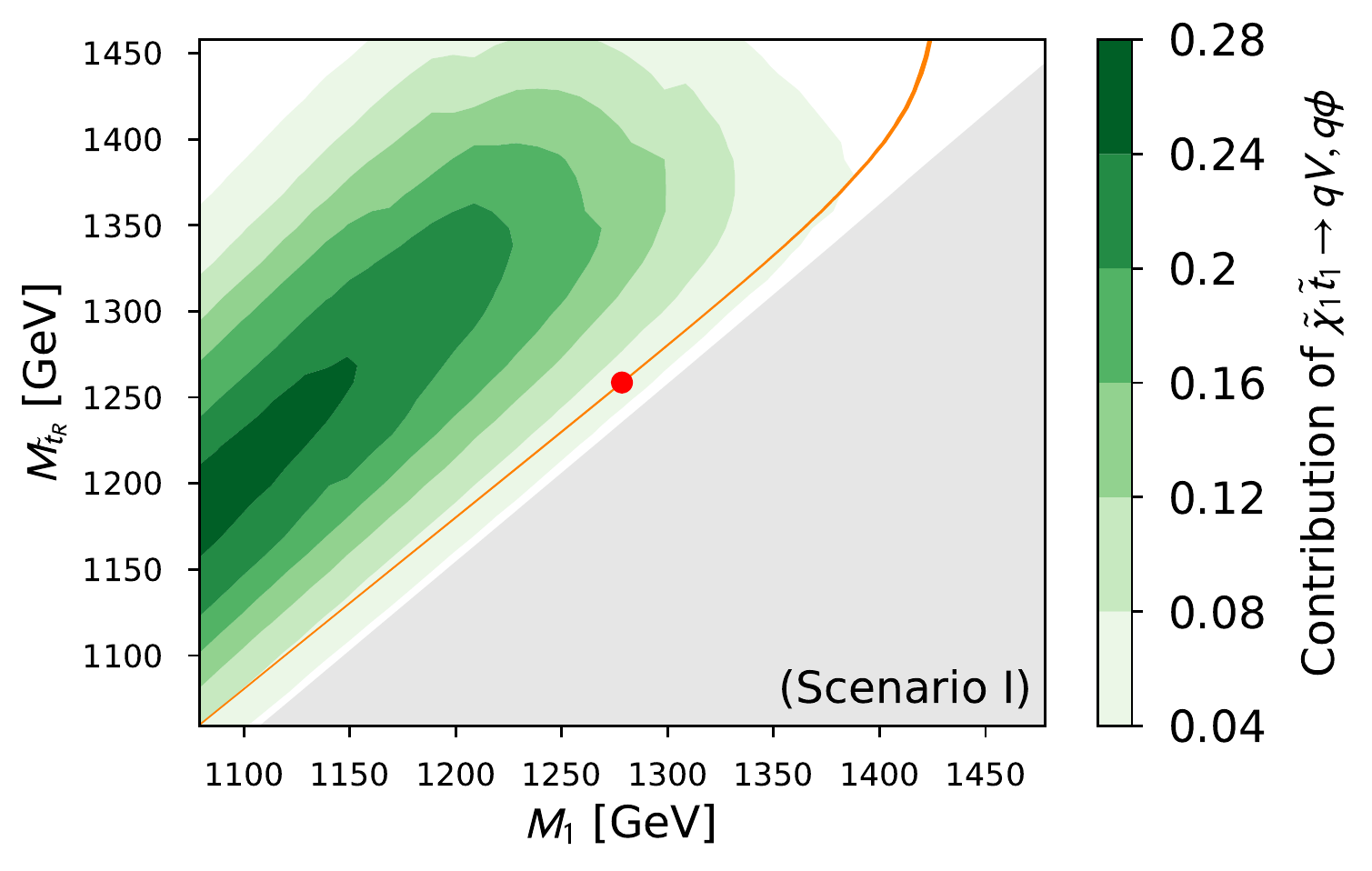}
%
\\[5mm]
	    \includegraphics[width=0.485\textwidth]{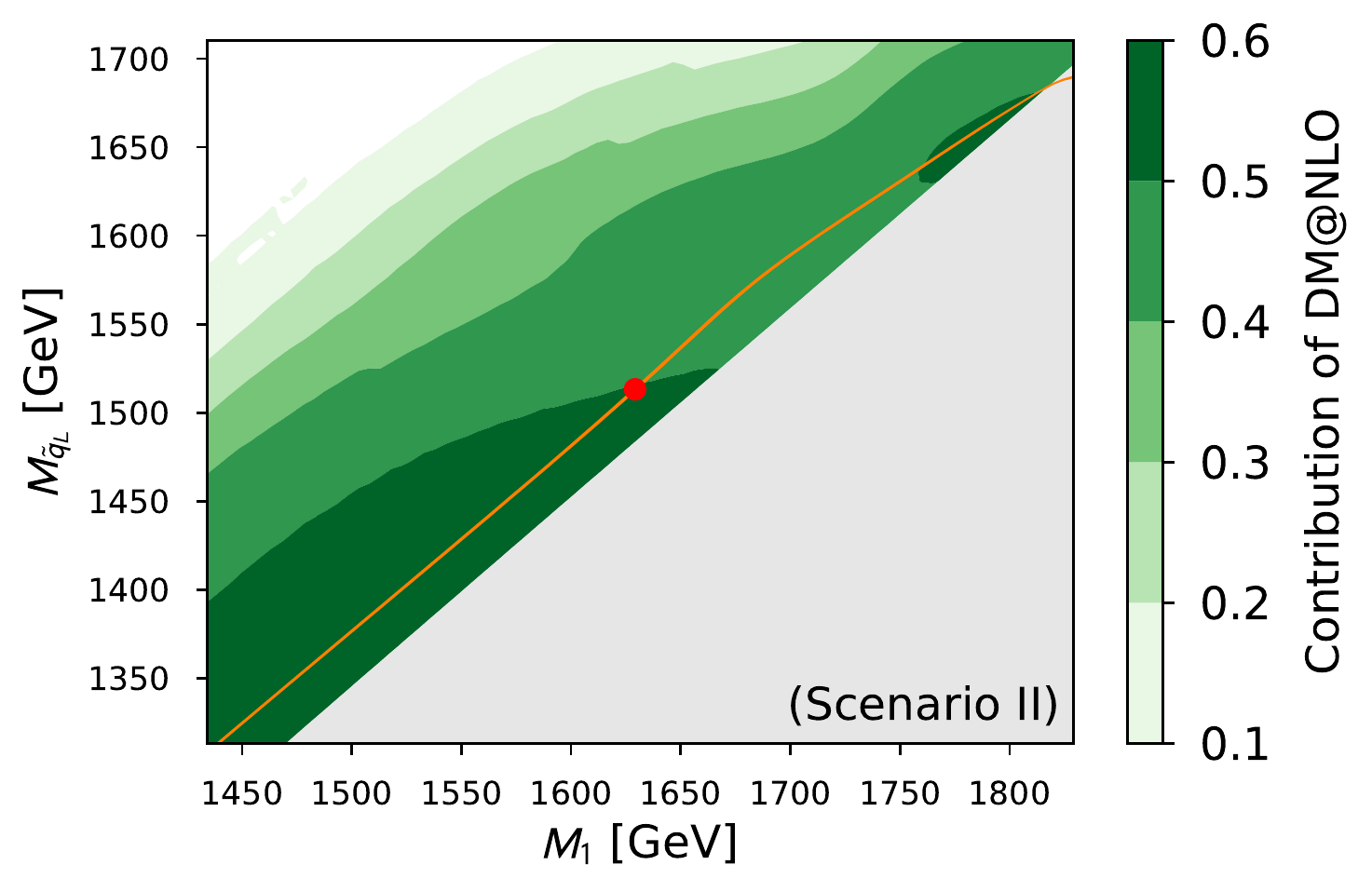}~~~~~
	    \includegraphics[width=0.495\textwidth]{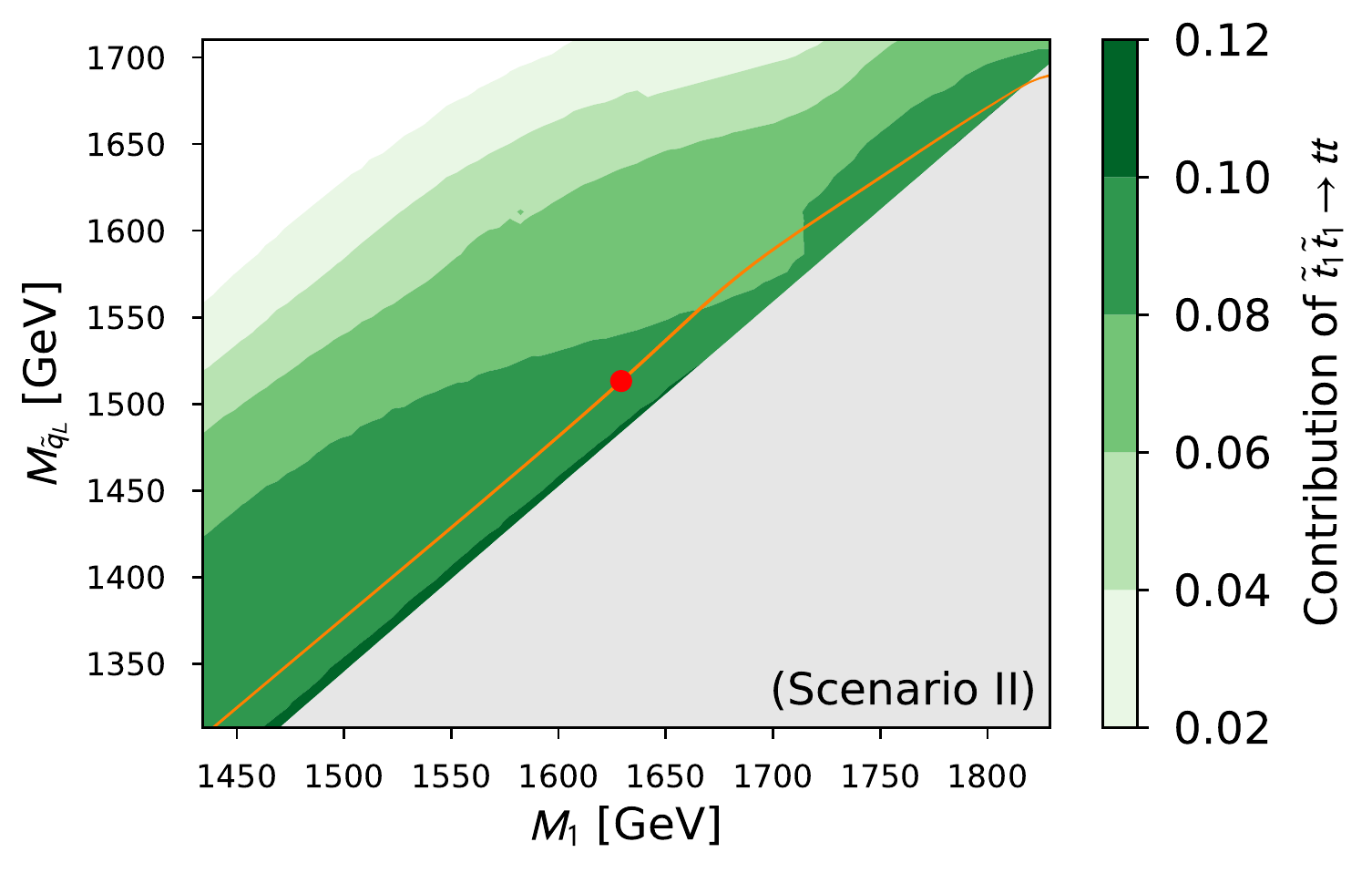}\\
	    \includegraphics[width=0.495\textwidth]{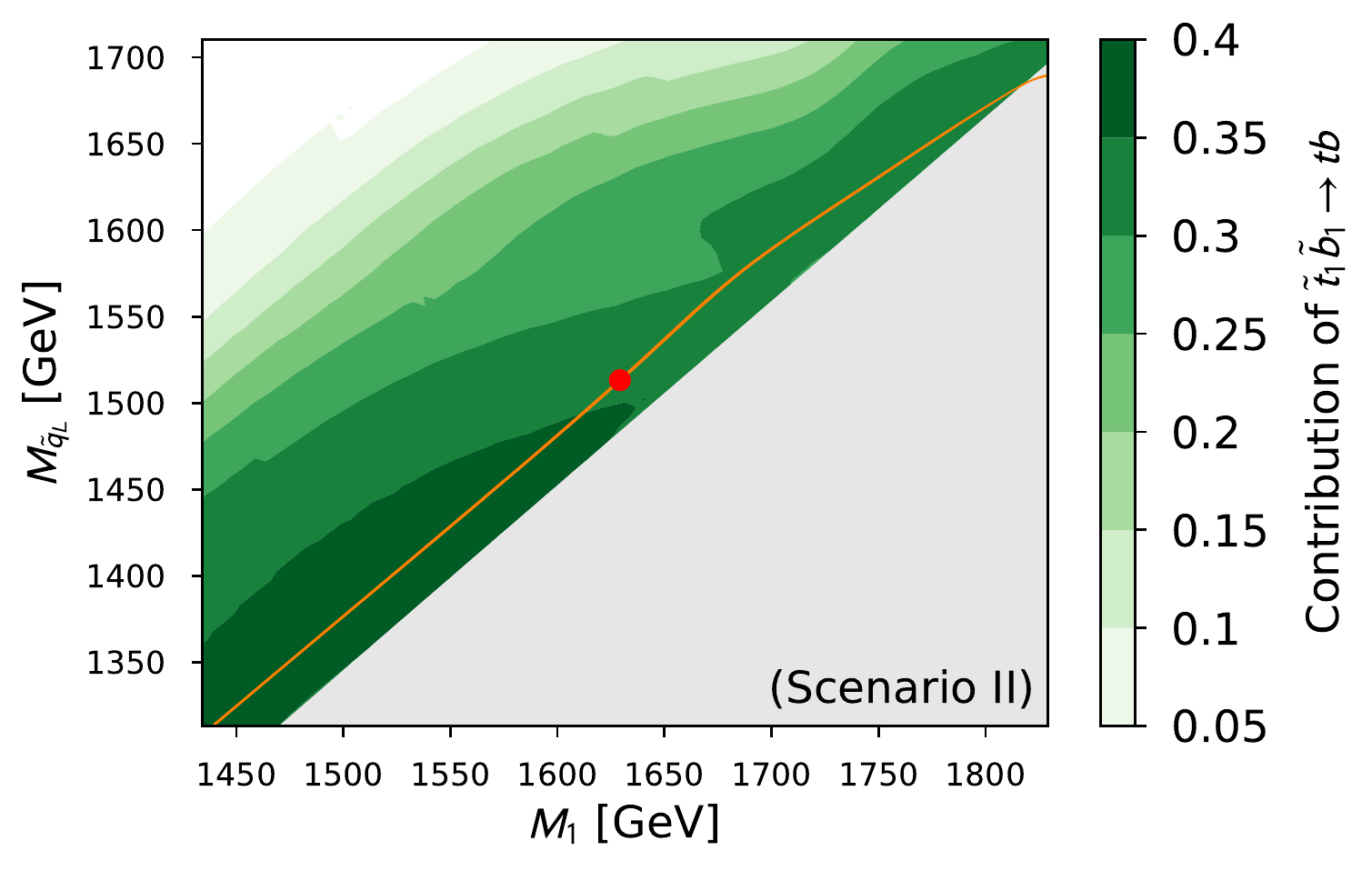}~~~
	    \includegraphics[width=0.495\textwidth]{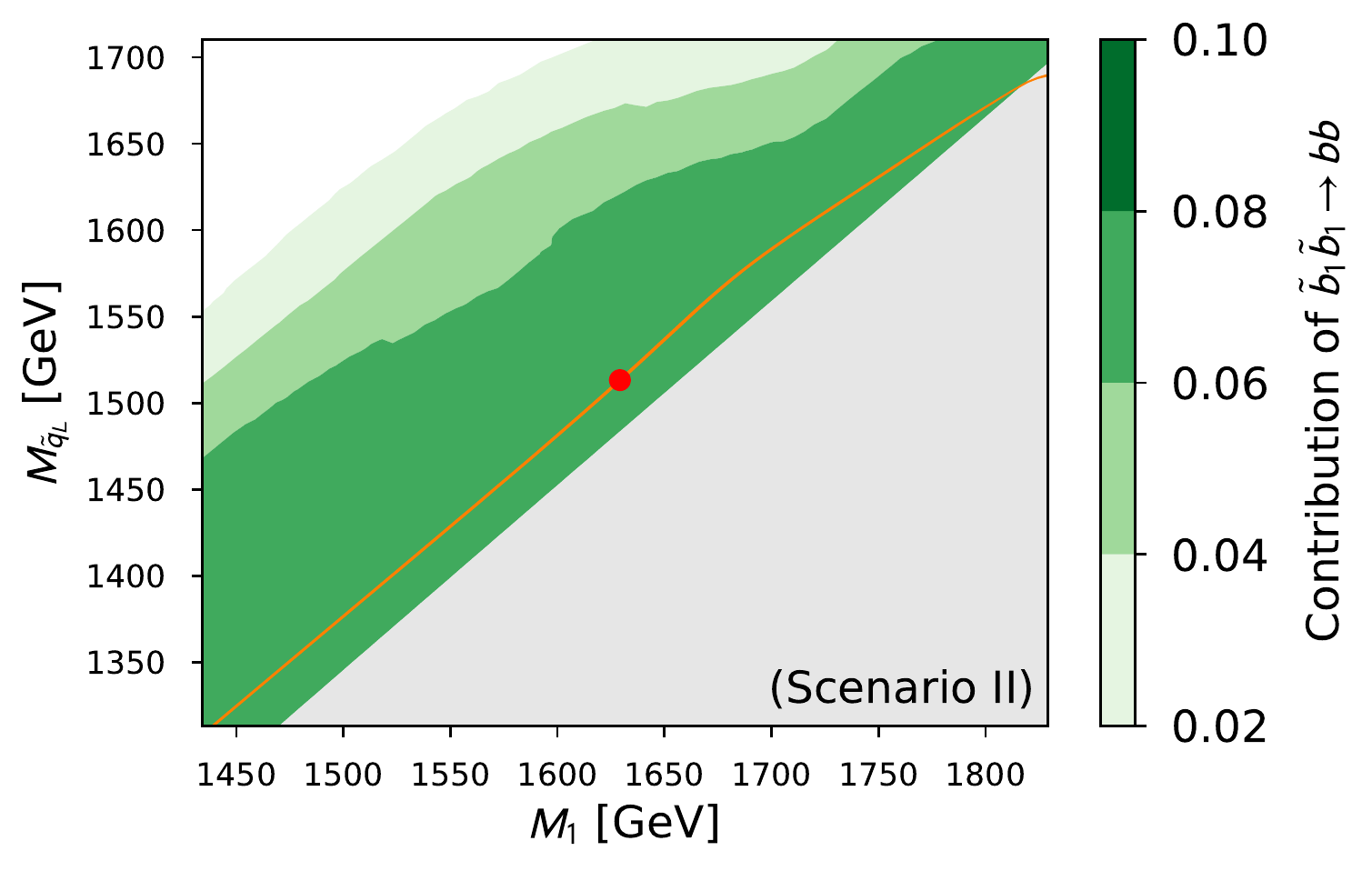}
	\end{center}
	\vspace*{-5mm}
    \caption{Contribution of selected processes to the total annihilation cross section $\sigma_{\rm ann}$ in the $M_1$-$M_{\tilde{t}_R}$ or $M_1$-$M_{\tilde{q}_L}$ plane around reference Scenarios I or II, respectively. The orange band indicates the parameter region in agreement with the {\it Planck} limit given in Eq.\ \eqref{eq:omh2} at the $2\sigma$ confidence level. The green levels indicate the relative importance of the processes that can be corrected by {\tt DM@NLO} (first and fifth plot) and of selected individual processes (remaining plots). The grey region corresponds to $m_{\tilde{t}_1}<m_{\tilde{\chi}^0_1}$. The red dots indicate Scenarios I and II of Table \ref{tab:scenarios}.}
    \label{fig:channelsI}
\end{figure*}
\begin{figure*}[t!]
	\begin{center}
	\begin{picture}(400,80)
	\put(0,0){\includegraphics[scale=1.0]{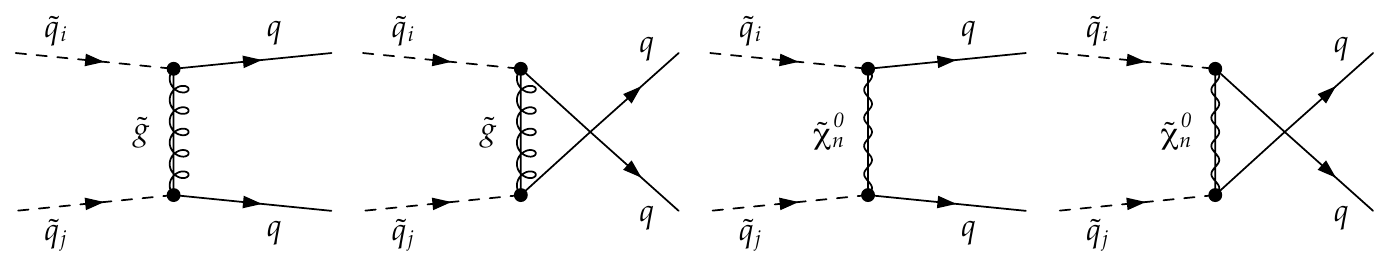}}
	\end{picture}
	\begin{picture}(300,80)
	\put(0,0){\includegraphics[scale=1.0]{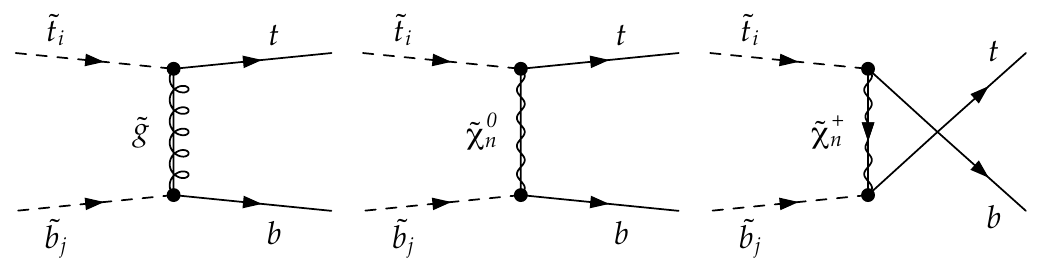}}
	\end{picture}
	\end{center}
	\vspace*{-2mm}
    \caption{Tree-level Feynman diagrams associated with the squark pair-annihilation into quark pairs for the case of squarks of identical (upper row) or different (lower row) type.}
    \label{fig:diagrams_lo}
\end{figure*}

\subsection{Reference scenarios}
\label{Sec:Scenarios}

As mentioned above, the numerical part of the present study will be based on two reference scenarios inspired by the findings presented in Ref.\ \cite{ATLAS2015pMSSM}. More precisely, we will focus on two pMSSM scenarios, whose most relevant soft-breaking parameters and particle masses are presented in Tab.~\ref{tab:scenarios}. It is to be noted that, although the input soft mass parameters of the two scenarios are identical to those of two actual scenarios given in Ref.\ \cite{ATLAS2015pMSSM}, the resulting physical masses slightly differ from those associated with the ATLAS study due to the fact that we are using a different computational setup. The actual shift in the physical masses is small so that all experimental constraints are still satisfied and the phenomenology is not altered. 

Both scenarios feature bino-like neutralinos, the bino mass parameter $M_1$ being smaller than the wino and higgsino mass parameters $M_2$ and $|\mu|$. The key parameters of the third-generation squarks of our interest are the ``left-handed'' stop and sbottom mass parameter $M_{\tilde{q}_L}$, and the ``right-handed'' stop and sbottom mass parameters $M_{\tilde{t}_R}$ and $M_{\tilde{b}_R}$. In both scenarios, squarks of the first and second generation, the sleptons, and other electroweak gauginos are heavier such that they do not influence the phenomenology discussed here. 

In our setup, starting from the soft-breaking terms defined at the scale $Q_{\rm SUSY}$ indicated in Tab.~\ref{tab:scenarios}, we obtain the physical mass spectrum using the spectrum generator {\tt SPheno 3.3.3} \cite{Porod:2003um, Porod:2011nf}. The mass spectrum is then handed over to {\tt micrOMEGAS 2.4.1} \cite{Belanger:2001fz, Belanger:2006is} making use of the SUSY Les Houches Accord 2 \cite{Allanach:2008qq}. In addition to the actual value of the relic density, {\tt micrOMEGAs} also provides the contributions of all individual channels contributing to $\sigma_{\rm ann}$ given in Tab.~\ref{tab:channels} for the two chosen reference scenarios.

As can be seen, the processes given in Eqs.\ \eqref{eq:process_tt} to \eqref{eq:process_tb} contribute in a significant manner for both scenarios. More precisely, in Scenario I, the scalar top pair annihilation is the second most important process and together with the processes previously analysed in Refs.\ \cite{Harz:2012fz, Harz:2014tma, Harz:2014gaa} makes up more than 45\% of the total annihilation cross section. In this scenario, the mostly ``right-handed'' scalar top $\tilde{t}_1$ is the NLSP, and the mass difference between the lightest neutralino and the NLSP is about $20{\rm\ GeV}$. Moreover, the process $\tilde{t}_1 \tilde{t}_1 \to t t$ is enhanced by the relatively low gluino mass. Scalar bottom quarks are heavy in this scenario, such that the corresponding annihilation channels are negligible. The process $\tilde{t}_1 \tilde{t}_1 \to tt$ amounts to about 30\% of the annihilation cross section, while neutralino-stop co-annihilation \cite{Harz:2012fz, Harz:2014tma} accounts for about 15\%, such that our next-to-leading SUSY-QCD corrections affect almost 50\% of the total annihilation cross section. 

The importance of different relevant contributions to the total annihilation cross section in and around this scenario is shown in the first four plots of Fig.\ \ref{fig:channelsI}. The part of the parameter space where the lightest neutralino is not the LSP and hence also not the dark matter candidate is indicated in grey. Different shades of green indicate the relative importance of several NLSP annihilation and LSP-NLSP co-annihilation processes. We see that in general the co-annihilations are most important when the mass splitting between the LSP and the NLSP (in our case the lightest neutralino and the lightest scalar top) is larger, i.e.\ about $150{\rm\ GeV}$, and the dominant contribution shifts to the NLSP annihilations as the mass splitting gets smaller. The parameter region where the dark matter relic density is within $2\sigma$ of the experimental value given in Eq.\ \eqref{eq:omh2} is highlighted in all plots in orange. The neutralino relic density is computed using {\tt micrOMEGAs}. For scenarios which contain lighter neutralinos and lighter stops than the reference Scenario I, the total annihilation cross section is dominated by the same processes as in the reference Scenario I, and the region where the relic density agrees with the experimental measurement follows an almost straight line where the mass splitting is constant. As we move along the region with the correct relic density towards scenarios with heavier LSP and NLSP, we reach a point where the LSP and NLSP are similar in mass to the light gluino ($m_{\tilde{g}}=1495.5$ GeV). In these scenarios, we have three particles with almost degenerate masses and gluino annihilations and co-annihilations with the stop dominate the total annihilation cross section.    
\begin{figure*}[t]
	\begin{center}
	\begin{picture}(550,150)
	\put(7,0){\resizebox{!}{5.2cm}{\includegraphics{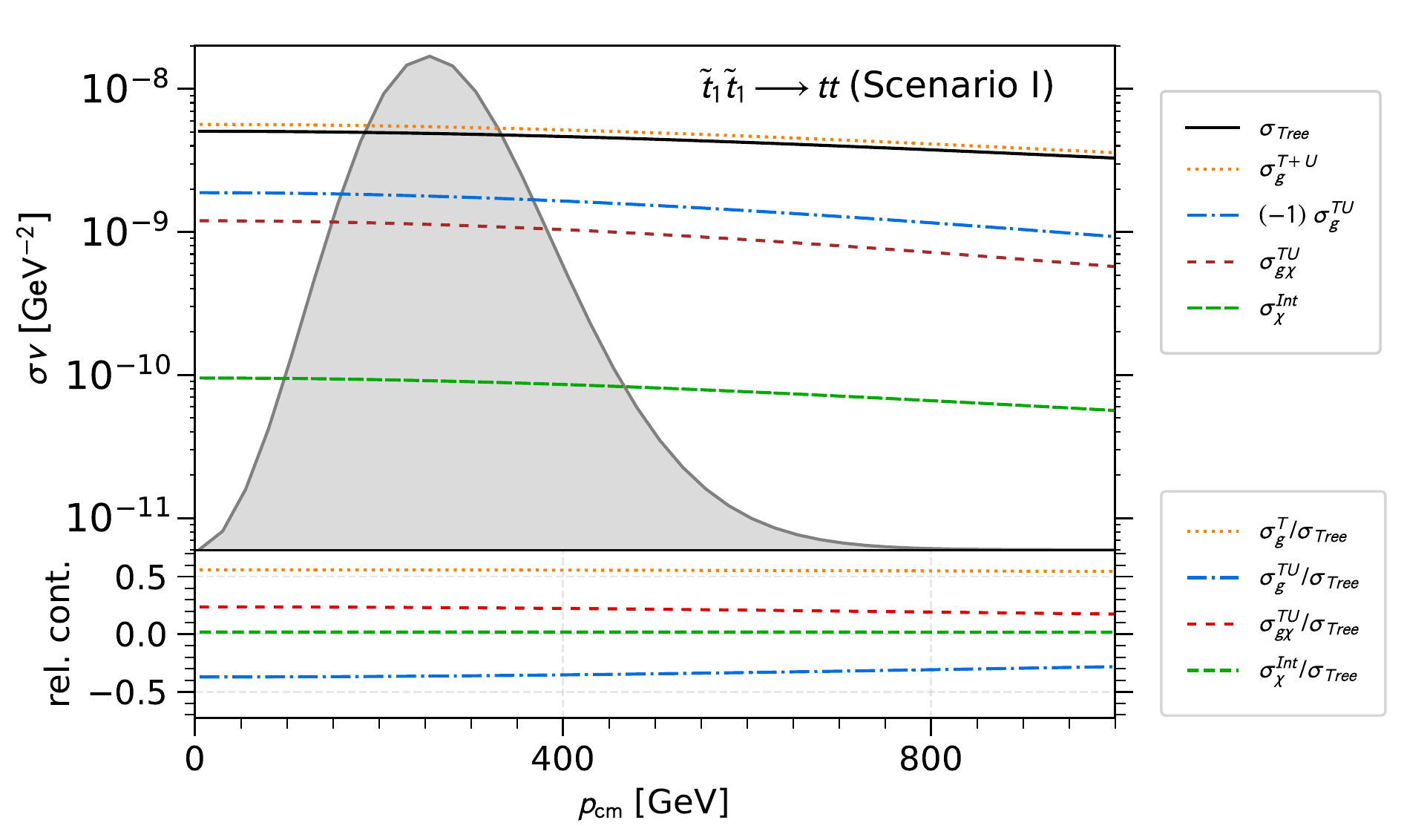}}}
	\put(260,0){\resizebox{!}{5.2cm}{\includegraphics{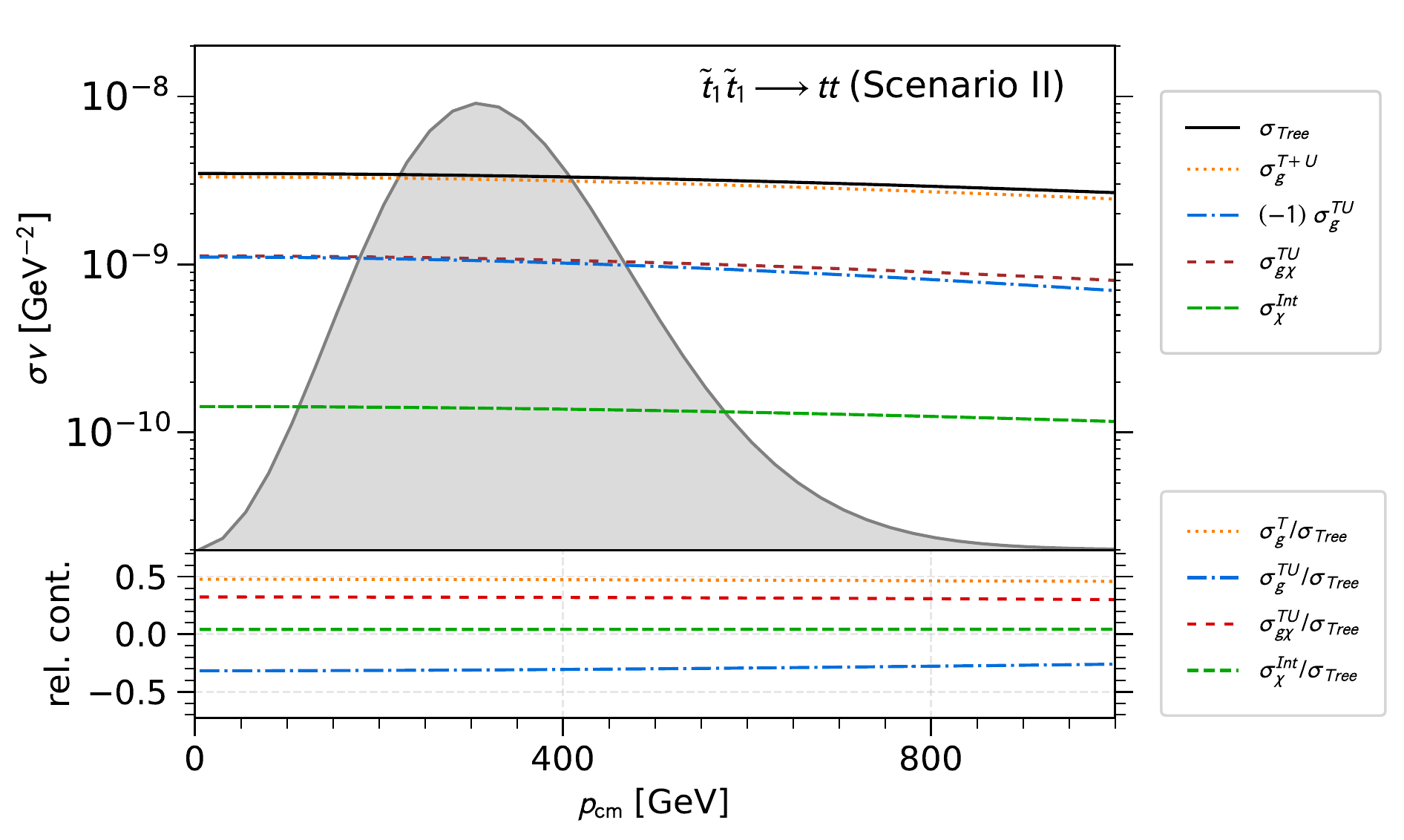}}}
	\end{picture}
	\begin{picture}(550,150)
	\put(10,0){\resizebox{!}{5.2cm}{\includegraphics{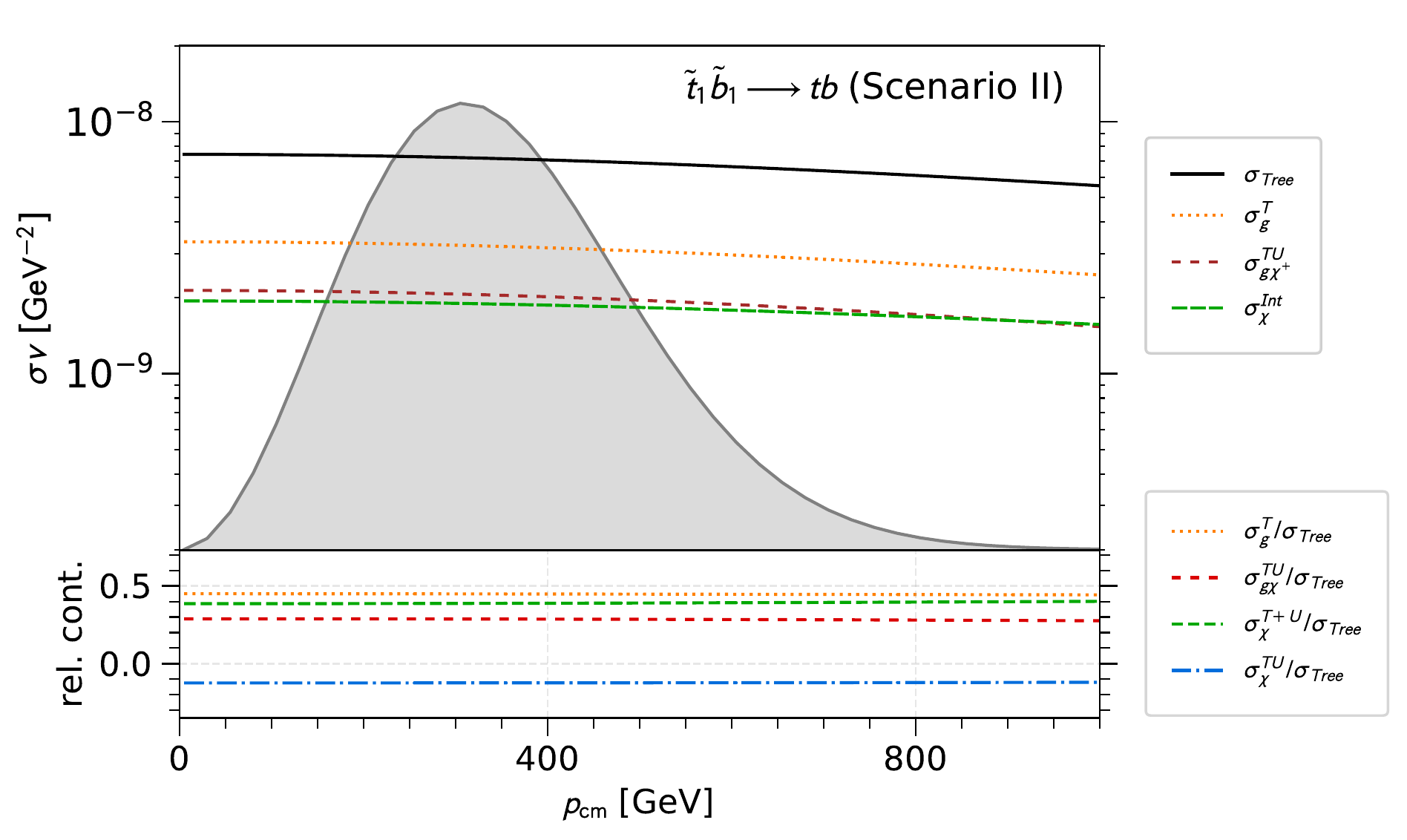}}}
	\put(260,0){\resizebox{!}{5.2cm}{\includegraphics{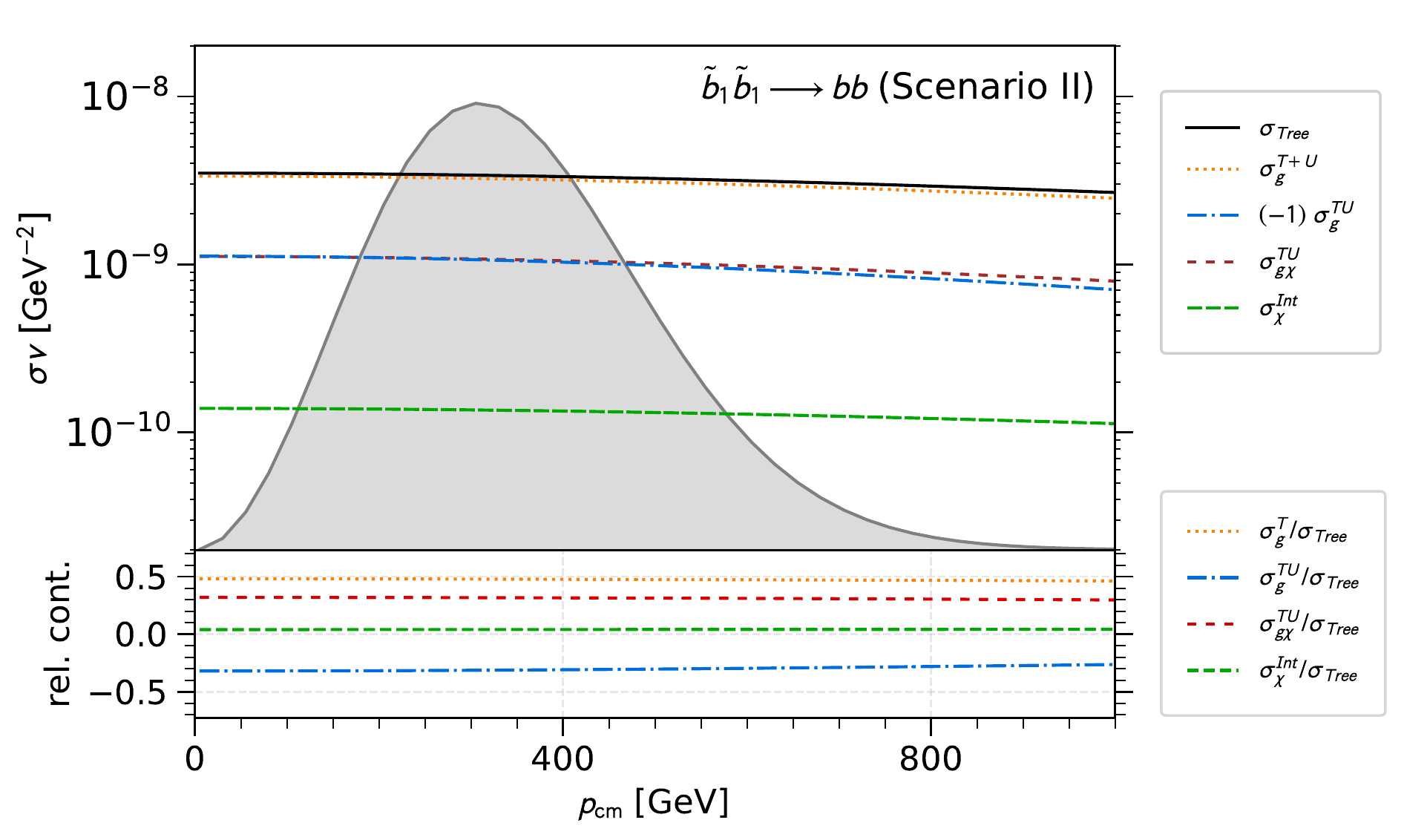}}}
	\end{picture}
	\end{center}
	\vspace*{-5mm}
\caption{Upper part: Leading-order cross section $\sigma v$ as a function of the center-of-mass momentum $p_{\rm cm}$ into different sub-channels according to Fig.\ \ref{fig:diagrams_lo} in the two pMSSM scenarios I and II of Table \ref{tab:scenarios}. In the legend, $\sigma_{\rm Tree}$ denotes the total tree-level cross section, the subscripts $g$, $\chi$ and $g\chi$ correspond to gluino exchange squared, gaugino exchange squared, and gluino-gaugino interference. The superscripts indicate squared $t$-channel (T), squared $u$-channel (U), the sum of both (T+U) and the $t$-$u$ interference contributions (TU). For the gaugino exchange, the superscript ``Int'' refers to the sum of all involved diagrams. Lower part: Contributions relative to the total tree-level result $\sigma_{\rm Tree}$.}
\label{fig:xsecLOcont}
\end{figure*}

The situation is different for Scenario II. Here, the ``left-handed'' mass parameter $M_{\tilde{q}_L}$ is much smaller than the ``right-handed'' masses $M_{\tilde{t}_R}$ and $M_{\tilde{b}_R}$, such that the relevant physical states $\tilde{t}_1$ and $\tilde{b}_1$ are mainly ``left-handed'' with almost degenerate masses. The mass difference between them and the lightest neutralino is about 30 GeV. As a consequence, processes containing both $\tilde{t}_1$ and $\tilde{b}_1$ contribute to the annihilation cross section $\sigma_{\rm ann}$, as can be seen in Tab.~\ref{tab:channels}. The three processes of our interest contribute to more than 50\% of the total annihilation cross section. As we shall discuss later in Sec.\ \ref{Sec:LO}, the mixed annihilation $\tilde{t}_1 \tilde{b}_1 \to t b$ dominates as compared to stop-pair or sbottom-pair annihilation. In the last four plots of Fig.\ \ref{fig:channelsI}, we show the relative importance of the channels of our interest in the vicinity of Scenario II. Again, the viable region of parameter space where the relic density is within $2\sigma$ of the experimental value determined by the {\it Planck} satellite closely follows the border between the neutralino and stop LSP regions. In most scenarios along this border the mixture of the contributing processes is similar to the one presented in Tab.~\ref{tab:channels}. However, around $M_{\tilde{q}_L}\sim 1600$ GeV, where the scalar top mass reaches about half of the heavy Higgs mass ($m_{H^0} = 3451.2$ GeV), the composition of the contributing processes changes. The stop-anti-stop annihilation processes enhanced by the Higgs exchange grow in importance. In contrast to the situation around Scenario I, for large masses of the neutralino dark matter and the scalar top NLSP, the annihilation and co-annihilation processes are not efficient enough to produce the required observed relic density which can be partly compensated by lowering the mass difference. For even larger masses the region where the relic density is compatible with the {\it Planck} measurement features a stop LSP, such that neutralino dark matter would be excluded for $M_1 \gtrsim 1800$ GeV.  

\subsection{Leading order}
\label{Sec:LO}

Having shown that the processes in Eqs.\ \eqref{eq:process_tt}, \eqref{eq:process_bb} and \eqref{eq:process_tb} are important in large regions around the two scenarios introduced in the previous section, we now turn to review important features of the leading-order cross sections of these processes. The Feynman diagrams for the processes in question are shown in Fig.\ \ref{fig:diagrams_lo}. The matrix elements of all three processes considered here have contributions from $t$-channel or $u$-channel exchanges of strongly interacting gluinos as well as from electroweak gauginos. Therefore the cross sections can be symbolically written as
\begin{equation}\label{treeASAE}
	\sigma = \sigma_s\big(\alpha_s^2\big) + \sigma_{se}\big(\alpha_s\,\alpha_{e}\big) + \sigma_{e}\big(\alpha^2_{e}\big)\,,
\end{equation}
where $\sigma_s$ is the cross section proportional to the square of the strong coupling constant $\alpha_s^2$, $\sigma_{se}$ is the cross section originating from the interference of the strong and electroweakly interacting parts of the scattering amplitude and $\sigma_e$ is the purely electroweak cross section proportional to the square of the electromagnetic coupling constant $\alpha_{e}^2$.

The decomposition of the total cross section into contributions from different channels and interferences of the three processes under consideration here, is shown in Fig.\ \ref{fig:xsecLOcont}. The cross sections for $\tilde{t}_1 \tilde{t}_1 \to t t$ and $\tilde{b}_1 \tilde{b}_1 \to b b$ in the top left, top right and bottom right panels in Fig.\ \ref{fig:xsecLOcont} show the expected hierarchy, in which the gluino $t$-channel and $u$-channel exchanges dominate the cross section and are about an order of magnitude larger than the next largest contribution which is the interference of the gluino exchange with the electroweak $t$- and $u$-channels. The contribution from the interference between the gluino exchange diagrams and the gaugino exchange diagrams is yet another order of magnitude larger than the purely electroweak contribution. As argued before, in scenarios where the processes in Eqs.\ \eqref{eq:process_tt} to \eqref{eq:process_tb} are important, the lightest neutralino is bino-like and the gluino mass is relatively small. These facts imply that the neutralino-squark-quark coupling and the gluino-squark-quark coupling differ mainly by the coupling constant. Therefore, the hierarchy observed in Fig.~\ref{fig:xsecLOcont} is simply due to the ratio of the different coupling constants $\alpha_s(\sqrt{m_{\tilde{t}_1}m_{\tilde{t}_2}})/\alpha_e(m_Z)$.  

\begin{figure*}[t]
	\begin{center}
	\begin{picture}(550,150)
	\put(10,0){\resizebox{!}{5.2cm}{\includegraphics{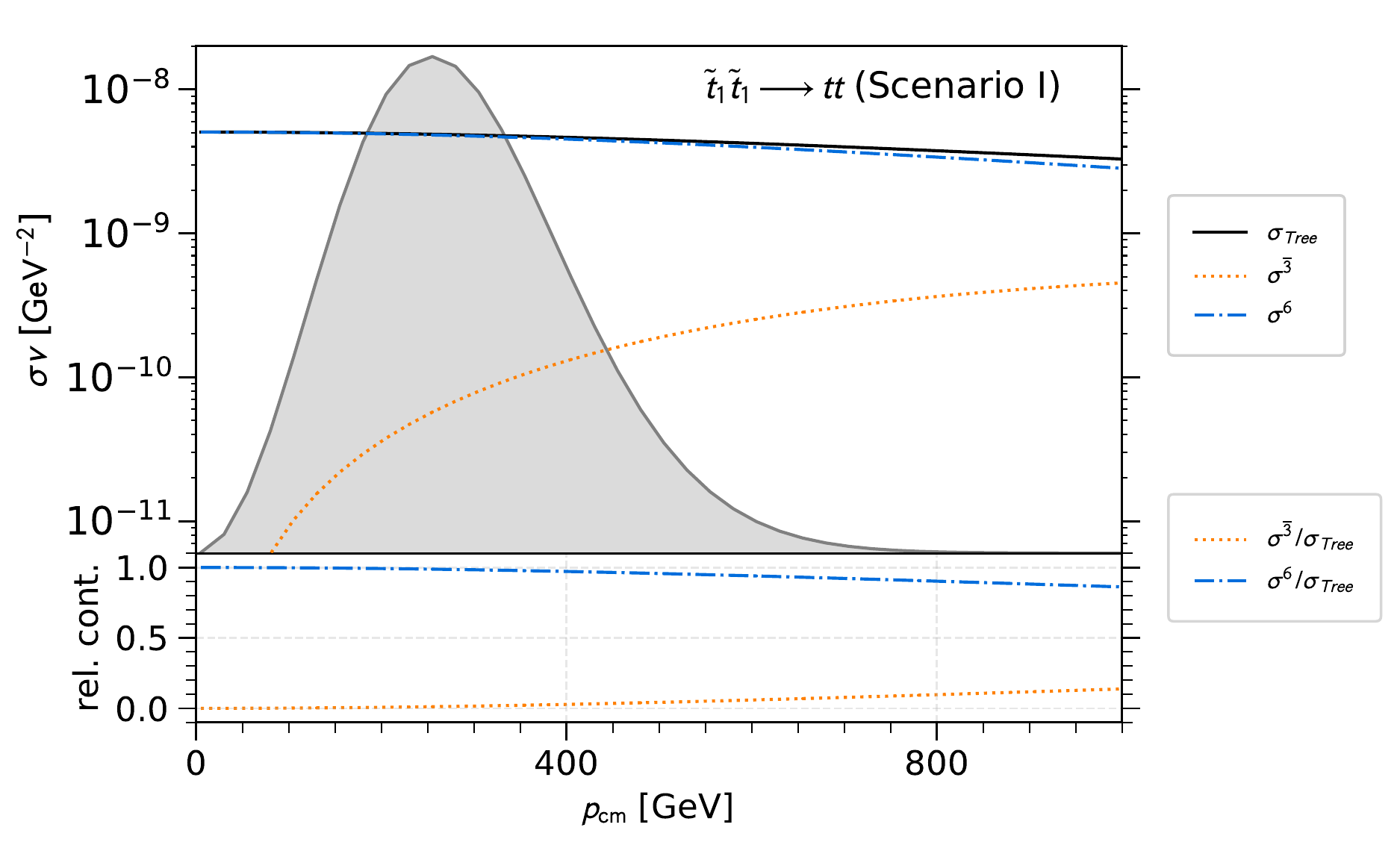}}}
	\put(260,0){\resizebox{!}{5.2cm}{\includegraphics{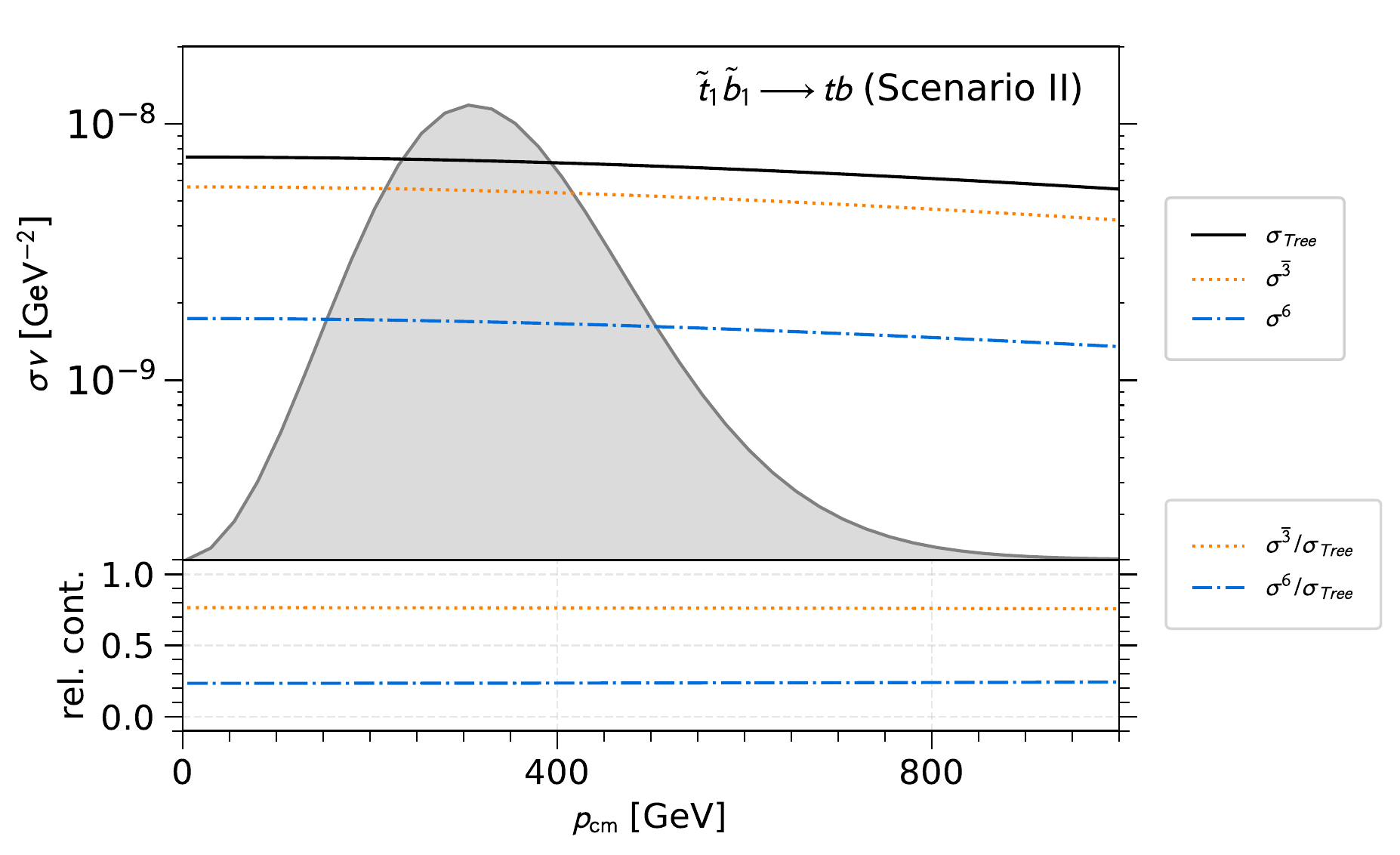}}}
	\end{picture}
	\end{center}
	\vspace*{-5mm}
\caption{Upper part: Decomposition of the leading order cross section into color basis in typical pMSSM scenarios for identical (left) and nonidentical (right) incoming particles. The superscripts $\overline{\mathbf 3}$ and $\mathbf 6$ refer to the respective color representation. Lower part: Contributions relative to the total tree-level cross section $\sigma_{\rm Tree}$.}
\label{fig:xsecLOcolor}
\end{figure*}

The only process where this hierarchy is not present is the annihilation $\tilde{t}_1 \tilde{b}_1 \to t b$. Here the hierarchy observed in the other two processes is modified due to a few factors. First, there is no gluino $u$-channel exchange. Then, the gluino mass in this scenario is larger, and this processes proceeds also through a chargino $u$-channel exchange. The larger gluino mass together with the missing $u$-channel suppresses the gluino contribution compared to the other two processes. Moreover, in the case of the higgsino-like chargino exchange, the Yukawa component of the chargino-squark-quark coupling is not suppressed as in the case of bino-like neutralino. The combination of these effects results in the interference between the gluino and the chargino exchange being suppressed with respect to the pure gluino contribution only by a factor of about two. On top of that, the electroweak contribution is comparable with the gluino-chargino interference.

\subsection{Color decomposition}
\label{Sec:color}

Another important aspect of the processes we investigate is the fact that both initial and final state particles carry color. The color structure of the initial and final state will be extremely relevant later in the discussion of the next-to-leading SUSY-QCD corrections and their resummation. Both scalar quarks in the initial state (and also the quarks in the final state) transform under the fundamental representation of the $SU(3)$ group (denoted here as ${\mathbf 3}$ due to the dimensionality of the representation). The two particle system however transforms under a tensor product of the corresponding representations ${\mathbf 3} \otimes {\mathbf 3}$ which can be decomposed via a Clebsch-Gordan decomposition into $SU(3)$-invariant subspaces as 
\begin{equation}
    {\mathbf 3} \otimes {\mathbf 3} = \mathbf{\overline{3}} \oplus \mathbf{6}\,.
\end{equation}
In order to construct a color basis adapted to our matrix element, we can use the Clebsch-Gordan coefficients of the decomposition
\begin{eqnarray}
    C^{\{\overline{\mathbf 3}\}}_{\alpha\, a_1a_2} &=& \frac{1}{\sqrt{2}}\epsilon_{\alpha a_1 a_2}\,,\qquad \alpha =1,2,3\,,\\ \nonumber
    C^{\{\mathbf 6\}}_{\alpha\, a_1a_2} &=& \frac{1}{2}\left(\delta_{\alpha_1 a_1}\delta_{\alpha_2 a_2}+\delta_{\alpha_1 a_2}\delta_{\alpha_2 a_1}\right)\,,\quad\ \alpha_i = 1,\ldots,6\,,\\
\end{eqnarray}
where the indices $a_{1,2}$ can take the values $1$ to $3$ (for details see Ref.\ \cite{Beneke:2009rj}). The basis is constructed by considering that $SU(3)$ color symmetry is an exact symmetry of the theory and so the color is conserved between the initial and final states. That means if a pair of initial state particles transforms in an irreducible representation of the $SU(3)$ group, the pair of final state particles must transform in the same representation. After proper normalization, we can combine the Clebsch-Gordan coefficients into the following basis relevant for our processes
\begin{equation}
    C^{\{\overline{\mathbf 3},\overline{\mathbf 3}\}}_{a_1a_2a_3a_4}=\frac{1}{\sqrt{2N_c(N_c-1)}}\left(\delta_{a_1 a_3}\delta_{a_2 a_4}-\delta_{a_1 a_4}\delta_{a_2 a_3}\right)\,,
\end{equation}
and
\begin{equation}
    C^{\{\mathbf 6,\mathbf 6\}}_{a_1a_2a_3a_4}=\frac{1}{\sqrt{2N_c(N_c+1)}}\left(\delta_{a_1 a_3}\delta_{a_2 a_4}+\delta_{a_1 a_4}\delta_{a_2 a_3}\right)\,.
\end{equation}
The matrix element can be expanded in this basis as
\begin{equation}\label{eq:MtreeCol}
    M_{stij} = M_{\overline{\mathbf 3}}\, C^{\{\overline{\mathbf 3},\overline{\mathbf 3}\}}_{stij} + M_{\mathbf 6}\, C^{\{\mathbf 6,\mathbf 6\}}_{stij}\,,
\end{equation}
where $s,t,i$ and $j$ are the color indices of the incoming and the outgoing particles. Given the orthonormality of the basis, the triplet and sextet parts of the amplitude can be determined as
\begin{equation}
	M_{\overline{\mathbf 3}} = M_{stij} C^{\{\overline{\mathbf 3},\overline{\mathbf 3}\}}_{stij}, 
	\qquad 
	M_{\mathbf 6} = M_{stij} C^{\{\mathbf 6,\mathbf 6\}}_{stij}\,. 
\end{equation}
In the case of the annihilation process $\tilde{t}_1 \tilde{t}_1 \to t t$ or $\tilde{b}_1 \tilde{b}_1 \to b b$, the triplet and the sextet matrix elements are a linear combination of the gluino and gaugino $t$-channel and $u$-channel exchanges. At tree-level the explicit expression for the triplet part of the matrix element is 
\begin{align}\label{eq:tripletLO}
    M_{\overline{\mathbf 3}} = \frac{(N_c^2-1)}{2\sqrt{2N_c(N_c-1)}}& (-M_{\tilde{g}}^t + M_{\tilde{g}}^u)\\ 
    & + \frac{N_c (N_c-1)}{\sqrt{2N_c(N_c-1)}} (M_{\tilde{\chi}}^t - M_{\tilde{\chi}}^u) \,. \nonumber
\end{align}
Analogously, the sextet part of the matrix element is
\begin{align}\label{eq:sextetLO}
    M_{\mathbf 6} = \frac{(N_c^2-1)}{2\sqrt{2N_c(N_c+1)}}& (M_{\tilde{g}}^t + M_{\tilde{g}}^u)\\ &+ \frac{N_c(N_c+1)}{\sqrt{2N_c(N_c+1)}}(M_{\tilde{\chi}}^t + M_{\tilde{\chi}}^u)\,.\nonumber
\end{align}
The same decomposition can be performed for the process $\tilde{t}_1 \tilde{b}_1 \to t b$ and the explicit results given in Eqs.\ \eqref{eq:tripletLO} and \eqref{eq:sextetLO} can be used after setting $M_{\tilde{g}}^u = 0$ and interpreting $M_{\tilde{\chi}}^u$ as the $u$-channel chargino exchange. The squared amplitude is then in all cases given simply by
\begin{equation}
    |M|^2 = |M_{\overline{\mathbf 3}}|^2+|M_{\mathbf 6}|^2\,,
\end{equation}
where due to the orthonormality of the color basis, there is no interference between the triplet and the sextet matrix elements.

The leading order triplet and sextet cross sections for the relevant processes are shown in Fig.\ \ref{fig:xsecLOcolor}. The general behavior of the color decomposed cross sections for the processes $\tilde{t}_1 \tilde{t}_1 \to t t$ and $\tilde{b}_1 \tilde{b}_1 \to b b$ is very similar. Both processes contain identical particles in the initial state and are symmetric with respect to their interchange. Given that the color basis vector $C^{\{\overline{3},\overline{3}\}}$ is anti-symmetric with respect to the same interchange, the partial wave of the the triplet cross section is a $p$-wave, making its contribution to the relic density subdominant. For these two processes only the sextet color combination contributes. In the case of the last process $\tilde{t}_1 \tilde{b}_1 \to t b$, the symmetry argument does not apply and both color combinations contain an $s$-wave and contribute equally to the relic density.

%
\section{Next-to-leading order}
\label{Sec:NLO}
In this section we will discuss the details of our analytical calculation of the full SUSY-QCD corrections to squark pair annihilation into a pair of quarks. We first concentrate on the virtual corrections and treatment of the UV divergencies in the case of squark pair annihilation. We continue with the discussion of the treatment of IR divergencies. Finally, we address the Sommerfeld enhancement, its treatment and impact on the full SUSY-QCD correction to the squark pair annihilation.

Before we discuss specific details of the next-to-leading order calculation, we will address the systematics of SUSY-QCD corrections to processes which at leading order have both strong and electroweak contributions (see Eq.~\eqref{treeASAE}). If we consider any radiative corrections (SUSY-QCD or electroweak) to the processes in question, the cross section including the next-to-leading order corrections can be symbolically written as
\begin{align}\label{nloASAE}
    \sigma^{\rm NLO} = \sigma^{\rm Tree} &+ \Delta\sigma_s^{\rm NLO}\big(\alpha_s^3\big) + \Delta\sigma_{se}^{\rm NLO}\big(\alpha_s^2\alpha_e\big)\\ &+ \Delta\sigma_{e}^{\rm NLO}\big(\alpha_s\alpha_e^2\big) + \Delta\sigma_{ee}^{\rm NLO}\big(\alpha_e^3\big)\,.\nonumber
\end{align}
The SUSY-QCD corrections contribute to the $\Delta\sigma_s^{\rm NLO}$, $\Delta\sigma_{se}^{\rm NLO}$ and $\Delta\sigma_{e}^{\rm NLO}$ parts of the NLO cross section whereas the electroweak corrections would contribute to the $\Delta\sigma_{se}^{\rm NLO}$, $\Delta\sigma_{e}^{\rm NLO}$ and $\Delta\sigma_{ee}^{\rm NLO}$ parts. Both classes of corrections, the SUSY-QCD and the electroweak, are ultraviolet and infrared finite and gauge independent by themselves making them formally consistent. 

The first and leading term in the NLO correction is $\Delta\sigma_s^{\rm NLO}\big(\alpha_s^3\big)$ which receives contributions only from SUSY-QCD corrections. In particular these are the SUSY-QCD corrections to the gluino exchange diagrams interfered with the gluino tree-level contribution. These corrections are the main result of this analysis.

The following term $\Delta\sigma_{se}^{\rm NLO}\big(\alpha_s^2\alpha_e\big)$ receives contributions from three sources - from the interference of the SUSY-QCD corrected gluino exchange with the electroweak gaugino exchange, from the interference of the SUSY-QCD corrected electroweak gaugino exchange with the gluino diagrams and the last contribution would come from electroweak corrections to the gluino exchange interfered with the gluino tree-level. The last contribution is not included in this analysis and even though it is formally of the same order, due to the small size of the electroweak corrections which are typically a factor 10 smaller than SUSY-QCD ones this last contribution is the smallest of the three. This way our analysis provides also the leading corrections in the term $\Delta\sigma_{se}^{\rm NLO}\big(\alpha_s^2\alpha_e\big)$.  

The third term $\Delta\sigma_{e}^{\rm NLO}\big(\alpha_s\alpha_e^2\big)$ contains the interference of the SUSY-QCD corrected electroweak gaugino exchange with the leading order electroweak gaugino diagrams as well as electroweak corrections to both parts of the interference between the gluino and the electroweak gaugino exchange.

The last term $\Delta\sigma_{ee}^{\rm NLO}\big(\alpha_e^3\big)$ is not considered here as it contains only electroweak corrections to the electroweak parts of the cross section.

The analysis presented here does not consider electroweak corrections as they are for the most part subleading and contribute about 1\% to 3\% correction \cite{Baro:2007em, Baro:2009na}. In some instances however, the electroweak corrections and specifically the Yukawa corrections can become important \cite{Carena:1999py}. Even though we do not calculate electroweak corrections in this analysis, the leading effects of the enhanced Yukawa corrections are taken into account as described in \cite{Harz:2012fz}. In particular, these become relevant in the case of chargino exchange in the Scenario II (neutralino exchanges are not as enhanced due to the lightest neutralino being a pure bino in both scenarios).

As the discussion below shows, the SUSY-QCD corrections presented here are the dominant corrections even in scenarios with large $\tan\beta$ and are even more dominant owing to the presence of the Sommerfeld enhancement.

\begin{figure*}
	\includegraphics[scale=1.2]{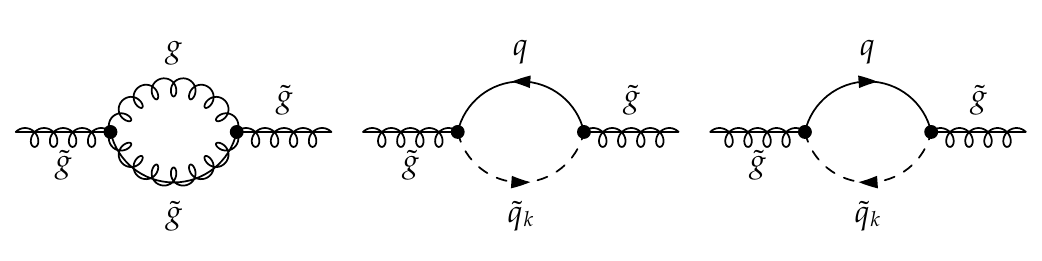}
	\vspace*{-4mm}
    \caption{Gluino self-energy diagrams relevant for the gluino mass and wave function renormalization.}
    \label{fig:diagrams_self}
	\begin{center}
	\begin{picture}(400,80)
	\put(0,0){\subfloat{\includegraphics[scale=1.0]{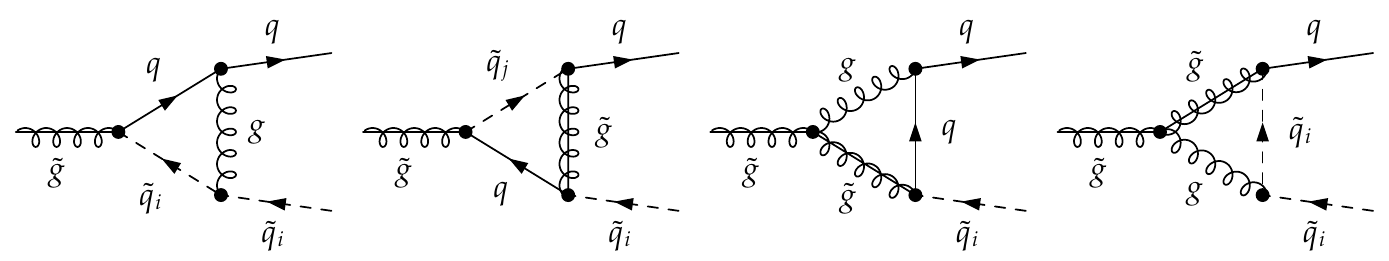}}}
	\end{picture}
	\begin{picture}(200,80)
	\put(0,0){\subfloat{\includegraphics[scale=1.0]{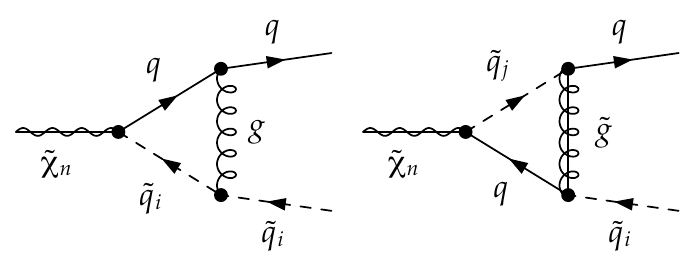}}}
	\end{picture}
	\end{center}
	\vspace*{-4mm}
    \caption{Vertex corrections diagrams associated with the squark pair-annihilation into quark pairs depicted in Fig.\ \ref{fig:diagrams_lo}.}
    \label{fig:diagrams_vert}
	\begin{center}
	\begin{picture}(400,80)
	\put(0,0){\includegraphics[scale=1.0]{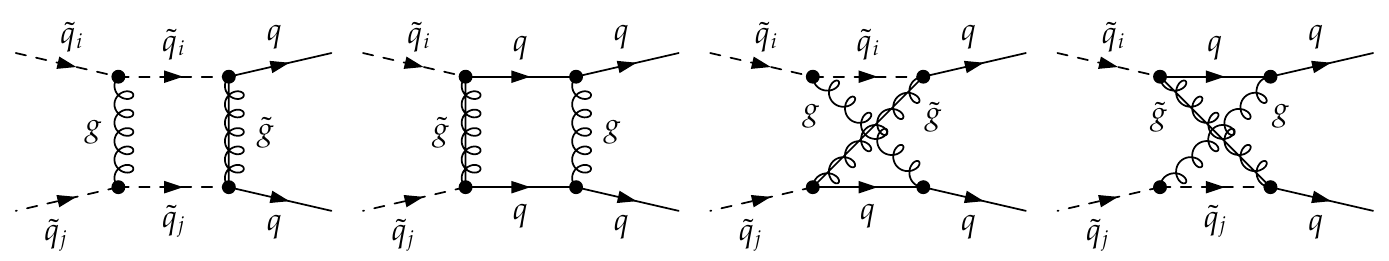}}
	\end{picture}
	\begin{picture}(200,80)
	\put(0,0){\includegraphics[scale=1.0]{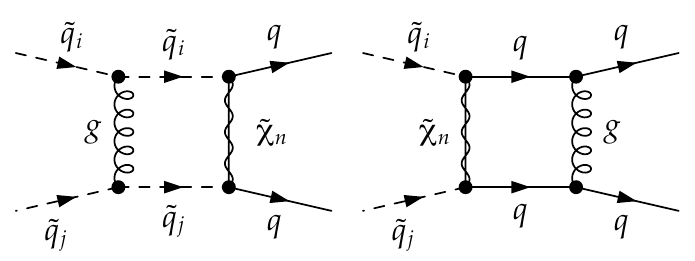}}
	\end{picture}
	\end{center}
	\vspace*{-4mm}
    \caption{Box diagrams associated with the squark pair-annihilation into quark pairs depicted in Fig.\ \ref{fig:diagrams_lo}.}
    \label{fig:diagrams_box}
	\begin{center}
	\begin{picture}(500,80)
	    \put(0,0){\includegraphics[scale=0.9]{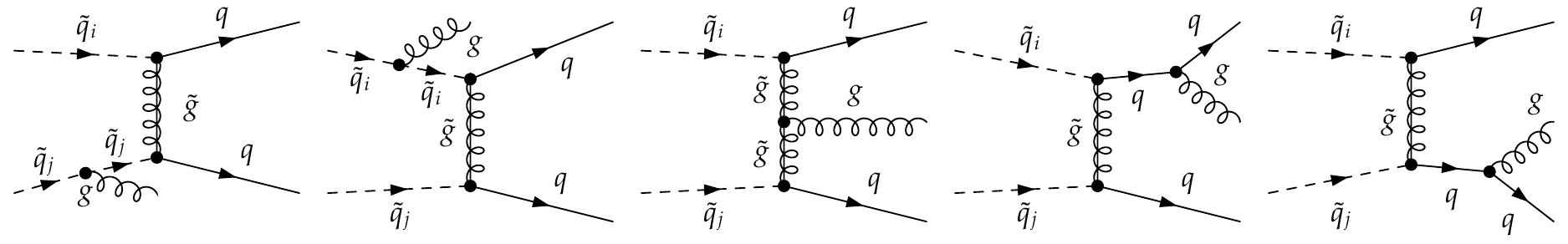}}
	\end{picture}
	\begin{picture}(400,80)
	    \put(0,0){\includegraphics[scale=0.9]{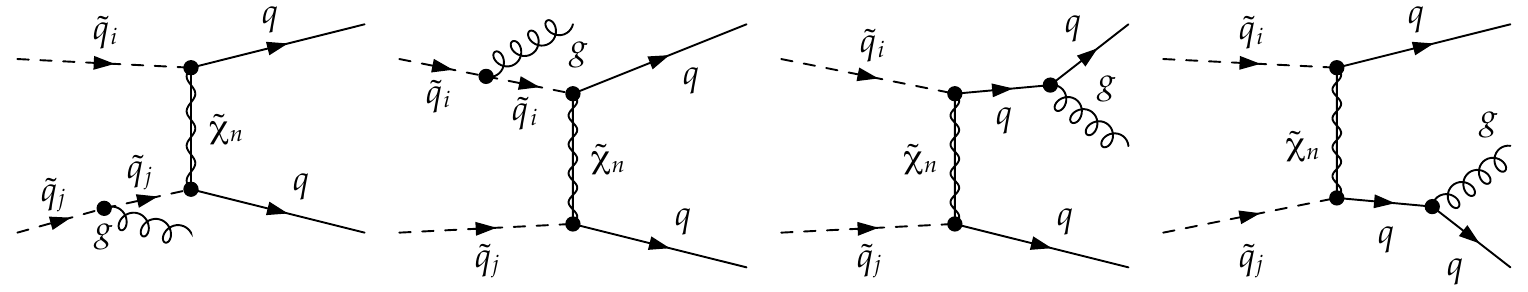}}
	\end{picture}
	\end{center}
	\vspace*{-4mm}
    \caption{Real gluon radiation diagrams associated with squark pair-annihilation into quark pairs depicted in Fig.\ \ref{fig:diagrams_lo}.}
\label{fig:diagrams_real}
\end{figure*}

\subsection{Virtual corrections and renormalization}

The class of processes considered here -- the squark pair annihilation to a pair of quarks -- include strongly interacting particles in the initial state, in the final state, and even in the intermediate state (in the case of the gluino $t$- and $u$-channel exchanges). As a consequence, the next-to-leading order SUSY-QCD corrections include contributions from vertex corrections, propagator corrections, and box corrections. The corresponding diagrams are displayed in Figs.\ \ref{fig:diagrams_self}, \ref{fig:diagrams_vert} and \ref{fig:diagrams_box}, respectively. The next-to-leading order corrections to the squark-pair annihilation contain one-loop diagrams which are ultraviolet (UV) and infrared (IR) divergent. The UV divergencies are cancelled by renormalization of the parameters of the theory and the fields. In order to cancel the IR divergences, one has to properly define an infrared safe cross section, which is done by including also $2\rightarrow 3$ processes with an additional gluon being radiated (see Fig.~\ref{fig:diagrams_real}). 

All one-loop diagrams have been calculated in the SUSY-invariant dimensional reduction scheme (\DRbar) \cite{Siegel:1979wq,Stockinger:2005gx} where, similar to the minimal subtraction scheme (\MSbar), the number of space-time dimensions is set to $D=4-2\varepsilon$ in order to regularize otherwise divergent loop integrals. We have used the standard Passarino-Veltman reduction \cite{Passarino:1978jh, Denner:1991kt} in order to reduce the tensor loop integrals in the one-loop amplitudes to only a few scalar integrals which have then been evaluated using well-known results, e.g., Refs.\ \cite{Dittmaier:2003bc,Denner:2010tr}. Our analytical calculations were performed and verified with the help of publicly available tools \FeynArts\ \cite{FeynArts}, \FeynCalc\ \cite{FeynCalc}, and \FORM\ \cite{FORM}.

In order to cancel UV divergencies of the one-loop amplitude, we renormalize the MSSM by introducing counterterms to the relevant parameters and fields. Because we consider SUSY-QCD corrections to a process involving scalar quarks, quarks and intermediate gluinos, the relevant parameters are the ones that receive corrections proportional to the strong coupling constant $\alpha_s$. Every renormalization scheme is characterized by a careful selection and definition of its input parameters. In a series of previous analyses \cite{Harz:2012fz, Harz:2014tma}, we have put forward a renormalization scheme which combines the advantages of both on-shell and \DRbar\ renormalization schemes and treats consistently the renormalization of the quark and the squark sector. In these sectors the input parameters are chosen to be the on-shell masses $m_t$, $m_{{\tilde t}_1}$, $m_{{\tilde b}_1}$, $m_{{\tilde b}_2}$, the mass of the bottom quark $m_b$, and trilinear couplings of the third generation $A_t$ and $A_b$. The last three parameters are defined in the \DRbar\ renormalization scheme. We define our renormalization scale as $\mu_R=Q_{\mathrm{SUSY}}=\sqrt{m_{\tilde{t}_1} m_{\tilde{t}_2}}$. In the following, we will comment on new aspects of the renormalization relevant for our analysis presented here such as the renormalization of the mass of the gluino and its wave function. 

\subsubsection{Gluino mass and wave-function renormalization}

In all our analyses we adopt a convention where not only the complete next-to-leading order corrections to the cross section should be rendered UV finite, but also all building blocks such as the $n$-particle irreducible Green's functions should be UV finite as well. This choice requires to introduce wave-function renormalization constants not only to the fields that correspond to the initial and final state particles but also to fields that give rise to internal propagators. In our case, the only strongly interacting particle that appears in a propagator in our amplitude is the gluino which has not yet been treated within the \DMNLO\ analysis.

In the case of the gluino, both wave function and mass have to renormalized in order for the vertex corrections and propagator corrections to be separately UV finite. To this end, we introduce counterterms to the gluino wave function $\delta Z^{L,R}_{\tilde{g}}$ and the gluino mass $\delta m_{\tilde{g}}$ as
\begin{eqnarray}
	\psi_{\tilde{g}} &\rightarrow & \Big(1+ \tfrac12 \delta Z^L_{\tilde{g}}P_L+ \tfrac12 \delta Z^R_{\tilde{g}}P_R\Big)\,\psi_{\tilde{g}}\,,\\
	m_{\tilde{g}} &\rightarrow & m_{\tilde{g}} + \delta m_{\tilde{g}}\,.
\end{eqnarray}
All gluino counterterms are determined by considering the gluino two-point Green's function. The one-loop contribution to the two-point Green's function is given by the gluino self-energy diagrams shown in Fig.\ \ref{fig:diagrams_self}. This contribution can be parametrized as
\begin{align}
	\Pi(k) = \not\!k\, \big(P_L \Pi^L(k^2) &+ P_R \Pi^R(k^2)\big)\\ &+  m_{\tilde{g}} \big(P_L \Pi^{SL}(k^2) +  P_R \Pi^{SR}(k^2)\big)\,,\nonumber
\end{align}
where $\Pi^{L,R}(k^2)$ and $\Pi^{SL,SR}(k^2)$ are form factors which receive contributions from the corresponding self-energy diagrams. 

Even though the gluino is not an external particle in the processes considered in this analysis, we still require that the residue of the propagator at one-loop order is set to unity. This condition fixes the wave-function renormalization constant using the form-factors as
\begin{align}
	\delta Z^{L/R}_{\tilde{g}} &= -\Pi^{L/R}(m_{\tilde{g}}^2) + \tfrac12 \Big(\Pi^{SL/SR}(m_{\tilde{g}}^2)\\ \nonumber &-\Pi^{SR/SL}(m_{\tilde{g}}^2)\Big)
	- m_{\tilde{g}}^2  \Big(\dot{\Pi}^{L/R}(m_{\tilde{g}}^2)+\dot{\Pi}^{R/L}(m_{\tilde{g}}^2)\\ \nonumber &+ 
	\dot{\Pi}^{SL/SR}(m_{\tilde{g}}^2)+\dot{\Pi}^{SR/SL}(m_{\tilde{g}}^2)\Big)\,,
\end{align}
where $\dot{\Pi}^{i}(m_{\tilde{g}}^2) = \frac{\partial}{\partial k^2}\left.\Pi^{i}(k^2)\right|_{k^2=m_{\tilde{g}}^2}$. Using the gluino wave-function counterterm renders both the propagator and vertex corrections separately UV finite. Moreover, given that the gluino is not an external particle, renormalization of its wave-function is not necessary for UV finiteness of the full next-to-leading order amplitude and so the full amplitude is independent of the gluino wave-function counterterm. This constitutes another consistency check of our analytical calculation.

The mass counterterm is determined from the on-shell condition which requires that the gluino mass $m_{\tilde{g}}$, which is an input parameter, is identical with the position of the pole of the gluino propagator. It is given as
\begin{align}
	\delta m_{\tilde{g}} = \tfrac12 m_{\tilde{g}}\Re &\left(\Pi^L(m_{\tilde{g}}^2)+\Pi^R(m_{\tilde{g}}^2)\right. \nonumber \\
	&\left.+\Pi^{SL}(m_{\tilde{g}}^2)+\Pi^{SR}(m_{\tilde{g}}^2)\right)\,.
\end{align}

\subsubsection{Some remarks on the renormalization of the squark sector}

Even though we have discussed the details of the renormalization of squark parameters in Ref.\ \cite{Harz:2012fz}, we would like to remark here on one feature of the renormalization scheme relevant for evaluating the results of this analysis. As discussed in Ref.\ \cite{Harz:2012fz}, we use the relation between the non-diagonal squark mass matrices for up-type and down-type squarks
\begin{equation}
	U^{\tilde q} 
	\left( \begin{array}{cc} m^2_{LL} & m^2_{LR} \\  
	m^2_{RL} & m^2_{RR} \end{array} \right) 
	(U^{\tilde q})^\dag
	~=~
	\left( \begin{array}{cc} m^2_{\tilde{q}_1} & 0 \\ 0 & m^2_{\tilde{q}_2} \end{array} \right)	\,,\label{Eq:massMatrix}
\end{equation}
where
\begin{eqnarray}
	m^2_{LL} &=& M_{\tilde Q}^2 + (I^{3L}_q \!-\! e_q\,s_W^2)\cos2\beta\,m_{Z}^{\,2} + m_{q}^2\,,\\
	m^2_{RR} &=& M_{\{\tilde U,\,\tilde D\}}^2 + e_{q}\,s_W^2 \cos2\beta\,m_{Z}^{\,2} + m_q^2\,,\\
	m^2_{LR} &=& m^2_{RL} = m_q\big( A_q - \mu \,(\tan\beta)^{-2 I^{3L}_q}\big)\,,
\end{eqnarray}
to relate the input parameters in the whole squark sector, which are defined in different renormalization schemes. In the next step, we determine the dependence of the soft supersymmetry-breaking squark mass parameters $M_{\tilde Q}^2$ and $M_{\{\tilde U,\,\tilde D\}}^2$ of the three on-shell masses $m_{\tilde{b}_1}$, $m_{\tilde{b}_2}$, $m_{\tilde{t}_1}$ and the input parameters contained in $m^2_{LR}$ ($A_q$, $m_q$). In the sbottom sector the parameters $M_{\tilde Q}^2$ and $M_{\tilde D}^2$ are contained in the matrix elements $m^2_{\tilde{b},LL}$ and $m^2_{\tilde{b},RR}$, which are given by the on-shell masses as
\begin{eqnarray}\nonumber
    m^2_{\tilde{b},LL} &=& \tfrac12 \big(m^2_{\tilde{b}_1}+m^2_{\tilde{b}_2}\big) \pm \tfrac12 \sqrt{\big(m^2_{\tilde{b}_1}-m^2_{\tilde{b}_2}\big)^2 - 4 m^4_{\tilde{b},LR}}\,,\\ \nonumber
    m^2_{\tilde{b},RR} &=& \tfrac12 \big(m^2_{\tilde{b}_1}+m^2_{\tilde{b}_2}\big) \mp \tfrac12 \sqrt{\big(m^2_{\tilde{b}_1}-m^2_{\tilde{b}_2}\big)^2 - 4 m^4_{\tilde{b},LR}}\,.\\
\end{eqnarray}
One notices that there are two possible values for the parameters $M_{\tilde Q}^2$ and $M_{\tilde D}^2$ and consequently also for the third parameter $M_{\tilde U}^2$ which can be found in one of the diagonal elements of the non-diagonal scalar top mass matrix and is related to the first two parameters
through
\begin{align}
    m^2_{\tilde{t},RR} &= \frac{1}{m^2_{\tilde{t},LL}-m^2_{\tilde{t}_1}}\nonumber\\
    &\times \big(m^2_{\tilde{t}_1} m^2_{\tilde{t},LL}-m^4_{\tilde{t}_1} + m^4_{\tilde{t},LR}\big)\,.
\end{align}
The parameter $M_{\tilde{Q}}^2$ is common to both elements $m^2_{\tilde{b},LL}$ and $m^2_{\tilde{t},LL}$. Given the freedom to choose from two possible solutions for the squark soft supersymmetry-breaking mass parameter, we can end up with two possibly very different mass matrices and two different sets of mixing matrices (three out of four masses of the squarks would be the same in both cases as they are used as input). In order to ensure a naturally small correction to the mixing matrices when changing between our and the \DRbar\ renormalization scheme, we always select the solution which preserves the hierarchy between the mass matrix elements in the scalar top quark sector $m^2_{\tilde{t},LL}$ and $m^2_{\tilde{t},RR}$ which was present in the pure \DRbar\ scheme.

\subsection{Real corrections}

After treating the ultraviolet divergencies and removing them by renormalization, we are left with the infrared (IR) divergencies in the one-loop amplitude. The IR divergencies are also regularized by performing calculations in a general dimension $D=4-2\varepsilon$ and are subsequently removed by considering an IR-safe observable, which in our case is a cross section with one additional gluon in the final state. The corresponding diagrams are depicted in Fig.\ \ref{fig:diagrams_real}. 

The cross section for the radiation of an additional gluon cannot be calculated analytically. On the other hand, the cancellation of the IR divergencies between the virtual corrections and the real radiation cross section has to be performed analytically. One of the methods how to extract the IR divergencies out of the real radiation cross section is the phase-space slicing method \cite{Denner:1991kt, GieleGlover, HarrisOwens}. This method is based on the fact that the infrared divergence is connected to a specific configuration of the momentum of the gluon. The soft infrared divergence arises when the additional gluon's energy vanishes whereas the collinear infrared divergence arises when the additional gluon is radiated collinearly to the momentum of an external massless particle. The phase-space slicing method uses kinematical cuts to divide the three-particle phase-space into regions where either one or both of the aforementioned configurations of the gluon 4-momentum occur and the remainder where the cross section is IR finite. In the singular regions the full matrix element is replaced by an approximation which can be integrated analytically making the IR divergence explicit. In the non-singular region the full matrix element can be integrated numerically without any obstacles.

In our case, all external particles are massive so the $2\rightarrow 3$ cross section contains only a soft IR divergence. In the singular region where the gluon's energy is smaller than an arbitrary small cutoff $\Delta E$, we use the soft gluon approximation which factorizes the differential cross section as
\begin{equation}
    \left( \frac{\dd\sigma}{\dd\Omega} \right)_{\textnormal{soft}} = F \times \left( \frac{\dd\sigma}{\dd\Omega} \right)_{\textnormal{Tree}},
\end{equation}
where $F$ contains the integral over the phase-space of the gluon
\begin{equation}
    I_{ab} =  \mu^{4-D}\int_{|\vec{k}| \le \Delta E} \frac{\dd^{D-1} k}{(2\pi)^{D-4}}\frac{1}{k^0} \frac{(a.b)}{(k.a)(k.b)},
\label{Eq:SoftIntegral}
\end{equation}
and therefore also the divergence. Here, $k$ is the $4-$momentum of the gluon and $a$ and $b$ are $4-$momenta of two external particles which can emit a gluon. These integrals are given in Refs.\ \cite{Denner:1991kt, tHooft:1978xw}.

The cutoff $\Delta E$ introduced to separate the singular from the non-singular region in the three particle phase-space enters into the calculation of the soft-gluon radiation as well as into the integration of the $2\rightarrow 3$ cross section over the non-singular phase-space. In principle the dependence on this cutoff should disappear in the sum of the contributions from both phase-space regions, but in practice the independence on the cutoff is limited by the numerical stability of the integration over the non-singular region. We have investigated the dependence on the cutoff and found that the integration in our case is stable and independent of the cutoff for a relatively large interval of cutoffs around $\Delta E = 10^{-5}\sqrt{s}$.

The complete result after we have included all virtual corrections, the counterterms and the real radiation $2\rightarrow 3$ cross section is UV and IR finite. In contrast to the leading order result which consists of a cross section with two particles in the final state and is implemented in \MO, the complete result is a consistent combination of a one-loop corrected cross section with two particles in the final state and a leading order cross section with three particles in the final state. In \DMNLO,\ the complete result replaces the leading order result of \MO.
\subsection{Sommerfeld resummation}
\label{Sec:Resum}

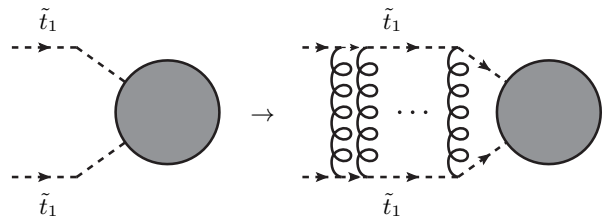
\begin{figure}
\centering
\begin{picture}(230,90)(0,0)
    \put(0,15){\mbox{\resizebox{!}{2.0cm}{\includegraphics{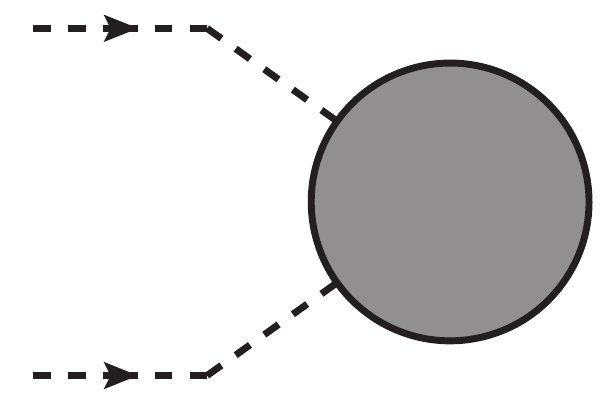}}}}
    \put(95,40){$\rightarrow$}
    \put(110,15){\mbox{\resizebox{!}{2.0cm}{\includegraphics{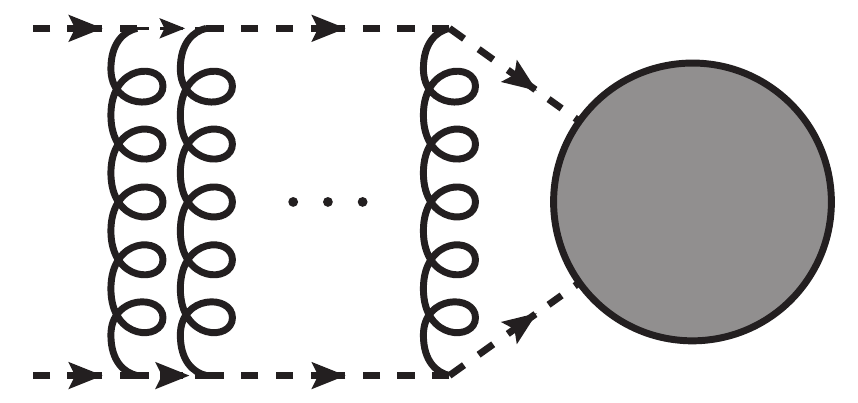}}}}
    \put(15,75){$\tilde{t}_1$}
    \put(15,5){$\tilde{t}_1$}
    \put(145,75){$\tilde{t}_1$}
    \put(145,5){$\tilde{t}_1$}
\end{picture}
\caption{Ladder diagram for a LO Coulomb potential.}
\label{fig:Sommerfeld}
\end{figure}

When calculating the relic density in our case, an important contribution comes from the annihilation of squarks moving with non-relativistic velocities. If annihilating, non-relativistic particles couple to much lighter force mediators which in our case are the gluons, the annihilation cross section is modified due to the well-known Sommerfeld effect \cite{Sommerfeld:1931}. The reason for this modification is that the exchange of $n$ gluons between the initial state squarks (see Fig.~\ref{fig:Sommerfeld}) contains a correction proportional to $(\alpha_s/ v_{\mathrm{rel}})^n$. This correction becomes significant and can spoil the perturbative expansion when the relative velocity of the squark pair $v_{\mathrm{rel}}$ is comparable to the strong coupling constant $\alpha_s$. In such a case these contributions have to be resummed to all orders leading to the Sommerfeld effect. 

Small relative velocities occur naturally in the freeze-out regime, $E_{\mathrm{kin}} \sim T_{\mathrm{FO}} \sim m_{\tilde{\chi}^0_1}/25$ and therefore the Sommerfeld resummation is expected to be relevant in the case of dark matter annihilation in general and in our case in particular. 

As for our processes of interest, $\tilde{q}_i \tilde{q}'_j \rightarrow qq'$, the cross section is dominated by the $s$-wave component (see Fig.~\ref{fig:xsecLOcolor}). We can factorize the resummed cross section as
\begin{equation}
\label{eq:sigmares}
    (\sigma v)_{\mathrm{resum}} = S_{0,\bf{[\overline{3}]}}\,(\sigma v)^{\mathrm{Tree}}_{\tilde{q} \tilde{q} \rightarrow qq, \bf{[\overline{3}]}} + S_{0,\bf{[6]}}\,(\sigma v)^{\mathrm{Tree}}_{\tilde{q} \tilde{q} \rightarrow qq, \bf{[6]}} \,,
\end{equation}
where we have split the leading-order cross section according to its color contribution to the triplet and sextet configurations (see Sec.\ \ref{Sec:color}). $S_{0,\{\bf{[\overline{3}],[6]}\}}$ indicate the corresponding $s$-wave Sommerfeld factors, whose evaluation we discuss in the following. In the non-relativistic limit, the resummation of the gluon exchange diagrams as shown in  Fig.~\ref{fig:Sommerfeld} amounts to solving the Schr\"odinger equation with the corresponding Coulomb potential. The Coulomb potential including gluon loops at next-to-next-to-leading order was evaluated in \cite{Fischler:1977yf} and extended by fermion loops in Ref.\ \cite{Billoire:1979ih}. In the \MSbar\ scheme, the Coulomb potential reads \cite{Kauth:2011bz}

\begin{align}
\label{eq:Coulombpotential}
	V^{[\bf{R}]}(r) &= C^{[\bf{R}]}\,\frac{\alpha_{s}(\mu^{[\bf{R}]}_C)}{r} \\  &\times \bigg\{ 1 + \frac{\alpha_s(\mu^{[\bf{R}]}_C)}{4\pi} \bigg[2 b_0 \bigg(\ln(\mu^{[\bf{R}]}_C r) + \gamma_E \bigg)+a_1 \bigg]  \bigg\}\,, \nonumber
\end{align}
with $\gamma_E=0.5772$ being the Euler-Mascheroni constant. Furthermore, we have defined 
\begin{align}
	b_0 &= \frac{11}{3}C_A - \frac{4}{3}T_f n_f\,, \\
	a_1 &= \frac{31}{9}C_A - \frac{20}{9}T_f n_f\,,
\end{align}
where $b_0$ corresponds to the one-loop $\beta$-function coefficient with $C_A=3$ and $T_F=1/2$. We treat the top as the only massive quark, such that we set the number of massless quarks to five ($n_f=5$). The Coulomb potential, given in Eq.~\eqref{eq:Coulombpotential}, describes the interaction of any non-relativistic colored particles transforming in general $SU(3)$-representations $\bf{R_1}$ and $\bf{R_2}$. The color structure of such a scattering process can be decomposed as
\begin{equation}
    \bf{R_1}\otimes \bf{R_2} = \bf{R^{\prime}} \oplus \bf{R^{\prime\prime}}\,.
\end{equation}
The color factor $C^{[\bf{R}]}$ is given in terms of the quadratic Casimir operators of the relevant $SU(3)$-representations as
\begin{align}
&C^{[\bf{R^{i}}]} = T_1^aT_2^a = \frac12\,[(T_1^a + T_2^a)^2 - (T_1^a)^2 - (T_2^a)^2]\\ \nonumber &= \frac12\,(C_2^{\bf{R^i}} - C_2^{\bf R_1} - C_2^{\bf R_2} )\qquad \mbox{where}\quad\bf{R^{i}} = \bf{R^{\prime}},\bf{R^{\prime\prime}}\,.    
\end{align}
In the case considered here, the two squarks in the initial state both transform under the fundamental representation of $SU(3)$ and the color decomposition is $\bf 3\otimes 3 = \overline{3} \oplus 6$.
Using the quadratic Casimir operators for the fundamental and the sextet representations, we obtain \cite{Beneke:2010da}
\begin{align}\label{eq:Somcol1}
C^{\bf \overline{3}} &= -\frac{1}{2} \left(1 + \frac{1}{N_c} \right) =  -2/3\,,  &\qquad 
\\ \label{eq:Somcol2}
C^{\bf 6} &=  \frac{1}{2} \left(1 - \frac{1}{N_c} \right)= 1/3\,. &\qquad 
\end{align}
The Sommerfeld factors are then obtained by solving the Schr\"odinger equation
\begin{align}
\label{eq:SchroedingerEQ}
	\Big[ -\frac{2}{m_{\mathrm{red}}} \nabla^2 + V^{[\bf R]}(\mathbf{r}) - \big( \sqrt{s}+i\Gamma_{\tilde{t}_1} \big) \Big]
		&\mathcal{G}^{[\bf R]} \big( \mathbf{r};\sqrt{s}+i\Gamma_{\tilde{t}_1}\big)\nonumber \\
		&= \delta^{(3)}(\mathbf{r})
\end{align}
with the reduced mass $m_{\mathrm{red}}=(m_{\tilde{q}}\, m_{\tilde{q}^\prime})/(m_{\tilde{q}} + m_{\tilde{q}^\prime})$ of the two annihilating particles $\tilde{q}$ and $\tilde{q}^{\prime}$. The solution of the Schr\"odinger equation with the NLO Coulomb potential defined in Eq.~\eqref{eq:Coulombpotential} is given by the Green's function $\mathcal{G}^{\bf{[R]}}(\mathbf{r},E+i\Gamma_{\tilde{t}_1}) = \mathcal{G}^{\bf{[R]}}(\mathbf{r}, \mathbf{r}^\prime=0,E+i\Gamma_{\tilde{t}_1})$. 
The Sommerfeld factor which is used to correct the cross section in Eq.~\eqref{eq:sigmares} is given by a ratio of two Green's functions at the origin ($\bf r = 0$) \cite{Cassel:2009wt,Hagiwara:2008df} 
\begin{align}
\label{eq:Sommerfeldfactor}
	S_{0,\bf{[R]}} = \frac{\Im[\mathcal{G}^{\bf{[R]}}(\mathbf{0},E+i\Gamma_{\tilde{t}_1})]}{\Im[\mathcal{G}_0(\mathbf{0},E+i\Gamma_{\tilde{t}_1})]}\,,
\end{align}
where the Green's function $\mathcal{G}_0(\mathbf{0},E+i\Gamma_{\tilde{t}_1})$ stands for the solution of the Schr\"odinger equation without any Coulomb potential.
The solution to Eq.~\eqref{eq:SchroedingerEQ} at the origin is well known \cite{Kiyo:2008bv}, and we consider here all terms up to NLO,
\begin{align}
	G^{\bf{[R]}}({\bf{0}};\sqrt{s}+i\Gamma_{\tilde{t}_1})&= \frac{i m^2_{\mathrm{red}} v_{\rm s}}{\pi} + \frac{C^{\bf{[R]}}\alpha_{s}(\mu^{[\bf{R}]}_C) m^2_{\mathrm{red}}} {\pi}\nonumber \\
	&\times \Big[ g_{\mathrm{LO}} + \frac{\alpha_{s}(\mu^{[\bf{R}]}_C)}{4\pi}g_{\mathrm{NLO}} \Big]\,,
\end{align}
where the LO and NLO contributions are given by
\begin{align}\label{eq:glo}
	&g_{\mathrm{LO}}\hspace{2mm} = L - \psi^{(0)}, \\ \label{eq:gnlo}
	&g_{\mathrm{NLO}}= 	\beta_0 \Big[ L^2 - 2L(\psi^{(0)} - \kappa\psi^{(1)}) + \kappa\psi^{(2)} + 
		(\psi^{(0)})^2  \nonumber \\
		&-3\psi^{(1)} - 2\kappa\psi^{(0)}\psi^{(1)} 
		+ 4\hspace{2mm} _4F_3(1,1,1,1;2,2,1-\kappa;1)\Big] \nonumber \\
	& + a_1 \Big[L-\psi^{(0)}+\kappa\psi^{(1)} \Big]\,.
\end{align}
In Eqs.~\eqref{eq:glo}-\eqref{eq:gnlo} and in the following, we use 
the short-hand notation
\begin{align}\label{eq:Somparam1}
	\kappa&=\frac{i C^{\bf{[R]}}\alpha_{s}(\mu^{[\bf{R}]}_C)}{2v}, \\ \label{eq:Somparam2}
	v_{\rm s} & =  \sqrt{\frac{\sqrt{s}+i\Gamma_{\tilde{t}_1}-2m_{\mathrm{avg}}}{2 m_{\mathrm{red}}}},\\
	L & =  \ln\frac{i \mu^{[\bf{R}]}_C}{4 m_{\mathrm{red}}v_{\rm s}}\,.
	\label{eq:Somparam3}
\end{align}
Moreover, $\psi^{(n)}=\psi^{(n)}(1-\kappa)$ is the $n$-th derivative of $\psi(z) = \gamma_{\mathrm{E}} + \mathrm{d}/\mathrm{d}z\ln\Gamma(z)$ with the argument $(1 - \kappa)$, $_4F_3(1,1,1,1;2,2,1-\kappa;1)$ a hypergeometric function, and $m_{\mathrm{avg}}=(m_{\tilde{q}}+ m_{\tilde{q}^\prime})/2$ the average mass of the two incoming particles. Note that in the case of identical initial state particles the parameter $v_{\rm s}$ in Eq.~\eqref{eq:Somparam2} is the non-relativistic velocity of one of the incoming particles, and should not be confused with $v_{\mathrm{rel}}=2v_{\rm s}$, the relativistic, relative velocity of the two annihilating particles. In order to calculate the Sommerfeld factor in Eq.~\eqref{eq:Sommerfeldfactor}, we also need the Green's function that solves the system without any potential term in Eq.~\eqref{eq:SchroedingerEQ}, which is given by
\begin{align}
    \Im[\mathcal{G}_0(\mathbf{0},E+i\Gamma_{\tilde{t}_1})] = \frac{m^2_{\mathrm{red}} v_{\rm s}}{\pi}\,.
\end{align}
Finally, we need to fix the scale $\mu_C$ that appears in the potential and has an impact on the evaluation of $\alpha_s$ in the Sommerfeld factor. We follow here the treatment presented in Ref.~\cite{Beneke:2010da} and set 
\begin{align}
    \label{eq:Coulombscale}
    \mu^{[\bf{R}]}_C = \mathrm{max} \left\{ 4 m_{\mathrm{red}} v_{\rm s}, \mu^{[\bf{R}]}_B \right\}\,,
\end{align}
where $4 m_{\mathrm{red}} v_{\rm s}$ is motivated by the typical momentum exchange of the gluons in the ladder diagram and the scale $\mu_B^{[\bf{R}]}$ corresponds to twice the inverse Bohr radius $r_B$. It is defined via 
\begin{align}\label{eq:muc}
\mu_B^{[\bf{R}]} \equiv 2/r_B = 2 C^{[{\bf{R}}]} m_{\mathrm{red}} \alpha_s(\mu_B^{[\bf{R}]})\,.
\end{align}
In order to obtain $\mu_B^{[\bf{R}]}$, we solve Eq.~\eqref{eq:muc} iteratively. 

As the box diagrams in the full NLO calculation also contain the velocity-enhanced part of the one-gluon exchange, which is at the same time already included in the Sommerfeld resummation, we have to subtract this contribution in order to avoid any double counting.  

To isolate the velocity-enhanced term from the box contribution, we expand the box contribution in the relative velocity (for details see App.~\ref{App}). We then construct the subtracted cross section $(\sigma v)^{\mathrm{sub}}_{\mathrm{NLO}}$ based on the expanded matrix element of the box diagrams given in Eq.~\eqref{eq:boxsub}. 
The leading velocity-enhanced term of the subtracted cross section 
is
\begin{align}
    (\sigma v)^{\mathrm{sub}}_{\mathrm{NLO}} \sim \sum_{\bf R}\left(\frac{\alpha_s({\mu_R}) \pi}{v_{\mathrm{rel}}} \right)\,  C^{[{\mathbf R}]}_{\mathrm{box}}\, (\sigma v)^{\bf R}_{\mathrm{Tree}} \,.
    \label{eq:subNLO}
\end{align}
Comparing Eq.~\eqref{eq:subNLO} with the next-to-leading order part of the Sommerfeld resummation which arises from the imaginary part of $g_{\mathrm{LO}}$, namely $g^{\mathrm{sub}}_{\mathrm{LO}} = i \pi/2$, and reads 
\begin{align}
    (\sigma v)^{\mathrm{NLO}}_{\mathrm{resum}}=  \sum_{\bf R}\left(\frac{C^{\bf{[R]}}\alpha_{s}(\mu^{[\bf{R}]}_C) \pi}{2 v_{\rm s}} \right) \, (\sigma v)^{\bf R}_{\mathrm{Tree}}\,,
\end{align}
we see, that given $C^{\bf{[R]}} =  C^{[{\mathbf R}]}_{\mathrm{box}}$ and $v_{\rm s}=v_{\mathrm{rel}}/2$, the two expressions differ in the scale at which the strong coupling constant is being evaluated. While in the perturbative NLO calculation $\alpha_s$ is evaluated at the renormalization scale $\mu_R = \sqrt{m_{\tilde{q}}m_{\tilde{q}^\prime}}$, in Sommerfeld resummation the characteristic scale $\mu_C$ is used. By choosing to use Eq.~\eqref{eq:subNLO} to avoid the double counting, we make use of the fact that the natural scale used in the description of the interaction between incoming particles at small velocities is $\mu_C$. This is consistently used to all orders in the resummed cross section $(\sigma v)_{\mathrm{resum}}$ given by Eq.~\eqref{eq:sigmares}.
%
%
The full next-to-leading order cross section including consistently also the Sommerfeld resummation reads
\begin{align}
    (\sigma v)^{\mathrm{full}}= (\sigma v)_{\mathrm{NLO}} + (\sigma v)_{\mathrm{resum}} - (\sigma v)^{\mathrm{sub}}_{\mathrm{NLO}}\,.
\end{align}
Given the large trilinear couplings in both scenarios, it might be interesting to study the Sommerfeld enhancement coming from the exchange of Higgs bosons \cite{Harz:2017dlj}. In the regime of Sommerfeld enhancement, also bound state formation can potentially occur, giving rise to new annhilation channels and thus altering the relic density prediction. This has been previously studied for stop-antistop annihilation for both gluon \cite{Harz:2018csl} and Higgs exchange \cite{Harz:2019rro}. Such studies, however, are far beyond the scope of this work and are left for future analyses.

%
\begin{figure*}[t]
	\begin{center}
	    \includegraphics[width=0.495\textwidth]{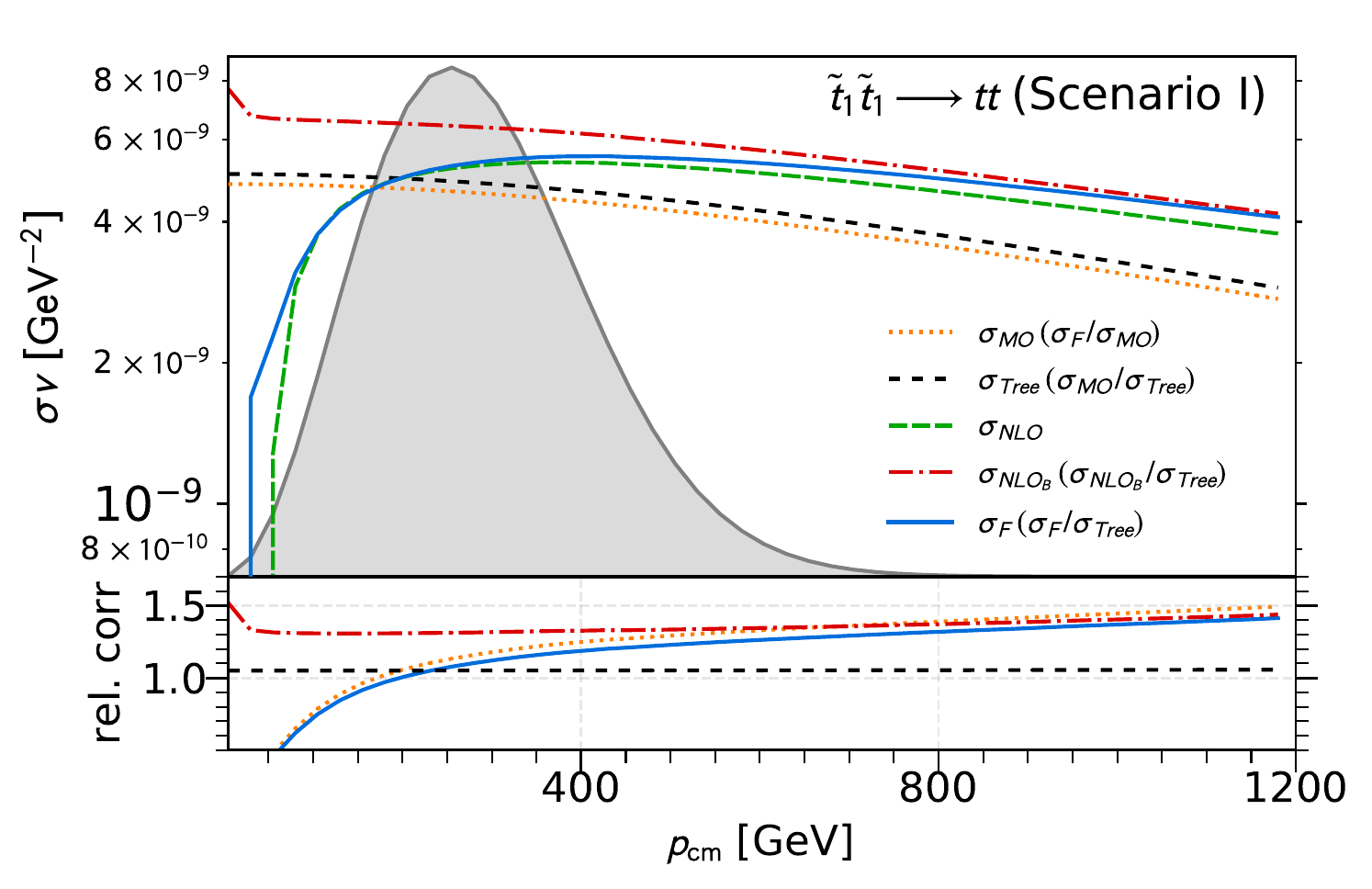}~~~
	     \includegraphics[width=0.495\textwidth]{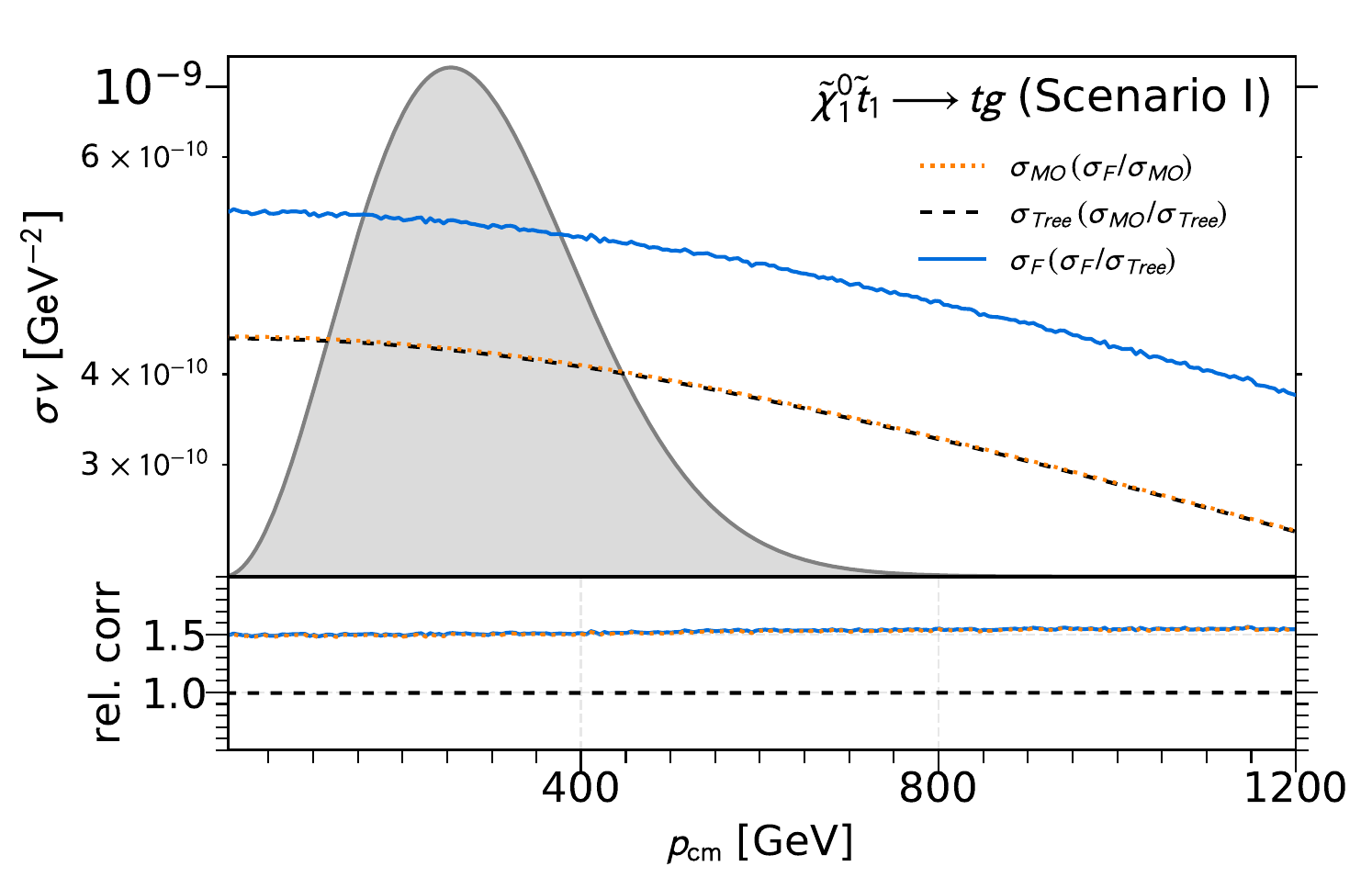}\\
	    \includegraphics[width=0.495\textwidth]{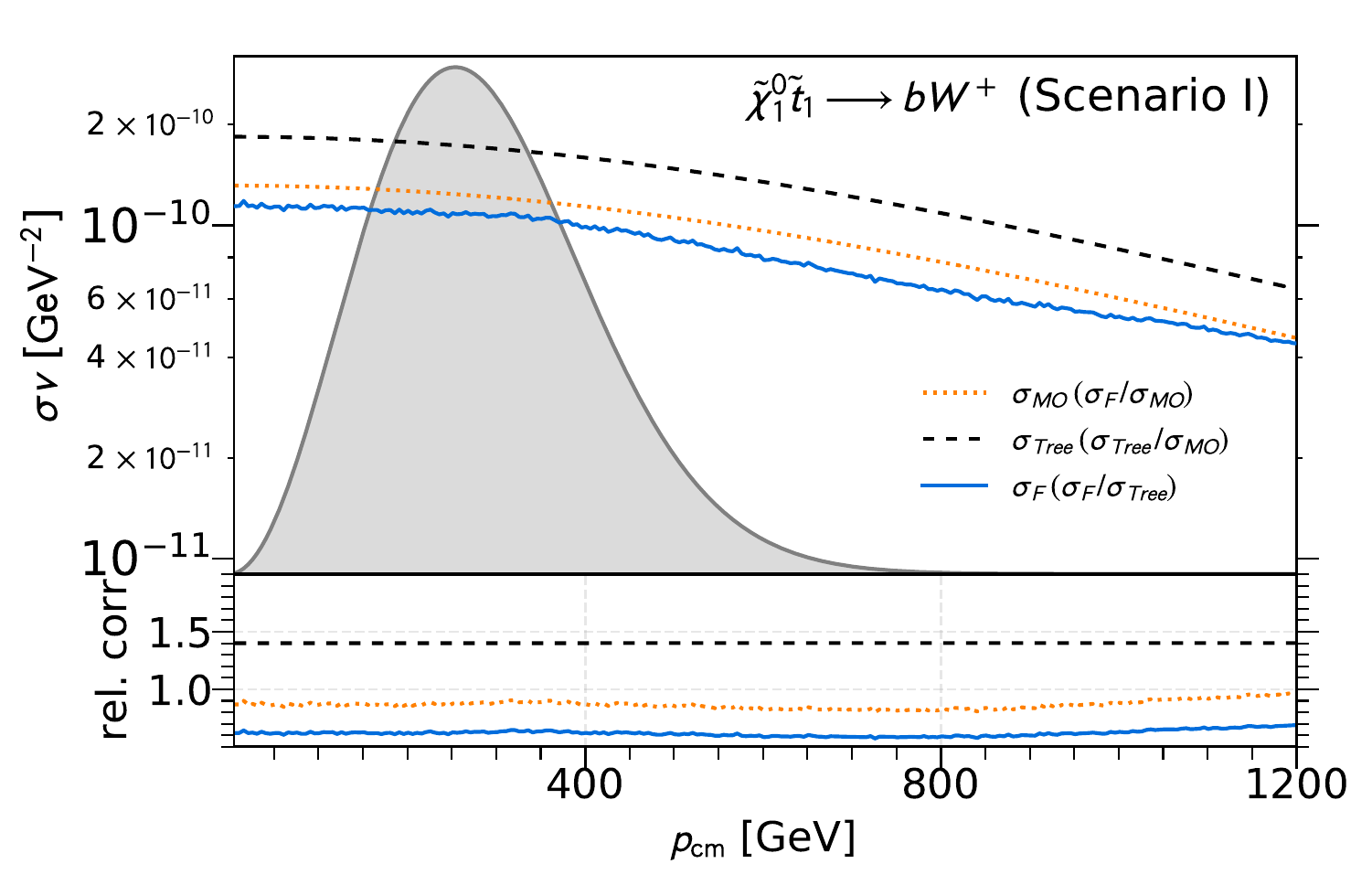}~~~
	    \includegraphics[width=0.495\textwidth]{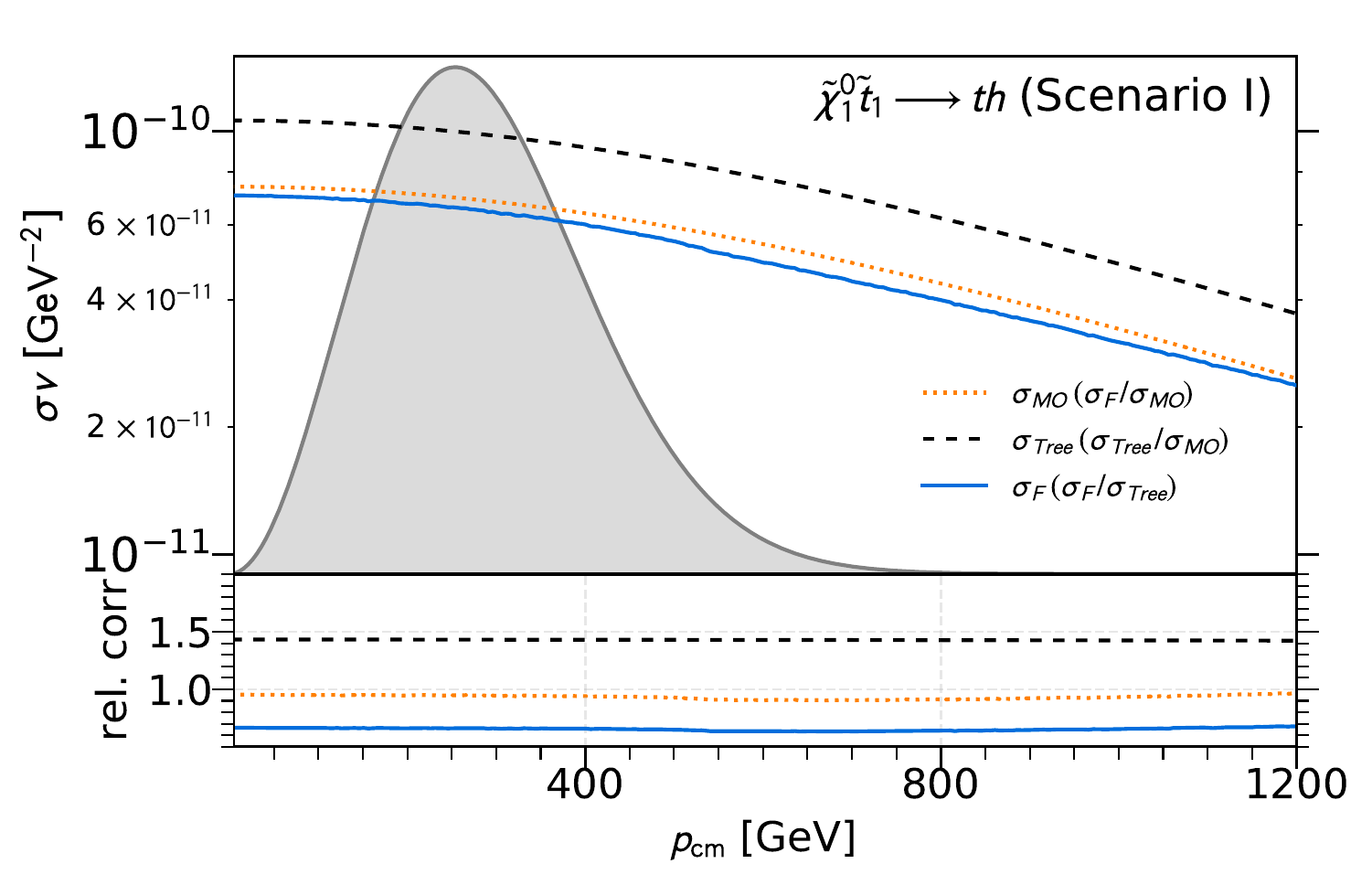}
	\end{center}
	\vspace*{-5mm}
    \caption{Annihilation cross section $\sigma v$ for the stop annihilation into top quarks (first panel) and neutralino-stop co-annihilation (three remaining panels) for Scenario I, computed using the {\tt micrOMEGAs} tree-level calculation (MO), our leading-order calculation (Tree), our fixed-order NLO calculation (NLO, only first panel), our fixed-order NLO calculation without the velocity enhanced part of the box contributions (NLO$_{\rm B}$, only first panel), and our full NLO calculation including resummation (F). The lower part shows various relative cross sections according to the second part of the legend.}
    \label{fig:visgmaNLOS1}
\end{figure*}

\subsection{The NLO cross section results}
\label{sec:NLOres}

In this section we present the first result of our analysis, which is the impact of SUSY-QCD next-to-leading order corrections on the annihilation cross sections of scalar top or bottom pairs. Apart from the cross section (or more precisely $\sigma v$) we also show in arbitrary units the Boltzmann distribution function which is involved in the calculation of the thermal average $\langle \sigma v \rangle$ at the freeze-out temperature (grey shaded area). It should serve as a reminder that the cross section contributes to the determination of the relic density only in a limited range in the center-of-mass momentum $p_{\rm cm}$. 

\subsubsection{Scenario I}

In the first scenario introduced in Sec.~\ref{Sec:Pheno}, the mass splitting between the lightest neutralino and the lightest stop quark is relatively
large. As a result the dark matter annihilation cross section receives important contributions not only from the stop pair
annihilation into top quarks but also from the neutralino-stop co-annihilation into a top quark and a gluon and other final states. 
The results of SUSY-QCD corrections for these processes are shown in Fig.~\ref{fig:visgmaNLOS1}. As described in the previous section, the 
next-to-leading order cross section consists of vertex corrections, propagator corrections, box corrections, counterterm contributions and the real radiation
cross section, which has to be added to render the prediction infrared finite. The corrections from each contribution are not shown in 
Fig.~\ref{fig:visgmaNLOS1} due to the cancellations of the ultraviolet and infrared divergencies between the contributions, which make each contribution on its own ill defined. In addition to the NLO cross section, we include also the enhanced higher-order contributions stemming from the non-relativistic Coulomb correction.

\begin{figure}[t]
	\begin{center}
	    \includegraphics[width=0.495\textwidth]{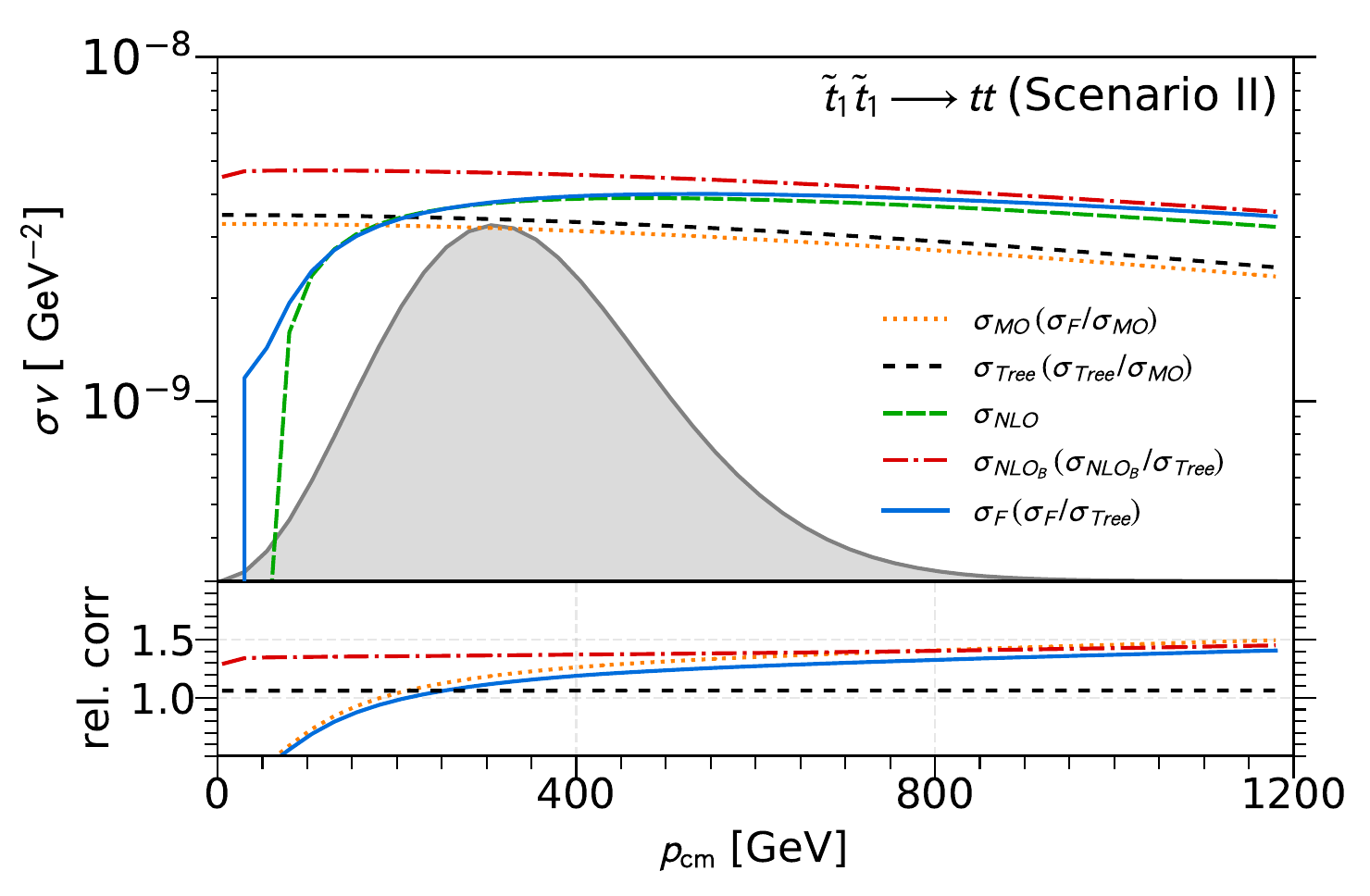}\\
	    \includegraphics[width=0.495\textwidth]{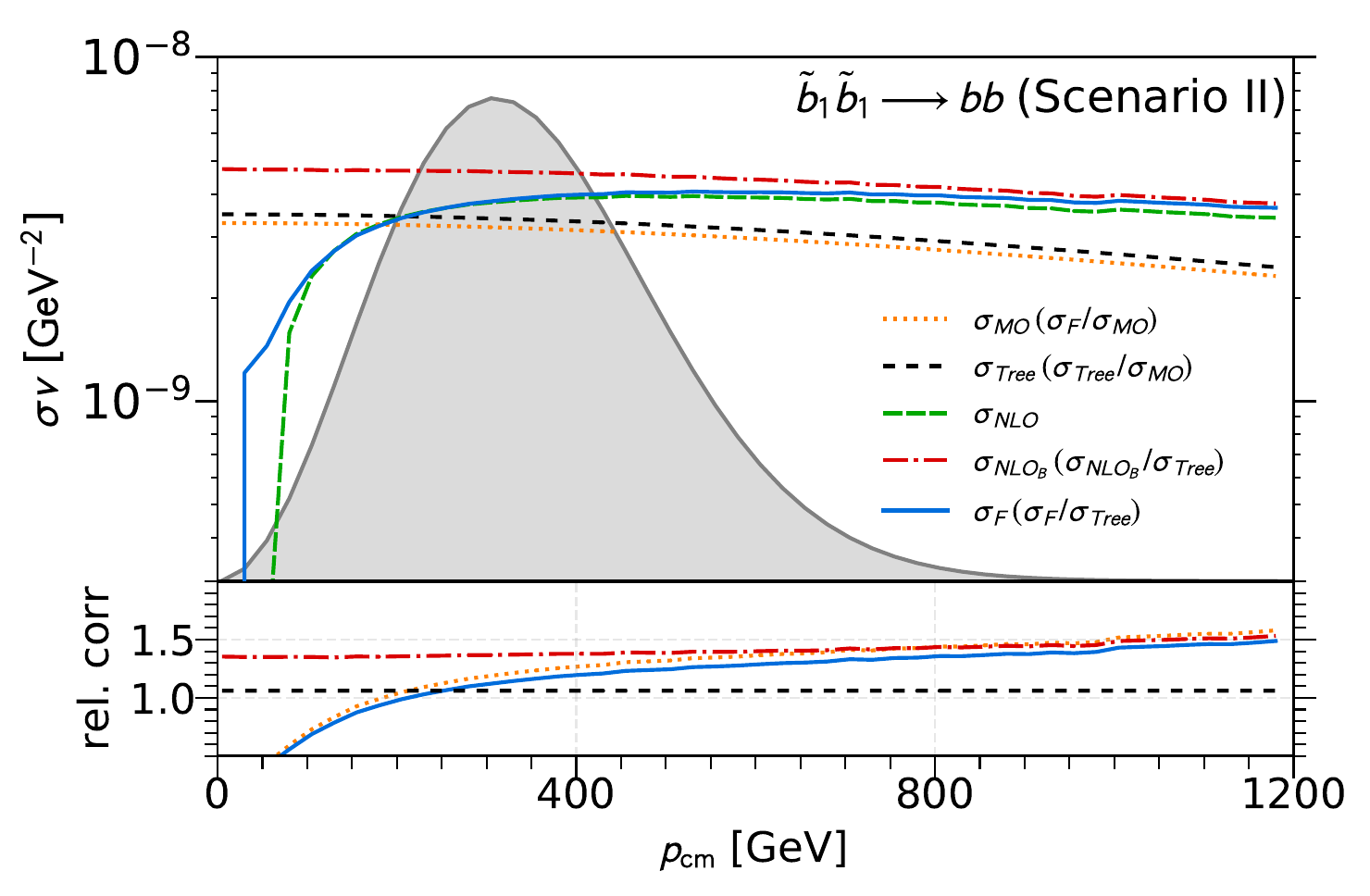}\\
	    \includegraphics[width=0.495\textwidth]{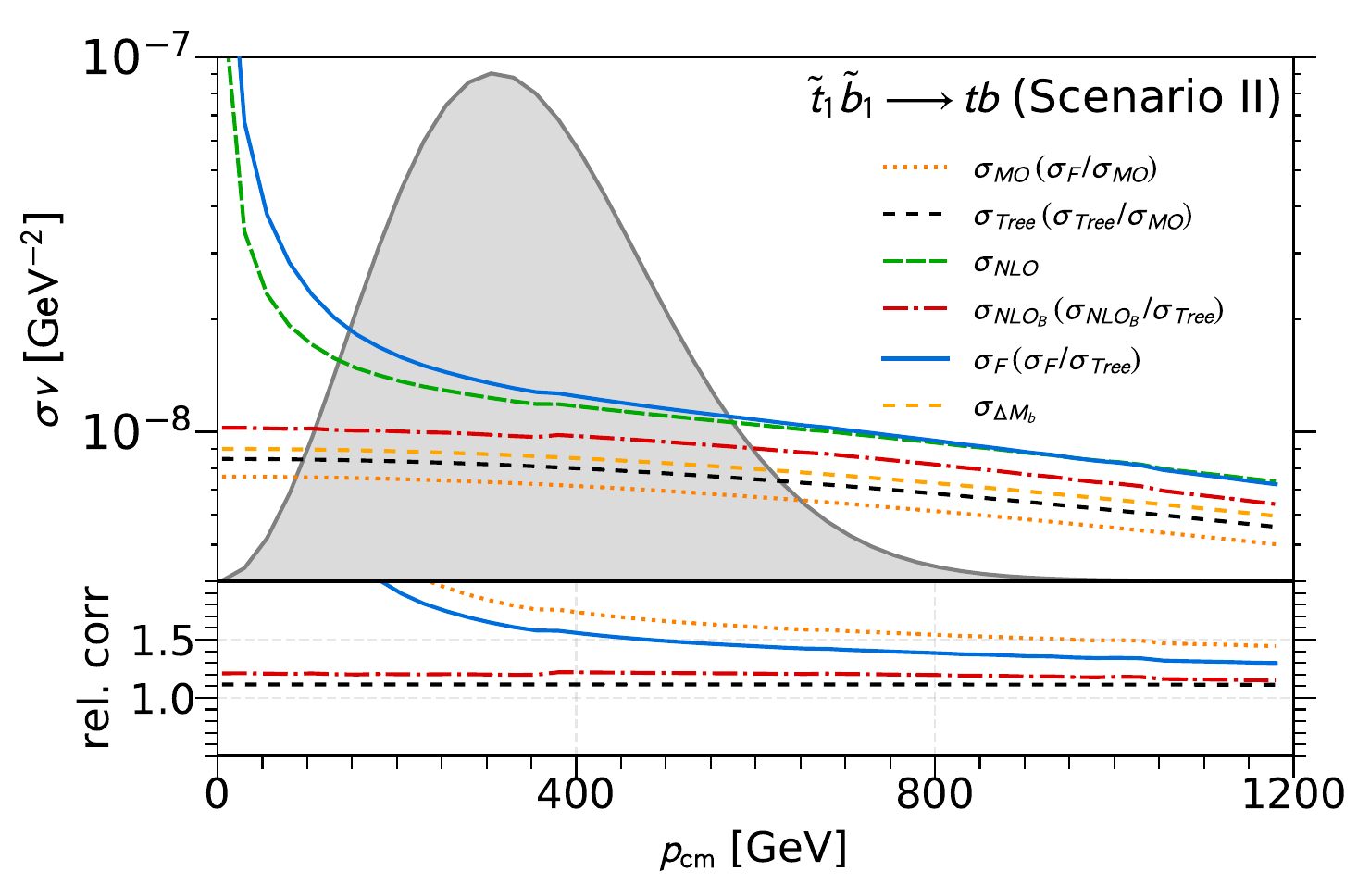}
	\end{center}
	\vspace*{-5mm}
    \caption{Same as Fig.\ \ref{fig:visgmaNLOS1} for the three quark-annihilation processes relevant in Scenario II. In the last panel, we show in addition the cross section for chargino $u$-channel exchange with $\beta$-enhanced Yukawa corrections ($\sigma_{\Delta M_b}$).}
    \label{fig:visgmaNLOS2}
\end{figure}

In case of the annihilation of a pair of scalar top quarks, both initial particles are colored and in the limit of vanishing relative velocity of the squark pair, the Coulomb corrections dominate the full corrected cross section. The origin of these corrections is the exchange of multiple gluons between a pair of slowly moving squarks in the initial state and the details were discussed in Sec.~\ref{Sec:Resum}. The effect of the Coulomb corrections strongly depend on the color multiplet, in which the pair of squarks transform. Based on the color decomposition presented in Sec.\ \ref{Sec:color}, the annihilation cross section $\tilde{t}_1 \, \tilde{t}_1 ~\to~ t \, t$ in Scenario I is dominated by the contribution where the squark pair forms a $SU(3)-$sextet (see Fig.~\ref{fig:xsecLOcolor}). In this representation the multiple exchange of the gluons can be described by a repulsive non-relativistic QCD potential, as discussed in detail in Sec.~\ref{Sec:Resum}. That is why the Coulomb corrections in this case cause a reduction of the cross section. As already discussed in the previous section, the next-to-leading order cross section contains the one-loop contribution included also in the Coulomb enhancement. This one-loop contribution can be traced back to all box diagrams in Fig.~\ref{fig:diagrams_box}, where one gluon is exchanged between the incoming squarks. The contribution from this class of diagrams dominates the one-loop cross section for small velocities and is so large that it causes the cross section with one-loop corrections to be negative (see the green dashed line in Fig.~\ref{fig:visgmaNLOS1}). As discussed in Sec.~\ref{Sec:Resum}, in order to prevent double counting, we remove the part of the box contribution which is already included in the Coulomb resummation. This allows us to quantify the pure one-loop correction to the annihilation cross section without any enhancement (red dash-dotted line in Fig.~\ref{fig:visgmaNLOS1}). We see that the one-loop correction without the enhancement is a large positive correction of about 30-40\% over a large range of $p_{\rm cm}$.

Comparing the result for the non-enhanced NLO cross section with the full result, which is the sum of the non-enhanced NLO cross section and the Sommerfeld corrections shows that the latter are important for all relevant values of $p_{\rm cm}$. Starting at the largest value of $p_{\rm cm} \sim 600\ {\rm GeV}$ which is still relevant for the determination of the relic density, we observe that the Coulomb corrections already reduce the constant 30\% NLO correction by a few percent. For smaller relative velocities corresponding to $p_{\rm cm} \sim 150\ {\rm GeV}$, the NLO correction is fully cancelled by the Coulomb corrections, and for very slow velocities the Coulomb corrections take over and the overall correction is large and negative. For almost vanishing velocities the total cross section becomes negative due to this large negative correction. Even though a negative cross section is unphysical, the fact that the cross section vanishes should not be very surprising. The dynamics of the squark pair in the regime when Coulomb corrections are very large (meaning for vanishing velocities) correspond to a motion of the pair in a highly repulsive QCD potential. This in turn means that large repulsive forces repel one squark from the other reducing the probability of annihilation and thereby reducing the cross section. The cross section becomes negative only for $p_{\rm cm} < 10\ {\rm GeV}$, which is irrelevant for the relic density determination, as for such small momenta it is multiplied by an almost vanishing Boltzmann distribution function.

In summary, we can conclude that SUSY-QCD corrections to $\tilde{t}_1 \, \tilde{t}_1 ~\to~ t \, t$ are sizeable either through the one-loop corrections for large $p_{\rm cm}$ or the enhanced Coulomb corrections for small $p_{\rm cm}$.

The co-annihilation processes important in this scenario were discussed in detail in Refs.\ \cite{Harz:2012fz, Harz:2014tma}. In Fig.~\ref{fig:visgmaNLOS1} we also show the effect of the SUSY-QCD corrections on the co-annihilation cross sections in Scenario I. We see that the next-to-leading order corrections in the case of co-annihilations are substantial ranging from -30\% in the case of the co-annihilation into top quark and Higgs-boson final state to +50\% in the case of the top gluon final state. There are a few substantial differences such as the fact that the corrections are negative in the case of co-annihilations with electroweak bosons or Higgs bosons in the final state or that there is a large difference between our leading order result and the \MO\ result for the co-annihilations with electroweak and Higgs bosons, which can be traced to a different definition of underlying parameters in our renormalization scheme. The next-to-leading order correction with respect to the \MO\ result is largely reduced to at most -10\%. Given that our leading order prediction for the co-annihilations into top quark and a gluon coincides with \MO\ prediction, the large next-to-leading order correction gives directly also the correction with respect to the \MO\ result. 

\subsubsection{Scenario II}

In the second scenario, the choice of parameters such as $\tan\beta$ and the gaugino and squark mass parameters cause the masses of the lightest neutralino, the lightest scalar top and bottom quarks to be almost degenerate. This leads to different processes contributing significantly to the total dark matter annihilation cross section. The smaller mass difference renders co-annihilations ineffective and the fact that three particles are mass degenerate leads to a larger number of annihilations. Moreover, the large value of $\tan\beta$ enhances the gaugino exchange in the case of the stop-sbottom annihilation, and this together with a different color structure makes this annihilation dominant in the case of Scenario II. 

The full next-to-leading order results for three dominant processes are shown in Fig.~\ref{fig:visgmaNLOS2}. The processes $\tilde{t}_1 \, \tilde{t}_1 ~\to~ t \, t$ and $\tilde{b}_1 \, \tilde{b}_1 ~\to~ b \, b$ have very similar features to the annihilation of a pair of scalar top quarks in Scenario I. The main difference in this scenario is the process $\tilde{t}_1 \, \tilde{b}_1 ~\to~ t \, b$, which has an entirely different decomposition of the leading order cross section in terms of the $t$- and $u$-channel exchanges combined with a different color decomposition, which is essential in explaining the behaviour of the NLO cross section. Similar to the already discussed case of $\tilde{t}_1 \, \tilde{t}_1 ~\to~ t \, t$, the NLO correction contains a velocity enhanced term, which is already resummed in the Sommerfeld correction. In order to avoid double counting, we define again the non-enhanced NLO correction $\sigma_{\mathrm{NLO_B}}$ where we subtract the term which is already accounted for by the Sommerfeld resummation (red dash-dotted curve in Fig.\ \ref{fig:visgmaNLOS2}). As one can see in Fig.\ \ref{fig:visgmaNLOS2}, the non-enhanced NLO correction is substantial in all processes in Scenario II. In the case of stop pair or sbottom pair annihilations, this NLO correction is compensated by a large and negative Sommerfeld correction which is here derived from a repulsive QCD potential. The color decomposition of the $\tilde{t}_1 \, \tilde{b}_1 ~\to~ t \, b$ shows (see Fig.~\ref{fig:xsecLOcolor}) that in contrast to the other processes, the cross section is here dominated by the part where the initial stop and sbottom quarks transform as a $SU(3)-$triplet. In this color configuration a pair of slowly moving squarks experiences an attractive strong force, which leads to a large enhancement of the annihilation cross section. Comparing the full result (solid blue line in Fig.~\ref{fig:visgmaNLOS2}) with the result containing just the non-enhanced NLO corrections, we see that the Sommerfeld enhancement is important over the whole region in $p_{cm}$ that is relevant for the calculation of the relic density. 

It is worth mentioning that due to the large value of $\tan\beta$ in Scenario II and to the fact that the chargino $u$-channel exchange gives an important contribution to the cross section $\tilde{t}_1 \, \tilde{b}_1 ~\to~ t \, b$, the Yukawa corrections to the chargino exchange can give a non-negligible contribution. We have included the $\tan\beta$ dependent Yukawa corrections even beyond next-to-leading order in the full result and show their effect separately in Fig.~\ref{fig:visgmaNLOS2} (yellow dashed line). Even though the Yukawa corrections are non-negligible, they are small (about 3\%) compared to the remaining SUSY-QCD corrections or to the Sommerfeld enhancement. 

The full correction to $\tilde{t}_1 \, \tilde{b}_1 ~\to~ t \, b$ is larger than 50\% over the whole relevant range of $p_{cm}$ and can even exceed 100\% (without threatening perturbativity as this correction originates from a resummation). We will show in the next section that using the fully corrected annihilation cross sections in Scenario II has a large impact on the relic density.


\begin{figure*}[t!]
	\begin{center}
	    \begin{picture}(550,160)
	        \put(5,0){\resizebox{!}{5.6cm}{\includegraphics{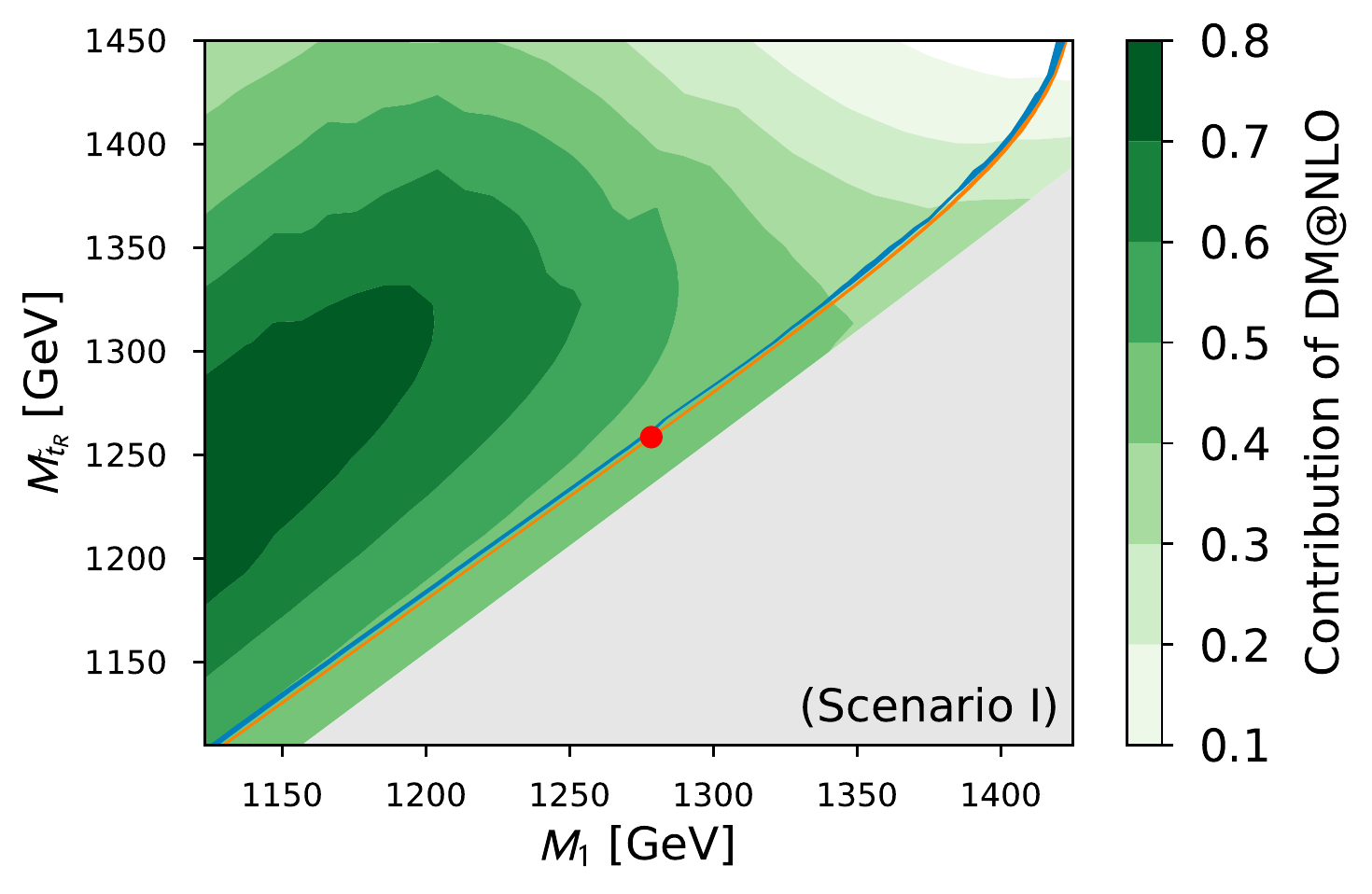}}}
	        \put(260,0){\resizebox{!}{5.6cm}{\includegraphics{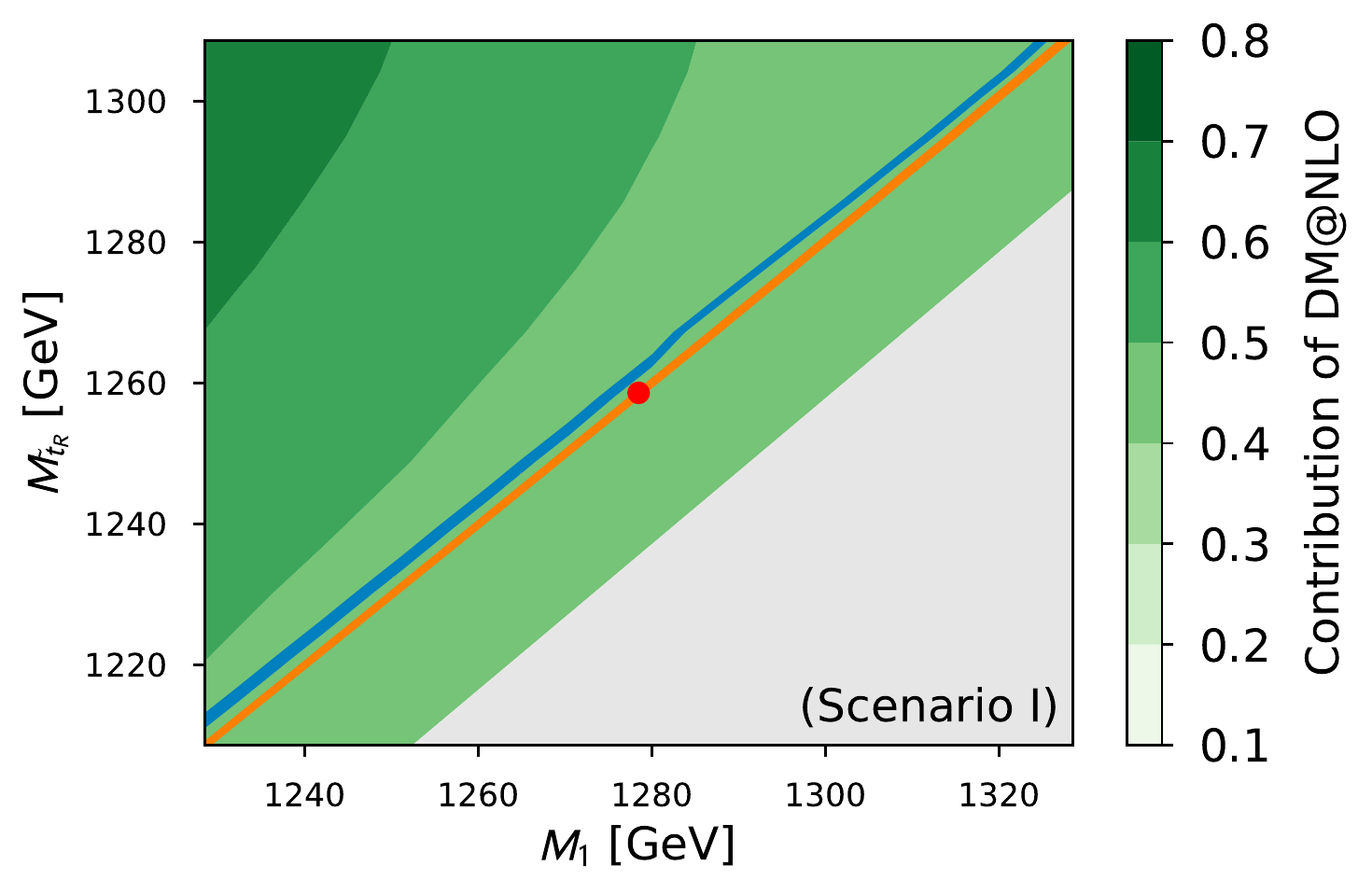}}}
	    \end{picture}
	    \begin{picture}(550,160)
	        \put(5,0){\resizebox{!}{5.6cm}{\includegraphics{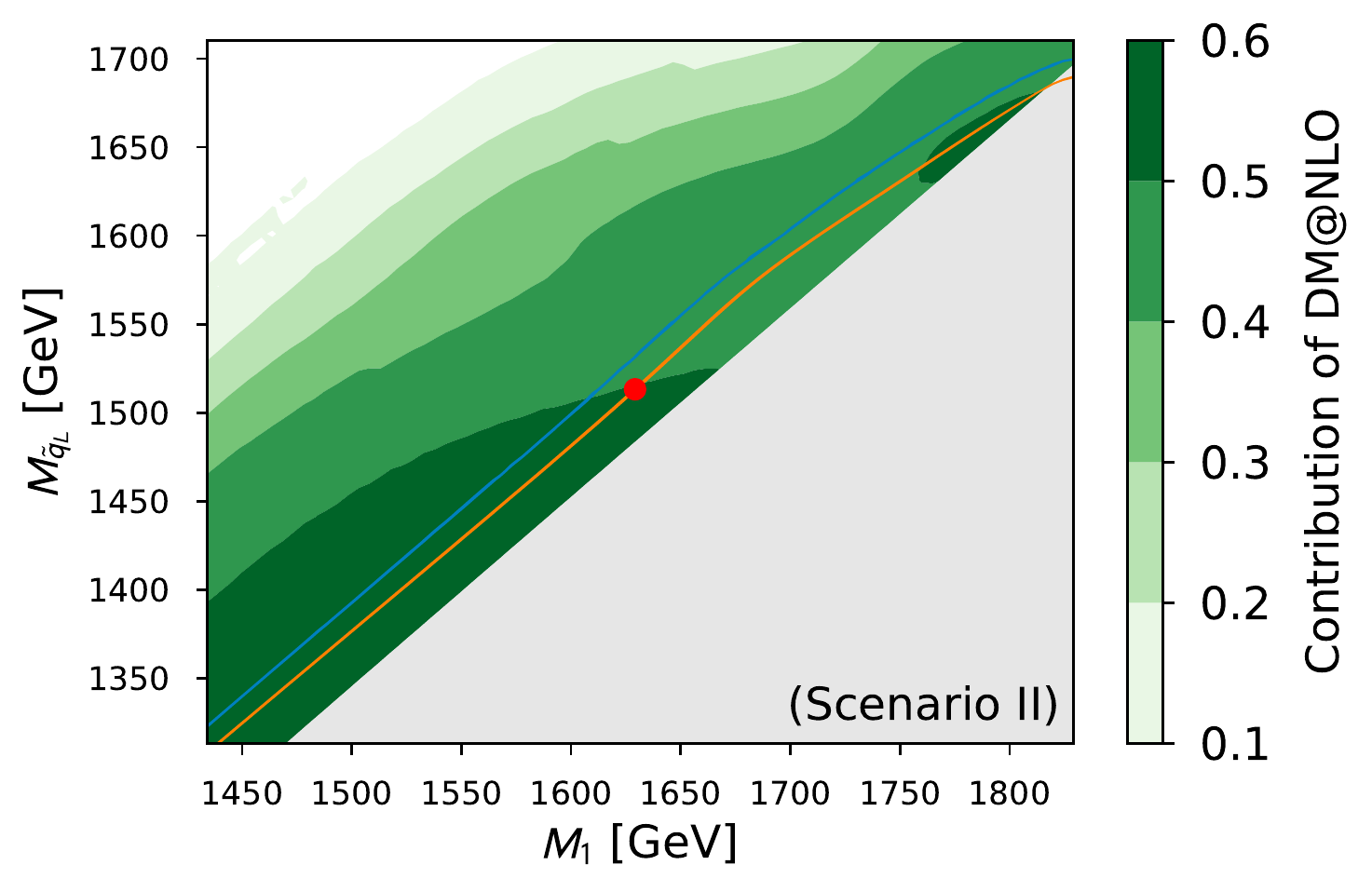}}}
	        \put(260,0){\resizebox{!}{5.6cm}{\includegraphics{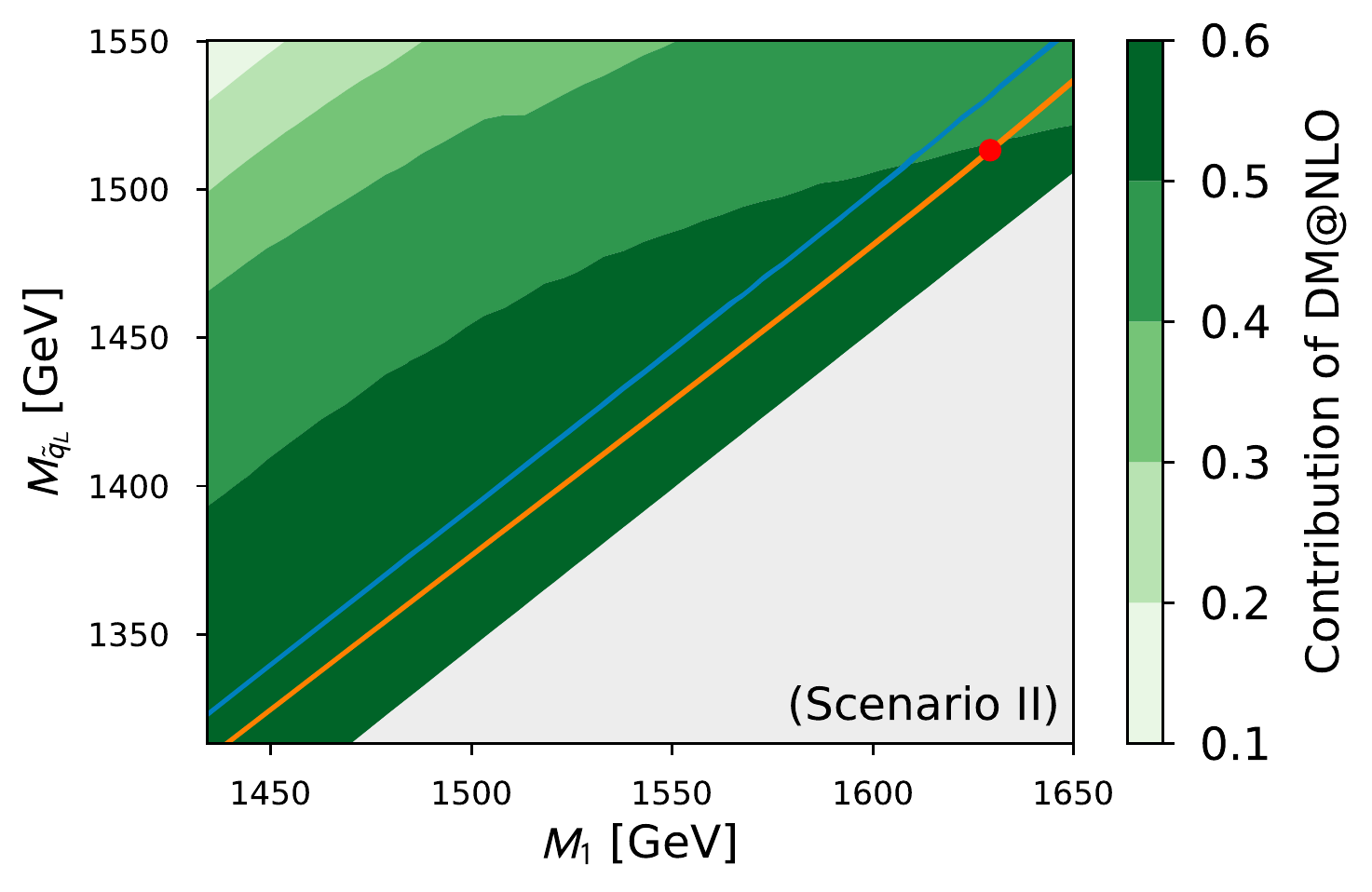}}}
	    \end{picture}
	\end{center}
	\vspace*{-5mm}
    \caption{Parameter regions compatible with the {\it Planck} limits given in Eq.\ \eqref{eq:omh2} presented in the $M_1$-$M_{\tilde{t}_R}$ plane around Scenario I and in the $M_1$-$M_{\tilde{q}_L}$ plane around Scenario II. The orange band corresponds to the {\tt micrOMEGAs} calculation, while the blue band stems from the full {\tt DM@NLO} one-loop calculation. The right panel corresponds to a zoom into the left panel around Scenario I (or Scenario II), which is indicated by the red dot. Grey regions are excluded due to stop LSP.}
    \label{fig:omh2NLOI}
\end{figure*}

\section{Impact on the neutralino relic density}
\label{Sec:Results}

We finally come to the discussion of the impact of the corrections presented in Sec.\ \ref{Sec:NLO} on the neutralino relic density $\Omega_{\tilde{\chi}^0_1}h^2$. To this end, we have implemented the corrections into the numerical code {\tt DM@NLO}, which is used as an extension to {\tt micrOMEGAs}. In practice, this means that the Boltzmann equation is still integrated using the latter, while the cross section calculation of the relevant processes (see Tab.\ \ref{tab:channels}) is performed by {\tt DM@NLO} instead of {\tt CalcHEP}. 

In the following, we will illustrate the impact of the corrections by comparing the relic density obtained using the full {\tt DM@NLO} NLO calculation to the values obtained using the tree-level calculation of {\tt micrOMEGAs} / {\tt CalcHEP}.

\subsection{Scenario I}
\label{Sec:ResultsI}

We start by examining the impact of NLO corrections in the vicinity of Scenario I, where we compare the relic density obtained from the {\tt micrOMEGAs} calculation to the one obtained using our full NLO result as presented in Sec.\ \ref{Sec:NLO}. The impact is illustrated in Fig.\ \ref{fig:omh2NLOI}, where we show the corresponding viable regions of parameter space in the $M_1$-$M_{\tilde{t}_R}$ plane. As can be seen, the favoured parameter region where the calculated relic density satisfies the experimental constraint, Eq.~(\ref{eq:omh2}), is shifted towards smaller mass parameters in order to compensate the increased annihilation cross section. It is important to note that this shift is larger than the width of the band which corresponds to the {\it Planck} $2\sigma$-uncertainties. 

The situation changes for higher masses, where the processes discussed in this work and corrected at the NLO level, are less relevant. The correction of the remaining processes relevant in this part of parameter space are left for future work.

\begin{figure*}[t]
	\begin{center}
	    \begin{picture}(550,150)
	        \put(5,0){\resizebox{!}{5.6cm}{\includegraphics{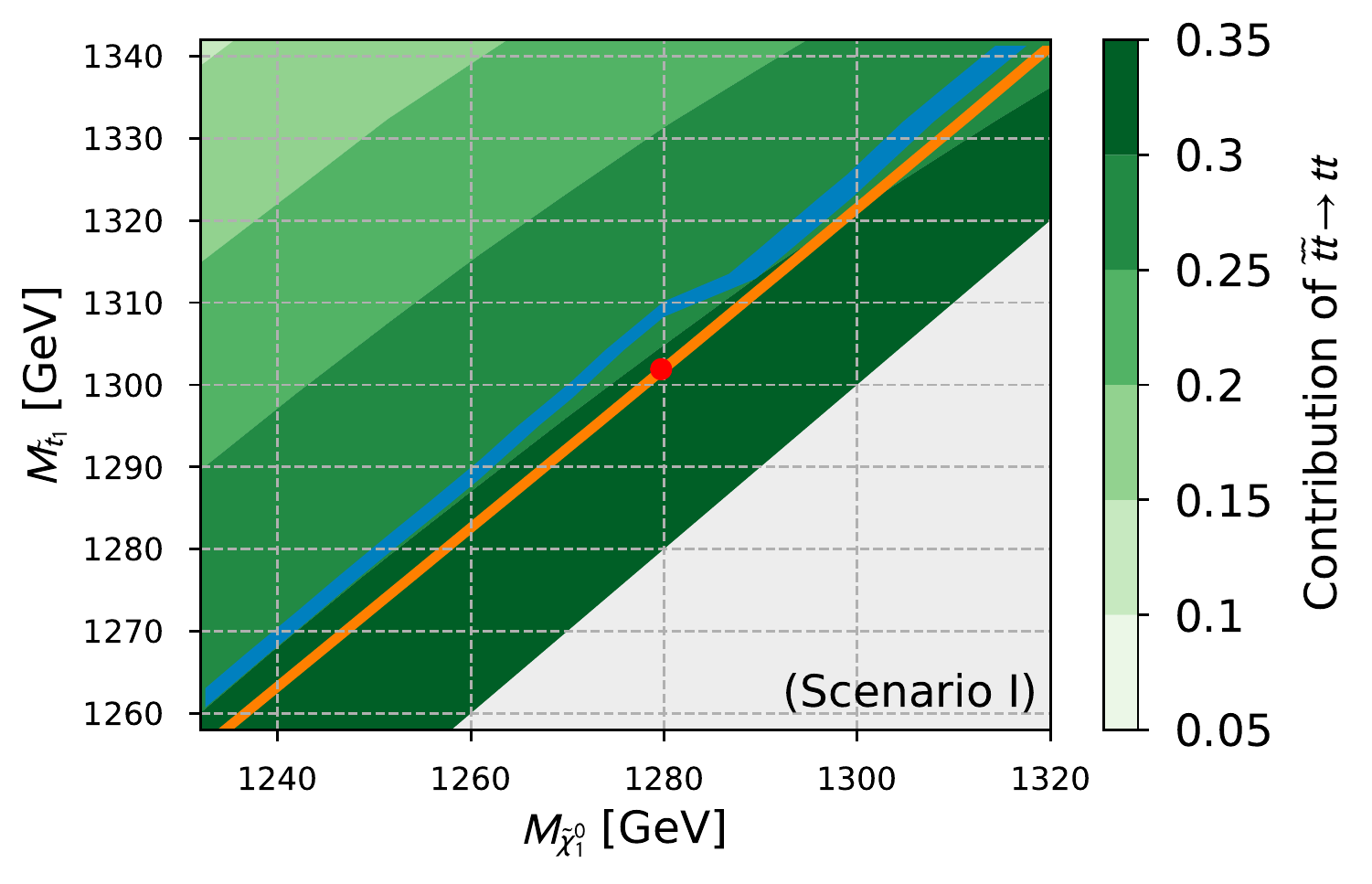}}}
            \put(260,0){\resizebox{!}{5.6cm}{\includegraphics{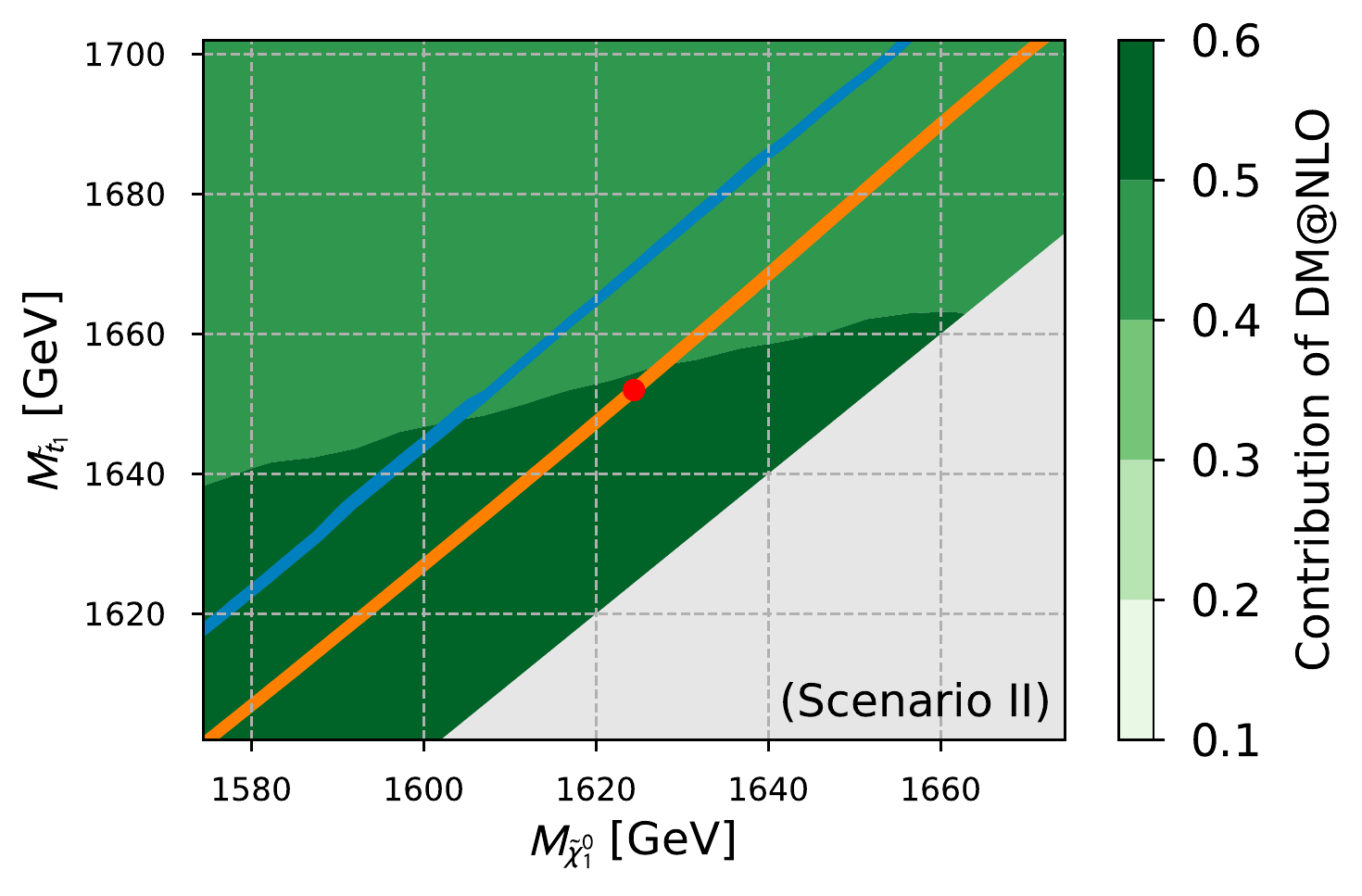}}}
	    \end{picture}
	\end{center}
	\vspace*{-5mm}
    \caption{Same as Fig.\ \ref{fig:omh2NLOI}, but projected into the plane of the physical neutralino and stop masses.}
    \label{fig:omh2NLOphysI}
\end{figure*}

In Fig.\ \ref{fig:omh2NLOphysI}, we show the same results in the vicinity of Scenario I, but this time projected onto the plane of the physical neutralino and stop masses. Note that, here, the variation of the physical masses solely stems from varying the parameters $M_1$ and $M_{\tilde{t}_R}$, respectively, while all other soft parameters, including those that in general may influence the neutralino and stop masses, are kept fixed to the values of Table \ref{tab:scenarios}. From Fig.\ \ref{fig:omh2NLOphysI}, we see that the cosmologically favoured region of parameter space is shifted by about 7 GeV in both the neutralino and the lighter stop mass. 

These results lead to the conclusion that the corrections presented in this work are relevant when performing an extraction of either physical masses or fundamental model parameters from cosmological data. Let us note that this conclusion is the same as for the previous analyses of other processes entering the calculation of $\Omega_{\tilde{\chi}_1^0}h^2$ \cite{Herrmann:2007ku, Herrmann:2009wk, Herrmann:2009mp, Herrmann:2014kma, Harz:2012fz, Harz:2014tma, Harz:2014gaa}.

\begin{figure*}[t]
	\begin{center}
	    \includegraphics[width=0.495\textwidth]{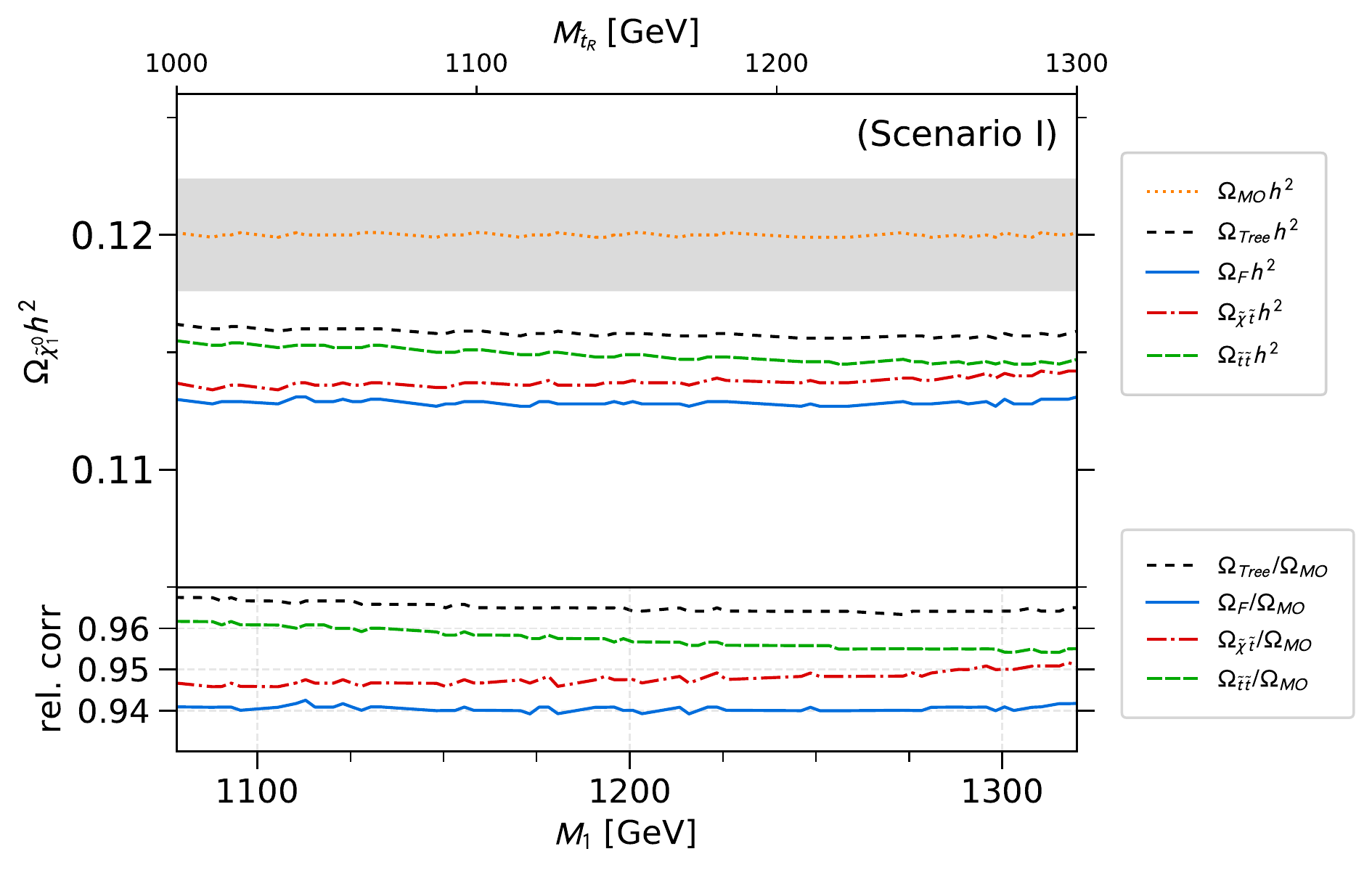}
	    \includegraphics[width=0.495\textwidth]{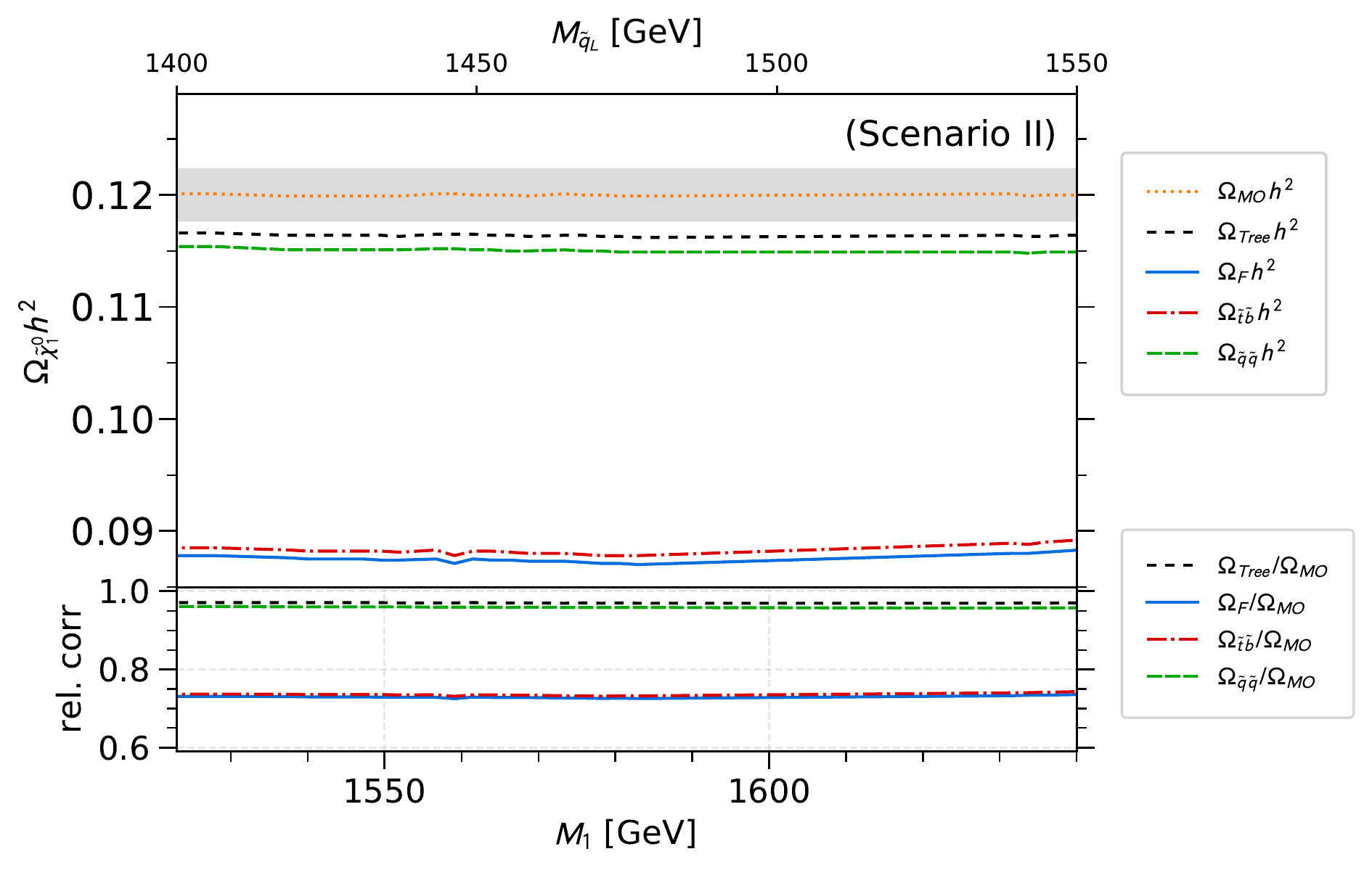}
	\end{center}
	\vspace*{-5mm}
    \caption{Upper part: Neutralino relic density $\Omega_{\tilde{\chi}^0_1}h^2$ along the parameter region satisfying the experimental constraints on the relic abundance. Both, the bino mass parameter $M_1$ and the squark mass parameter $M_{\tilde{t}_R}$ (or $M_{\tilde{q}_L}$) around Scenario I (or Scenario II). In both plots we show the values obtained using {\tt micrOMEGAs} ($\Omega_{\rm MO}h^2$), our tree-level calculation ($\Omega_{\rm Tree}h^2$), our full one-loop calculation including the resummation ($\Omega_{\rm F}h^2$). For Scenario I, we also show the value obtained correcting only neutralino-stop co-annihilation ($\Omega_{\tilde{\chi}^0_1 \tilde{t}_1} h^2$), and the value obtained correcting only stop-pair annihilation ($\Omega_{\tilde{t}_1 \tilde{t}_1} h^2$). Similar for Scenario II we show the effect of correcting only the stop and sbottom pair annihilations ($\Omega_{\tilde{q}_1 \tilde{q}_1} h^2$) and the stop-sbottom annihilations ($\Omega_{\tilde{t}_1 \tilde{b}_1} h^2$). Lower part: Impact of the different contributions relative to the relic density obtained by using {\tt micrOMEGAs}.}
\label{fig:omh2M1I}
\end{figure*}

In order to get a better understanding of the impact of the different contributions to the annihilation cross section $\sigma_{\rm ann}$, and consequently to the relic density $\Omega_{\tilde{\chi}^0_1} h^2$, we have performed a one-dimensional scan along the region where \MO\ predicts the correct relic density varying simultaneously the parameters $M_1$ and $M_{\tilde{t}_R}$. The result of this scan is shown in Fig.~\ref{fig:omh2M1I} as a function of both parameters while all other parameters were fixed to the values given in Tab.\ \ref{tab:scenarios}. 

First, it can be noticed that our tree-level prediction differs from the {\tt micrOMEGAs} result. This is a direct consequence of the corresponding difference in the annihilation cross sections, as discussed in Sec.\ \ref{Sec:LO} and shown, e.g., in Fig.\ \ref{fig:visgmaNLOS1}. Taking into account the corrections discussed in Sec.\ \ref{Sec:NLO}, it can be seen that the total correction is split into two parts associated with the relevant classes of processes, namely $\tilde{t}_1 \tilde{t}_1 \to tt$ and $\tilde{\chi}^0_1 \tilde{t}_1 \to qg,qV,q\phi$. Fig.~\ref{fig:visgmaNLOS1} shows that the correction to the relic density in Scenario I is dominated by the corrections to the co-annihilation processes even though at leading order these processes contribute about a factor two less than $\tilde{t}_1 \tilde{t}_1 \to tt$. This is a consequence of the Sommerfeld supression of the annihilation cross section as discussed in Sec.~\ref{sec:NLOres}. Moreover, we see that for lower bino/squark mass parameters $M_1$/$M_{\tilde{t}_R}$ the correction to the co-annihilation processes is numerically more important than for large mass parameters. This is explained by the fact that the relative importance of the co-annihilation processes is higher in this region of parameter space (see, e.g., Fig.\ \ref{fig:channelsI}). Moving towards higher values of $M_1$, the relative importance of the stop-pair annihilation increases and, consequently, the associated correction to the relic density becomes more important. 

Overall, the relic density obtained using our full (i.e.\ NLO including resummation) caluclation is about 6\% smaller than the one obtained by using {\tt micrOMEGAs}. Again, we emphasize that this shift is more important than the uncertainty given by the {\it Planck} measurement, which is, at the $2\sigma$ confidence level of about 2\%.

\subsection{Scenario II}
\label{Sec:ResultsII}

Let us now focus on Scenario II, where not only $\tilde{t}_1 \tilde{t}_1 \to tt$ is relevant, but also the related processes $\tilde{b}_1 \tilde{b}_1 \to bb$ and $\tilde{t}_1 \tilde{b}_1 \to tb$ give sizeable contributions to the annihilation cross section $\sigma_{\rm ann}$. Therefore they need as well to be corrected at the NLO level including the resummation, as discussed in Sec.\ \ref{Sec:NLO}.


Again, we start by depicting the parameter region compatible with the measured value for the relic density (see the second row of plots in Fig.~\ref{fig:omh2NLOI}) in the vicinity of Scenario II, in this case in the $M_1$-$M_{\tilde{q}_L}$ plane, which are the relevant neutralino and squark mass parameters. Similar to Scenario I, as discussed above, the viable regions with respect to the relic density is shifted towards lower masses. Here, the shift is again more important than the uncertainty and is much larger than in Scenario I. It corresponds to a shift of about 17 GeV in the bino mass parameter $M_1$ and about 15 GeV in the left-handed squark mass parameter $M_{\tilde{q}_L}$. 

In terms of physical masses, shown in Fig.\ \ref{fig:omh2NLOphysI}, this corresponds to a shift of about 17 GeV for the neutralino mass, and of about 15 GeV in the lighter stop mass. Once more, these findings underline the importance of the presented corrections in the light of precision cosmology. It is to be noted that for this part of the analysis only $M_1$ and $M_{\tilde{q}_L}$ have been varied and the results have been projected on the so obtained plane of the physical neutralino and stop masses, while all other parameters have remained fixed to the values given in Tab.\ \ref{tab:scenarios}.

In order to decompose our full NLO prediction for the relic density into the contributions from different processes, we show in Fig.\ \ref{fig:omh2M1I} the NLO corrected contributions to the relic abundance from individual processes along the region where \MO\ predicts the correct relic density varying simultaneously the parameters $M_1$ and $M_{\tilde{q}_L}$. We see that already using our tree-level annihilation cross section with differently defined input parameters shifts the relic density by a few percent (black dashed line). In all remaining contributions we use our tree-level annihilation cross section for all relevant processes. Starting from our tree-level, a very small correction of about 1\% in the whole region comes from including NLO corrections only to the $\tilde{t}_1 \tilde{t}_1 \to tt$ and $\tilde{b}_1 \tilde{b}_1 \to bb$. It is important to point out that the reason for this extremely small correction is the Sommerfeld enhancement, which in this case in fact suppresses the cross section. This is due to the repulsive nature of the dominant $SU(3)-$sextet contribution. Even if the full correction to the cross section was large and negative for small $p_{\rm cm}$ (see Fig.~\ref{fig:visgmaNLOS2}), interestingly the correction to the thermal averaged cross section is still positive leading to a drop in relic density. The largest contribution comes from the annihilation cross section of $\tilde{t}_1 \tilde{b}_1 \to tb$. The first reason is the large contribution of this process to the total annihilation cross section already at tree-level (see Tab.~\ref{tab:channels}). The second reason is the large Sommerfeld enhancement emerging from the attractive potential of the dominant $SU(3)-$triplet contribution,  which in this case makes the full correction to the cross section extremely large. We see that depending on the dominant contribution of the color decomposition, $SU(3)-$sextet or $SU(3)-$triplet, the Sommerfeld corrections either suppresses or enhances the cross section such that the total SUSY-QCD correction to the relic density over the whole range is about 25\%. This results in the visible shift of the preferred parameter region, which is much larger than the experimental uncertainty given in Eq.\ \eqref{eq:omh2}.

%
\section{Conclusion}
\label{Sec:Conclusion}
Scenarios in the MSSM with light stops are still very appealing due to their potential to address many problems that the MSSM with heavy particles might have. In such scenarios, squark pair annihilations into quarks are often very important processes that govern the annihilation of dark matter, which is typically the lightest neutralino. 

We have analyzed two such example scenarios, which pass all current experimental constraints. We focused on the SUSY-QCD corrections to squark pair annihilations into quarks. We have reviewed the details of the one-loop calculation and of the Sommerfeld enhancement. We have shown that the one-loop corrections of the cross sections are sizeable even without the Sommerfeld enhancement. The Sommerfeld corrections are shown to cause two different effects depending on the nature of the strong force between the pair of incoming scalar quarks. In the case of annihilations between the same type of squarks, the enhancement turns into a reduction of the cross section as here the squarks experience a strong repulsive force. If the scalar quarks are different, however, the cross section is strongly enhanced due to the attractive strong force. Finally, we have investigated the impact of these corrections on the predicted relic density. We have demonstrated in our typical scenarios that even with the Sommerfeld reduction, the corrections are larger than the experimental uncertainty. In case of an enhancement, the corrections cause a 25\% shift in the preferred parameter region where the relic density satisfies the experimental constraints.

\acknowledgments
The authors would like to thank A.~Pukhov for providing us with the possibility to interface our {\tt DM@NLO} code with {\tt micrOMEGAs}. J.H.~was  supported  by  the  DFG  Emmy  Noether  Grant  No.  HA  8555/1-1, and earlier by the Labex ILP (reference ANR-10-LABX-63) part of the Idex SUPER, and received financial state aid managed by the Agence Nationale de la Recherche (ANR), as part of the programme Investissements d'avenir under the reference ANR-11-IDEX-0004-02.  The work of B.\,H.\ is supported by {\it Investissements d'avenir}, Labex ENIGMASS, contrat ANR-11-LABX-0012. The work of S.\,S. was supported by the DFG through RTG 2149 {\it Strong and Weak Interactions - from Hadrons to Dark Matter}. The figures presented in this paper have been generated using {\tt MatPlotLib} \cite{Matplotlib}.

\appendix 
\section{Small velocity expansion of the box contribution}
\label{sec:tensor}

\begin{figure}
    \setlength{\unitlength}{1cm}
    \begin{picture}(8.3,4.)(-2,-1.5)
        \put(-1.5,-0.75){\mbox{\resizebox{!}{3.0cm}{\includegraphics{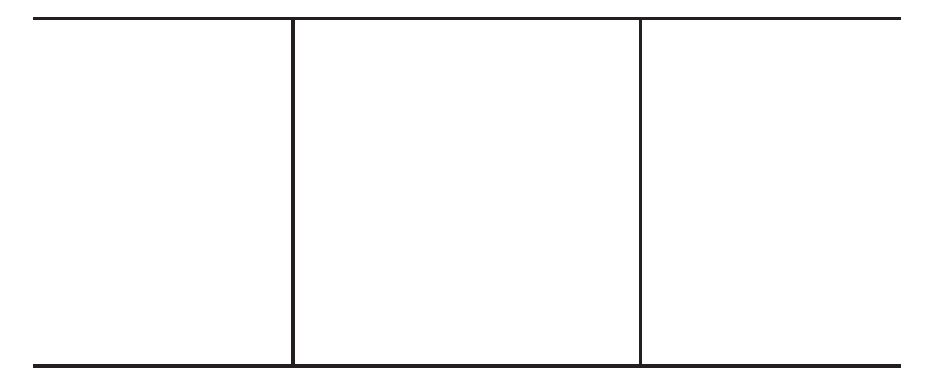}}}}
        \put(-1.8,2){$m_1$}
        \put(-1.8,-0.74){$m_2$}
        \put(5.7,2){$m_3$}
        \put(5.7,-0.74){$m_4$}
        \put(0.4,0.63){$0$}
        \put(2,1.6){$M_1$}
        \put(3.7,0.63){$M_2$}
        \put(2,-0.4){$M_3$}
    \end{picture}
    \caption{Box diagram corresponding to the gluon exchange.}
    \label{fig:D0}
\end{figure}
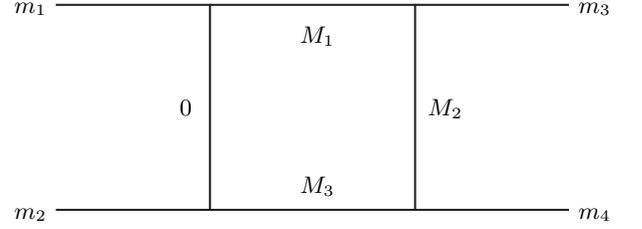

In order to subtract the velocity enhanced part of the NLO contribution that is already included in the Sommerfeld resummation, we expand the corresponding contribution of the box diagrams in the relativistic relative velocity \cite{Cannoni:2016hro}
\begin{align}
    v_{\mathrm{rel}} = \frac{\sqrt{\lambda(s,m_1^2,m_2^2)}}{s-m_1^2-m_2^2}\,,
    \label{eq:vrelrel}
\end{align}
with $\lambda(a,b,c) = a^2 + b^2 + c^2 - 2ab - 2 ac - 2bc$.

All box diagrams that contain the velocity enhanced contribution feature an exchange of a massless gluon between the incoming pair of scalar quarks. A generic diagram showing the masses of internal and loop particles is shown in Fig.\ \ref{fig:D0}. The matrix element contains tensor coefficients 
\begin{align}\label{eq:D0ind}
    D_i \big( p_{10}^2,p_{21}^2,&p_{32}^2,p_{30}^2,p_{20}^2,p_{31}^2,M_0^2,M_1^2,M_2^2,M_3^2 \big) ~=~ \\ & D_i \big( m_1^2,m_3^2,m_4^2,m_2^2,t_i,s,0,m_1^2,M_2^2,m_2^2 \big) \,, \nonumber
\end{align}
where $t_i$ are the Mandelstam variables $t$ or $u$ depending on the box diagram. The velocity enhanced terms in the box contribution are contained in the scalar 4-point integrals $D_0 \big( m_1^2,m_3^2,m_4^2,m_2^2,t_i,s,0,m_1^2,M_2^2,m_2^2 \big)$ \cite{Denner:2010tr}. The full enhanced box matrix element consisting of box diagrams, where different gauginos with mass $m_{\chi}$ are exchanged, can be written using the corresponding color decomposed tree-level matrix element (see Eq.~\eqref{eq:MtreeCol}) as
\begin{align}\nonumber
    M_{\mathrm{box}} &= \sum_{t_i}\sum_{\chi} M_{\mathrm{box}}^{\chi, t_i} = \\ \nonumber &\big(C^{[\overline{\mathbf 3}]}_{\mathrm{box}}\,M^{\mathrm{Tree},\chi, t_i}_{\overline{\mathbf 3}}\, C^{\{\overline{\mathbf 3},\overline{\mathbf 3}\}}_{stij} + C^{[{\mathbf 6}]}_{\mathrm{box}}\,M^{\mathrm{Tree},\chi, t_i}_{\mathbf 6}\, C^{\{\mathbf 6,\mathbf 6\}}_{stij}\big)\times\\
    & 2\frac{\alpha_s}{4\pi}(s-m_1^2-m_2^2)(t_i-m_{\chi}^2)\,D_0\,,
\end{align}
where the tensor integral has the arguments as in Eq.~\eqref{eq:D0ind} with $M_2=m_{\chi}$ and the color factors $C^{[{\mathbf R}]}_{\mathrm{box}}$ are given as
\begin{equation}
    C^{[\overline{\mathbf 3}]}_{\mathrm{box}} = -\frac{N_c+1}{2N_c}\,,\qquad\qquad C^{[{\mathbf 6}]}_{\mathrm{box}} = \frac{N_c-1}{2N_c}\,.
\end{equation}
The scalar integral for the specific arguments from Eq.~\eqref{eq:D0ind} can be written as
\begin{align}\nonumber
    D_0 &= \frac{x_{13}}{m_1m_2 (t_i-m_{\chi}^2)(1-x_{13}^2)} \times \\ \nonumber &\left\{2\ln(x_{13})\left[-\frac{c_\epsilon}{\epsilon}-\ln\left(\frac{\mu m_{\chi}}{m_{\chi}^2-t_i}\right)+\ln(1-x_{13}^2)\right]\right.\\ \nonumber
    &\left.+\ln^2(x_{12})+\ln^2(x_{23})+\mathrm{Li}_{2}(x_{13}^2)\right. \\
    &\left.+\sum_{k,l=\pm1}\mathcal{L}i_{2}(x_{13},x_{12}^k,x_{23}^l)-\frac{\pi^2}{6} \right\}\,.
\label{eq:D0exp}
\end{align}
The generalized polylogarithm in Eq.\ \eqref{eq:D0exp} is defined as \cite{Denner:2010tr}
\begin{widetext}
\begin{align}\nonumber 
    \mathcal{L}i_2 (x_1,\dots,x_n) =&\ \mathrm{Li}_{2}\left(1-\prod \limits_{i=1}^{n} x_i\right) + \left[\ln\left(\prod \limits_{i=1}^{n} x_i\right)-\sum_{i=1}^n\ln(x_i)\right]\left[\ln\left(1-\prod \limits_{i=1}^{n} x_i\right)
    -\theta\left(\left|\prod \limits_{i=1}^{n} x_i\right|-1\right)\times \right.
    \\ &\left.\left(\ln\left(-\prod \limits_{i=1}^{n} x_i\right)-\frac12 \ln\left(\prod \limits_{i=1}^{n} x_i\right)-\frac12 \sum_{i=1}^n\ln(x_i)\right)
    \right]\,.
\end{align}
\end{widetext}
The variables $x_{ij}$ are defined using the loop masses $M_i$ and $M_j$ as well as the invariant combinations of 4-momenta $p^2_{ij}$, as given in Eq.\ \eqref{eq:D0ind}, as
\begin{equation}
    x_{ij}= \frac{\sqrt{1-\frac{4M_iM_j}{p_{ij}^2-(M_i-M_j)^2}}-1}{\sqrt{1-\frac{4M_iM_j}{p_{ij}^2-(M_i-M_j)^2}}+1} \,.
    \label{eq:xij} 
\end{equation}
In our case the only velocity dependent $x_{ij}$ is $x_{13}$ which for $M_1=m_1$, $M_2=m_{\chi}$ and $M_3=m_2$ gives
\begin{align}
\begin{split}
    x_{13} &= -\frac{2(s-m_1^2-m_2^2)-2\sqrt{\lambda(s,m_1^2,m_2^2)}}{4m_1m_2}\\
&= -\frac{(1-v_{\mathrm{rel}})}{\sqrt{1-v_{\mathrm{rel}}^2}}\,,
\end{split}
\end{align}
having used the relative velocity as defined in Eq.~\eqref{eq:vrelrel}. Given the color factors $C^{[{\mathbf R}]}_{\mathrm{box}}$ are independent of the exchanged gaugino, we can simplify the enhanced matrix element as
\begin{align}\nonumber
    2\Re\big(M_{\mathrm{box}}M^\dag_{\mathrm{Tree}}&\big) =  \sum_{t_i,t_j}\sum_{\chi,\chi'} 2\Re\big(M_{\mathrm{box}}^{\chi,t_i}\big(M^{\chi', t_j}_{\mathrm{Tree}}\big)^{\dag}\big)  \\ \label{eq:boxsub} = \sum_{t_i,t_j}\sum_{\chi,\chi'} \Big(C^{[\overline{\mathbf 3}]}_{\mathrm{box}}\,&\big(M^{\chi,t_i}_{\mathrm{Tree},\overline{\mathbf 3}}\big(M^{\chi',t_j}_{\mathrm{Tree},\overline{\mathbf 3}}\big)^{\dag}\big) \\ \nonumber & + C^{[{\mathbf 6}]}_{\mathrm{box}}\,\big(M^{\chi,t_i}_{\mathrm{Tree},{\mathbf 6}}\big(M^{\chi',t_j}_{\mathrm{Tree},{\mathbf 6}}\big)^{\dag}\big)\Big) 2\Re (F^{\chi,t_i}_{\mathrm{box}})\,,
\end{align}
where $F^{\chi,t_i}_{\mathrm{box}}$ is given by
\begin{eqnarray}
    F^{\chi,t_i}_{\mathrm{box}} &=& \frac{\alpha_s}{\pi}\frac{(s-m_1^2-m_2^2)}{2m_1m_2}\,\frac{x_{13}}{(1-x_{13}^2)} \times \\ \nonumber &&\left\{2\ln(x_{13})\left[-\frac{c_\epsilon}{\epsilon}-\ln\left(\frac{\mu m_{\chi}}{m_{\chi}^2-t_i}\right)+\ln(1-x_{13}^2)\right]\right.\\ \nonumber &&\left.+\ln^2(x_{12})+\ln^2(x_{23})+\mathrm{Li}_{2}(x_{13}^2)\right. \\ \nonumber &&\left.+\sum_{k,l=\pm1}\mathcal{L}i_{2}(x_{13},x_{12}^k,x_{23}^l)-\frac{\pi^2}{6} \right\}\,.
\end{eqnarray}
We first expand the expression in the relative velocity $v_{\mathrm{rel}}$ retaining just the leading term. The pre-factor can be expressed in terms of the relative velocity using
\begin{equation}
    \frac{(s-m_1^2-m_2^2)}{2m_1m_2}\,\frac{x_{13}}{(1-x_{13}^2)} = -\frac{1}{2v_{\mathrm{rel}}}\,.
\end{equation}
Taking the real part of the $F^{\chi,t_i}_{\mathrm{box}}$ factor results in
\begin{align}\nonumber
    2\Re (F^{\chi,t_i}_{\mathrm{box}}) &= -\frac{\alpha_s}{\pi}\frac{1}{v_{\mathrm{rel}}}\   \Re\bigg\{\ln^2(x_{12})+\ln^2(x_{23})+\mathrm{Li}_{2}(1)\\  &+\sum_{k,l=\pm1}\mathcal{L}i_{2}(-1,x_{12}^k,x_{23}^l)-\frac{\pi^2}{6} \bigg\}\,.
\end{align} 
This expression seems to be implicitly dependent on the mass of the gaugino $m_{\chi}$ through the variables $x_{12}$ and $x_{23}$. However, this dependence vanishes after a more careful analysis, making the factor $2\Re (F^{\chi,t_i}_{\mathrm{box}})$ generic for all underlying hard processes.

We will show the universality explicitly for a simple case where $m_1=m_2$ (i.e. $x_{12} = x_{23}$) and where all $x_{ij}$ are purely real. In such a case $\mathcal{L}i_{2}(-1,x_{12}^k,x_{23}^l)$ reduces to a simple polylogarithm $\mathrm{Li}_{2}\left(1+x_{12}^{k+l}\right)$. The factor then reduces to 
\begin{align}\nonumber
    2\Re (F^{\chi,t_i}_{\mathrm{box}}) &= -\frac{\alpha_s}{\pi}\frac{1}{v_{\mathrm{rel}}}\  \Re\bigg\{2\ln^2(x_{12})+2\,\mathrm{Li}_{2}(2)\\  &+\,\mathrm{Li}_{2}\big(1+x_{12}^2\big)+\mathrm{Li}_{2}\big(1+x_{12}^{-2}\big)\bigg\}\,,
\end{align} 
which can be simplified using \cite{Lewin:1981}
\begin{align}
    \mathrm{Li}_{2}(z) = -\mathrm{Li}_{2}\left(\frac{z}{z-1}\right)-\frac12\ln^2(1-z)\quad z<1 \,.
\end{align}
The use of this identity eliminates all dependence on $x_{12}$ and the factor greatly simplifies to
\begin{equation}
    2\Re (F^{\chi,t_i}_{\mathrm{box}}) = -\frac{\alpha_s}{\pi}\frac{1}{v_{\mathrm{rel}}}\pi^2\,.
\end{equation}
Even though we have derived this particularly simple result in a special case, the same can be obtained in the most general case as well. With such a universal factor
the one-loop contribution to the squared matrix element from the enhanced box contribution is just
\begin{align}
    2\Re\big(M_{\mathrm{box}}M^\dag_{\mathrm{Tree}}\big) = \left(-\frac{\alpha_s}{v_{\mathrm{rel}}}\pi\right)\,\big(C^{[\overline{\mathbf 3}]}_{\mathrm{box}}&\,|M^{\mathrm{Tree}}_{\overline{\mathbf 3}}|^2 \\ \nonumber &+ C^{[{\mathbf 6}]}_{\mathrm{box}}\,|M^{\mathrm{Tree}}_{\mathbf 6}|^2\big)\,.
\end{align}
This expression is compatible with the next-to-leading part of the Sommerfeld enhancement after we realize that in the non-relativistic case the relative velocity can be easily related to the velocity used in the Sommerfeld enhancement for identical incoming particles (see Eq.\ \eqref{eq:Somparam2}) as $v_{\mathrm{rel}} = 2v_{\rm s}$.
\label{App}


\end{document}